\documentclass[aps,prd,twocolumn,showpacs,nofootinbib,floatfix]{revtex4}

\usepackage{amsmath,amssymb,amsfonts}
\usepackage{graphicx,epsfig,rotate,psfig}
\usepackage{dcolumn}
\usepackage{bm}
\newcommand{\dd}{\mathrm{d}}
\newcommand{\HH}{\mathcal{H}}
\newcommand{\cav}{\mathrm{cav}}
\newcommand{\eff}{\mathrm{eff}}
\newcommand{\ppn}{\mathrm{PPN}}
\newcommand{\nuc}{\mathrm{nuc}}
\newcommand{\deff}{\delta_\eff}
\newcommand{\deltaN}{\delta^{\rm N}}
\newcommand{\mat}{\mathrm{mat}}
\begin{document}
%%%%%%%%%%%%%%%%%%%%%%%%%%%%%%%%%%%%%%%%%%%%%%%%%%%%%%%%%%%%%%%%%%%%%%%%%%%%%%%%%%%%%%%%
\title{Weak lensing in scalar-tensor theories of gravity}

\author{Carlo Schimd}
 \email{schimd@iap.fr}
 \affiliation{Institut d'Astrophysique de Paris,
              GR$\varepsilon$CO, 98 bis bd Arago, 75014 Paris (France),\\
              and\\
              Dipartimento di Fisica, Universit\`a di Parma\\INFN-Gruppo Collegato di Parma
              \\Parco Area delle Scienze, 7/A - 43100 Parma (Italy)}
\author{Jean-Philippe Uzan}
 \email{uzan@iap.fr}
 \affiliation{Laboratoire de Physique Th\'eorique, CNRS-UMR 8627,
              B\^at 210,\\ Universit\'e Paris XI, 91415 Orsay cedex (France) \\
              and\\
              Institut d'Astrophysique de Paris, GR$\varepsilon$CO, 98 bis bd Arago, 75014 Paris (France).}
\author{Alain Riazuelo}
 \email{riazuelo@iap.fr}
 \affiliation{Institut d'Astrophysique de Paris,
              98 bis bd Arago, 75014 Paris (France).}
\date{December 6, 2004}
%%%%%%%%%%%%%%%%%%%%%%%%%%%%%%%%%%%%%%%%%%%%%%%%%%%%%%%%%%%%%%%%%%%%%%%%%%%%%%%%%%%%%%%%%

\begin{abstract}
This article investigates the signatures of various models of dark
energy on weak gravitational lensing, including the
complementarity of the linear and non-linear regimes. It
investigates quintessence models and their extension to
scalar-tensor gravity. The various effects induced by this
simplest extension of general relativity are discussed. It is
shown that, given the constraints in the Solar System, models such
as a quadratic nonminimal coupling do not leave any signatures
that can be detected while other models, such as a runaway
dilaton, which include attraction toward general relativity can
let an imprint of about 10\%.
\end{abstract}

 \maketitle
%%%%%%%%%%%%%%%%%%%%%%%%%%%%%%%%%%%%%%%%%%%%%%%%%%%%%%%%%%%%%%%%%%%%%%%%%%%%%%%%%%%%%%%%%
\section{Introduction}\label{sec1}

The growing evidences for the acceleration of the universe have led to
the formulation of numerous scenarios of {\it dark energy} (see
Ref.~\cite{peebles03} for reviews). In this construction various
routes have been investigated~\cite{uzan04}. They reduce mainly to the
introduction of new degrees of freedom in the cosmological scenario,
either as matter fields with negative pressure or in the gravitational
sector.

The simplest models accounting for this acceleration rely on the
introduction of a slow rolling scalar field whose expectation
value varies during the history of the universe. These
quintessence models~\cite{quintessence} are characterized by their
potential and numerous choices have been discussed in the
literature~\cite{peebles03}. Generically this scalar field may
also couple to matter fields~\cite{wette03}. As a consequence, the
values of the fundamental constants will depend on the value of
this field and may vary~\cite{urmp} and it will induce a fifth
force that will be long range if the scalar field is light, which
is usually the case if it is the source of the acceleration of the
universe.

In the simplest extension of the quintessence models, the
quintessence field universally couples to the other fields. It was
realized that the properties of this quintessence field were
conserved in that situation~\cite{uzan99,chibamen,cmb2}. This
implies that we are dealing with a scalar-tensor theory of gravity
in which the spin-0 partner of the graviton is also the
quintessence field. From a phenomenological point of view,
extended quintessence models are the simplest well defined
theories in which there is a modification of gravity. Various
observational signatures on the background evolution, the cosmic
microwave background (CMB)
anisotropies~\cite{ru02,cmb1,cmb2,cmb3,cmb4}, the big-bang
nucleosynthesis have been worked out~\cite{bp} and the reconstruction
problem was discussed in details~\cite{gilles01}. In particular,
it was shown that the attraction mechanism toward general
relativity~\cite{dn1} still hold~\cite{dp} in these extended
quintessence models, a crucial point since they have to satisfy
sharp constraints in the Solar System~\cite{will}. It was also
pointed out that a runaway dilaton~\cite{runaway1,runaway2} that
does not couple universally, is an appealing models with specific
signature such as a variation of some fundamental constants and a
violation of the universality of free fall.

Scalar-tensor theories are the most natural alternative to general
relativity~\cite{will,def}, preserving the universality of free
fall and constancy of all non gravitational constants. Gravity is
mediated not only by a massless spin-2 graviton, corresponding to
the spacetime metric, but also by a spin-0 scalar field. Many
theoretical motivations to consider such a scalar partner to the
graviton have been put forward, particularly in higher-dimensional
theories. In string theory, the supermultiplet of the graviton
contains a dilaton and moduli fields~\cite{polchinski}.

Scalar-tensor theories are well constrained in the Solar system
and one can try to extend these constraints to astrophysical and
cosmological scales. In the case of extended quintessence, the
effect of the scalar field is important particularly in the recent
universe when it starts to dominate. It was pointed
out~\cite{berben} that the modification of the equation of state
in the recent universe has striking effects on the growth of
density perturbation and on weak lensing observables such as the
convergence power spectrum.

Weak gravitational lensing has now proven to be a powerful tool to
study large scale structures~\cite{schneider92,mellier,bartelmann01}
and to gather information on the nature of the dark
energy~\cite{ludoben,simpsonbridle}. In particular, cross-correlation
techniques seem a very promising way to achieve this task (see
e.g.~\cite{cc}). Weak lensing can be detected by the deformation of
the shape of background galaxies. It was recently observed by various
groups~\cite{wldetect} and can probe the large scale structures of the
universe both in the linear and non-linear regimes. Recently, the weak
lensing has been studied in the context of generalized
cosmologies~\cite{acqua04}, specializing to quadratic non-minimally
coupled models and focusing on the effects on the CMB anisotropies,
hence just looking after the linear regime of structure
formation. However, as was shown on various
examples~\cite{berben,ludoben,tereno04}, the most stringent
constraints on cosmological models, able to distinguish dark energy
models, arise from the comparison of the linear and non-linear
regimes.

In this article, our main goal is to study in details the lensing
observable, focusing on 2-point functions to start with, in the
case of general relativity and scalar-tensor theories of gravity.
In particular, most of the previous studies (e.g.
Refs.~\cite{berben,ludoben,tereno04}) treats the matter power
spectrum independently of the CMB anisotropies. We choose to
normalize all our spectra on the CMB at low multipole and deduce
the lensing observables, both in the linear and non-linear
regimes, with the same normalization. Independently of any
considerations on the law of gravity, this tool will be of first
importance to deal with combined analysis of CMB and lensing
data~\cite{siprop}.

Our analysis will be applied to both quintessence and extended
quintessence models. We will show that the modification of the
equation of state of the universe leaves a detectable imprint on weak
lensing observables. Concerning scalar-tensor theories we will show
that, given the constraints in the Solar System, many models will let
very little signatures. This is the case for instance of a quadratic
coupling. Interestingly other models such as a runaway dilaton can
have a 10\% effect at that is likely to be constrained and/or detected
(see Fig.~\ref{fig9}). Note that both conclusions are of interest
since it will tell us for which class of modifications we have to
bother about this extension of the law of gravity.

The article is organized as follows. We first introduce
scalar-tensor theories of gravity in \S~\ref{sec1bis}. In
particular, we describe the Einstein and Jordan frames and argue
that the latter is the one in which observations take their
standard interpretations. We also recall (\S~\ref{IIB}-\ref{IIC})
the standard constraints on these models and discuss briefly the
properties of gravitational lensing (\S~\ref{IID}). In
\S~\ref{sec2}, we derive the distortion of a geodesic light bundle
due to large scale structures in a way that is valid for any
metric theory of gravity, in particular it does not assume that
gravity is described by general relativity and it is valid for
scalar-tensor theories. This allows us to define in \S~\ref{sec5}
the weak lensing observables such as the shear power spectrum and
the 2-point statistics of the shear field. We then describe our
numerical implementation (\S~\ref{subsec3b}) in details, as well
as the mapping to the non-linear regime. As a first check, we
investigate in \S~\ref{sec3} weak lensing in general relativity
both for a $\Lambda$CDM and quintessence models. Sec.~\ref{sec4}
discusses the various effects and differences that arise in
scalar-tensor theories of gravity and we then investigate in
\S~\ref{sec6} two families of models: a non-minimally coupled
scalar field and runaway dilaton-like models that include
attraction toward general relativity.

%----------------------------------------------------------------------------------------
\section{General results on scalar-tensor theories}\label{sec1bis}

In this article, we focus on scalar-tensor theories of gravity
described by the action
\begin{eqnarray}\label{action}
  S &=&\frac{1}{16\pi G_*}\int \dd^4 x \sqrt{-g}
     \left[F(\varphi)R-g^{\mu\nu}\varphi_{,\mu}\varphi_{,\nu} -
     2U(\varphi)\right]\nonumber\\ &&   \qquad\qquad +
     S_m[g_{\mu\nu};\mathrm{matter\; fields}]
\end{eqnarray}
where $G_*$ is the bare gravitational constant from which we
define $\kappa_*=8\pi G_*$. The coupling function $F$ multiplying
the Ricci scalar $R$ is dimensionless and needs to be positive to
ensure that the graviton carries positive energy. $S_m$ is the
action of the matter fields that are coupled minimally to the
metric, hence ensuring the universality of free fall. The metric
has signature $(-,+,+,+)$ and we work in units in which $c=1$.

\subsection{Field equations}

The variation of action~(\ref{action}) leads to the field
equations
\begin{eqnarray}
 F(\varphi)G_{\mu\nu} &=& 8\pi G_* T_{\mu\nu} \nonumber\\
       && + \partial_\mu\varphi\partial_\nu\varphi
          -\frac{1}{2}g_{\mu\nu}\left(\partial_\alpha\varphi\right)^2
          -g_{\mu\nu} U(\varphi)\label{fieleq1}
          \nonumber\\
       && + \nabla_\mu \partial_\nu F(\varphi)
          - g_{\mu\nu} \Box F(\varphi)\label{fieleq2}\\
 \Box\varphi &=& U_{,\varphi} - \frac{1}{2}F_{,\varphi} R\label{fieleq3}\\
 \nabla_\mu T^{\mu\nu} &=& 0.\label{fieleq4}
\end{eqnarray}
Here $\nabla_\mu$ is the covariant derivative associated to
$g_{\mu\nu}$ and the subscript ``${,\varphi}$" denotes the
functional derivative with respect to $\varphi$. The stress-energy
tensor is defined as
$$
T^{\mu\nu} \equiv \frac{2}{\sqrt{-g}}\frac{\delta S_m}{\delta g_{\mu\nu}}.
$$
Action (\ref{action}) has been written in the so-called {\it
Jordan frame} in which matter is universally coupled to the
metric. The Jordan metric defines the length and time as measured
by laboratory apparatus. In the following, we will be interested
in particular in the shape of background galaxies and all
observations will have their standard interpretation in this
frame~\cite{def}.

It is however useful to define an Einstein frame action through a
conformal transformation of the metric
\begin{equation}\label{jf_to_ef}
 g_{\mu\nu}^* = F(\varphi)g_{\mu\nu}.
\end{equation}
In the following all quantities labelled by a star (*) will refer
to Einstein frame. Defining the field $\varphi_*$ and the two
functions $A(\varphi_*)$ and $V(\varphi_*)$ (see e.g.~\cite{gilles01}) by
\begin{eqnarray}
 \left(\frac{\dd\varphi_*}{\dd\varphi}\right)^2
              &=& \frac{3}{4}\left(\frac{\dd\ln F(\varphi)}{\dd\varphi}\right)^2
                  +\frac{1}{2F(\varphi)}\label{jf_to_ef1}\\
 A(\varphi_*) &=& F^{-1/2}(\varphi)\label{jf_to_ef2}\\
 2V(\varphi_*)&=& U(\varphi) F^{-2}(\varphi)\label{jf_to_ef3},
\end{eqnarray}
the action (\ref{action}) reads as
\begin{eqnarray}
 S &=& \frac{1}{16\pi G_*}\int \dd^4x\sqrt{-g_*}\left[ R_*
        -2g_*^{\mu\nu} \partial_\mu\varphi_*\partial_\nu\varphi_*
        - 4V(\varphi_*)\right]\nonumber\\
   && \qquad + S_m[A^2(\varphi_*)g^*_{\mu\nu};\psi].
\end{eqnarray}
The kinetic terms have been diagonalized so that the spin-2
and spin-0 degrees of freedom of the theory are perturbations of
$g^*_{\mu\nu}$ and $\varphi_*$ respectively.  In this frame, the
field equations take the form
\begin{eqnarray}
 G^*_{\mu\nu} &=& 8\pi G_* T^*_{\mu\nu} \nonumber\\
       &+&  2\partial_\mu\varphi_*\partial_\nu\varphi_*
          - g^*_{\mu\nu}\left(\partial_\alpha\varphi_*\right)^2
          - 2g^*_{\mu\nu}V\label{einframe}\\
 \Box_*\varphi_* &=& V_{,\varphi_*} - 4\pi G_* \alpha(\varphi_*)
            T^*_{\mu\nu}g_*^{\mu\nu}\\
 \nabla_\mu T^{\mu\nu}_* &=& \alpha(\varphi_*)
                 T^*_{\sigma\rho}g_*^{\sigma\rho} \partial_\nu\varphi_*
\end{eqnarray}
where we have defined the Einstein frame stress-energy tensor
$$
 T^{\mu\nu}_* \equiv \frac{2}{\sqrt{-g_*}}\frac{\delta S_m}{\delta g^*_{\mu\nu}},
$$
related to the Jordan frame stress-energy tensor by
$T_{\mu\nu}^*=A^2T_{\mu\nu}$. The function
\begin{equation}\label{eqalpha}
 \alpha(\varphi_*)\equiv \frac{\dd\ln A}{\dd\varphi_*}.
\end{equation}
characterizes the coupling of the scalar field to matter (we
recover general relativity with a scalar field when it vanishes).
For completeness, we also define
\begin{equation}\label{eqbeta}
 \beta(\varphi_*)\equiv \frac{\dd\alpha}{\dd\varphi_*}.
\end{equation}
Note that in Einstein frame the Einstein equations
(\ref{einframe}) are the same as the ones obtained in general
relativity with a minimally coupled scalar field. Table~\ref{def:EJf}
summarize the notations used in Jordan and Einstein frames.

From these definitions, we can define an effective gravitational
constant as
\begin{equation}\label{geff}
 G_{\rm eff} = \frac{G_*}{F(\varphi)} = G_*A^2(\varphi_*)
\end{equation}
but, as we shall see, this does not correspond to the Newton
constant that will be effectively measured in a Cavendish-like
experiment.

\begin{table}
\caption{Notations in Jordan and Einstein frames.}
\begin{ruledtabular}
\begin{tabular}{lcc}
                 & Jordan frame & Einstein frame \\
  \hline
  coordinates    & $(x,t$)      & $(x_*,t_*)$ \\
  metric         & $g_{\mu\nu}$ & $g_{\mu\nu}^*$ \\
  scalar field   & $\varphi$    & $\varphi_*$ \\
  potential      & $U$            & $V$ \\
  coupling       & $F$            & $A$ \\
\end{tabular}
\end{ruledtabular}\label{def:EJf}
\end{table}

\subsection{Local constraints}\label{IIB}

Deviations of the theory of gravity from general relativity are
sharply constrained in the Solar System~\cite{will} as well as by
binary pulsars~\cite{gilles03}. The constraints are usually set on
the post-Newtonian parameters~\cite{will}. In the particular case
of scalar-tensor theories, they reduce~\cite{gilles01,def} to
\begin{eqnarray}
 \gamma^\ppn - 1 &=& \frac{F_{,\varphi}^{2}}{F + 2F_{,\varphi}^{2}} \\
            &=& -2\frac{\alpha^2}{1+\alpha^2} \label{gppn}\\
 \beta^\ppn - 1  &=& \frac{1}{2}\frac{\alpha^2}{(1+\alpha^2)^2}\frac{\dd\alpha}{\dd\varphi_*}\\
            &=& \frac{1}{4}\frac{FF_{,\varphi}}{2F + 3F_{,\varphi}^{2}}
                 \frac{\dd\gamma^\ppn}{\dd\varphi}\label{bppn}.
\end{eqnarray}
Solar System experiments set sharp constraints on the values of
the PPN parameters $(\gamma^\ppn_0,\beta^\ppn_0)$ today. The
observed value of the perihelion shift of Mercury implies the
bound~\cite{mercury}
\begin{equation}\label{boundmercury}
 \left|2\gamma^\ppn_0-\beta^\ppn_0-1\right|<3\times10^{-3}.
\end{equation}
The Lunar laser ranging experiment~\cite{llr} imposes
\begin{equation}\label{boundllr}
 4\beta^\ppn_0-\gamma^\ppn_0-3=-(0.7\pm1)\times10^{-3}
\end{equation}
and the measurements of the light deflection by Very Long Baseline
Interferometry~\cite{vlbi} improves the constraint on
$\gamma^\ppn_0$ to
\begin{equation}\label{boundvlbi}
 \left|\gamma^\ppn_0-1\right|=4\times10^{-4}.
\end{equation}
The recent analysis of the frequency shift of radio waves to and
from the Cassini spacecraft have set the even more stringent
bound~\cite{bertotti}
\begin{equation}\label{boundgamma}
 \gamma^\ppn_0-1=(2.1\pm2.3)\times10^{-5}.
\end{equation}
The previous bounds to those obtained in Refs.~\cite{vlbi,bertoti}
were (see Ref.~\cite{will} for a review)
\begin{equation}\label{boundold}
 \left|\gamma^\ppn_0-1\right| \leq2\times10^{-3},
 \qquad
  \left|\beta^\ppn_0-1\right| \leq6\times10^{-4}.
\end{equation}
These constraints can be transformed to constraints on the set
$(\alpha_0,\beta_0)$ (see Ref.~\cite{gilles03} for a summary). In
particular, we can note that binary pulsars imply that
$\beta_0>-4.5$. The bound (\ref{boundgamma}) implies that
\begin{equation}\label{systsolcon}
 \alpha_0^2 \sim \frac{F_0^{'2}}{F_0}<4\times10^{-5}.
\end{equation}

The local value of the gravitational constant is deduced from a
Cavendish-like experiment, i.e. from the measurement of the Newton
force between to masses $m_1$ and $m_2$, as $G_\cav = F
r^2/m_1m_2$. It can be shown~\cite{def} that its theoretical
expression is
\begin{eqnarray}\label{Gcav}
 G_\cav &=& \frac{G_*}{F}\left( 1 +
          \frac{F_{,\varphi}^2}{2F + 3F_{,\varphi}^2}\right)\\
        &=& G_* A^2(1+\alpha^2).
\end{eqnarray}
Note that today Eq.~(\ref{systsolcon}) implies that $G_\cav$ and
$G_\eff$ do not differ by more than some $10^{-3}$ percent, which
is anyhow larger than the accuracy of the measurement of the
gravitational constant. At higher redshift this difference may be
larger.

Current constraints~\cite{dickey} (see also Ref.~\cite{urmp} for
a review) imply that
\begin{equation}
 \left|\frac{\dot G_\cav}{G_\cav}\right|_0<6\times10^{-12}\,{\rm
 yr}^{-1}.
\end{equation}
Using Eqs.~(\ref{eqalpha}) and (\ref{eqbeta}), this implies that
\begin{equation}
 \left|\alpha_0 + \frac{\beta_0\alpha_0}{(1+\alpha_0^2)}\right|
 \left|\dot\varphi_{*0}\right|
 <3\times10^{-12}\,{\rm yr}^{-1}
\end{equation}

All these constraints are local and we will consider in this
article only models that satisfy them.

\subsection{Cosmological constraints}\label{IIC}

From a cosmological point of view, there are very few constraints
on scalar-tensor theories.

CMB observations may in principle give some constraints but they
are often degenerate with other parameters, such as e.g. the
cosmological parameters (see e.g.
Ref.~\cite{ru02,cmb1,cmb2,cmb3,cmb4} for some studies on the CMB
imprints of scalar tensor theories).

More stringent constraints arise from Big Bang nucleosynthesis
(BBN). BBN results, and in particular the helium-4 abundance, are
very sensitive to the weak interaction freeze-out. This
temperature depends on the strength of the gravitational
interaction which dictates the expansion rates. BBN mainly
requires that (i) the universe is dominated by radiation at the
time of nucleosynthesis and (ii) that the number of degrees of
freedom of relativistic particles does not vary by more than 20\%
with respect to its expected value $g_{\rm r}=10.75$. The
Friedmann equation (\ref{a6}) for a flat universe reduces to
\begin{equation}
 {\cal H}^2 = \kappa_0\frac{\pi^2}{90}a^2g_{\rm r}\left(1+\frac{\delta g_{\rm r}}{g_{\rm r}}\right)T^4
\end{equation}
with $\delta g_{\rm r}/g_{\rm r}=F_0/F(z_{\rm nuc}) -1$, if one
neglects the contribution of the scalar field to the energy
density. Requiring that $\delta g_{\rm r}/g_{\rm r}$ be smaller
than 0.2 implies~\cite{bp} that
\begin{equation}\label{bbnbound}
 0.8\leq
 \left|\frac{F_0}{F_\nuc}\right| =
 \left|\frac{A^2_\nuc}{A^2_0}\right|\leq1.2.
\end{equation}
Let us emphasize that large value of $|{F_0}/{F_\nuc}|$ were shown
to be consistent with the BBN constraints~\cite{dp} if $\beta$ is
large enough so that the naive limit~(\ref{bbnbound}) can be much
more stringent than a detailed study may show.

\subsection{A remark on gravitational lensing}\label{IID}

Gravitational bending of light by a single mass $M$ in
scalar-tensor theory can be derived easily in Einstein frame.
Since photons are coupled to the gravitational field only, they
are insensitive to the scalar field. One deduces that they are
deflected by an angle $\Delta\theta$ exactly in the same way as in
general relativity, that is
\begin{equation}
 \Delta\theta = 4 G_* A^2(\varphi_*)M/r_0,
\end{equation}
where $A^2M$ is the deflecting mass in Einstein frame. $G_*$
cannot be measured directly and what we know is only the value of
the gravitational constant (\ref{Gcav}) determined by a
Cavendish-like experiment today. Masses attract themselves due to
the exchange of both a graviton and a scalar. It follows that
\begin{equation}
 G_\cav  = G_* A^2_0 (1+\alpha_0^2)
         =  2 \frac{G_* A^2_0}{1+\gamma_0}
\end{equation}
so that the deflection angle turns out to be
\begin{equation}
 \Delta\theta = 2(1+\gamma_0)G_\cav M/r_0.
\end{equation}
In conclusion, the deflected angle is different from the
prediction of general relativity, not because photons are
deflected differently but because there is a  scalar interaction
between the test masses which contributes to the value of the
effective Newton constant~\cite{def}.

This argument can be straightforwardly generalized to the
cosmological context. If the lens seats at a redshift $z$ and if
we work in the thin lens approximation then the deflection angle
is given by
\begin{equation}
 \Delta\theta = 4 G_* A^2(z_{\rm lens})M_{\rm lens}/r_0.
\end{equation}
again, $G_*$ cannot be measured and we have access to
\begin{equation}
 G_{\cav,0}  =  2 \frac{G_* A^2_0}{1+\gamma_0}
\end{equation}
where the subscript 0 means that it is determined today. It
follows that
\begin{equation}\label{estim}
 \Delta\theta = 2(1+\gamma_0)G_{\cav,0} \frac{M_{\rm lens}}{r_0}\frac{A^2(z)}{A^2(0)}.
\end{equation}
The mass of the lensing galaxy, $M_{\rm lens}$, cannot be measured
directly.

We mainly expect to have two major effects on lensing
observations: (i) an effect of the background dynamics through the
angular distances, (ii) a gravitational effect due to the $z$
dependence deflection angle. Eventually, observations of lenses
located at different redshift should enable to give an information
on $A(z)/A(0)$ as a function of the redshift.

Before we go into a detailed computation, we can give an upper
bound on the amplitude of this effect. From the constraint
(\ref{bbnbound}), we deduce that the factor depending on $A$ in
Eq.~(\ref{estim}) has to lie in the range $F_0/F(z_\nuc)\simeq
[0.8,1.2]$ so that effects larger than 20\% are unlikely to
happen, independently of any model (see however Ref.~\cite{dp}).
Let us stress that, as we shall see, the dominant effect will
arise from the growth of the density field and not from the two
effects we have just mentioned.

\vskip0.5cm
To finish this general discussion let us emphasize that we
could have added a coupling of the scalar field to the
electromagnetic tensor of the form
\begin{equation}
 B(\varphi) F_{\mu\nu}F^{\mu\nu}
\end{equation}
in the action (\ref{action}). Such a term will induce a variation
of the fine structure constant (see e.g. Ref.~\cite{urmp} for a
review), that has to be smaller than $10^{-5}$ between $z\sim3$
and today, and a violation of the universality of free fall.
Interestingly, such a term will not affect the equation of
propagation of photons in the eikonal approximation at first order
and thus weak lensing observables that are considered in this
article.

%----------------------------------------------------------------------------
\section{Light propagation in a perturbed FLRW spacetime}\label{sec2}

This section is devoted to the general theory of gravitational
lensing. Our construction is based on the geodesic deviation of a
bundle of null geodesics and will then be valid as soon as light
travels on such null geodesics, which is the case in particular in
general relativity and scalar-tensor theories of gravity.
\S~\ref{subsec2a} reviews the standard derivation on the
propagation of a light bundle and defines the shear and
convergence. In \S~\ref{subsec2b} and \S~\ref{subsec2c}, we apply
this formalism to a Friedmann-Lema\^{\i}tre spacetime and then to
a perturbed spacetime in order to get the final expression of the
shear.

\subsection{General derivation}\label{subsec2a}

Let us consider the evolution of a light bundle in a spacetime
with metric $g_{\mu\nu}$ following the original work of
Ref.~\cite{sachs62} along the lines of Ref.~\cite{ub00}. The
worldline of each geodesic can be decomposed as
\begin{equation}
x^\mu(\lambda)=\bar x^\mu(\lambda)+ \xi^\mu(\lambda)
\end{equation}
where $\bar x^\mu(\lambda)$ is a fiducial null geodesic of the
bundle and $\lambda$ is an affine parameter along this geodesic.
$\xi^\mu$ is a displacement vector that labels the other geodesics
of the bundle. The tangent vector $k^\mu$ along the fiducial
geodesic satisfies
\begin{equation}\label{null_geo}
 k_\mu k^\mu = 0, \quad k^\nu\nabla_\nu k^\mu= 0.
\end{equation}

We assume that the geodesic bundle is converging at the observer's
position $O$. We assume $\lambda$ to vanish in $O$ and to increase
toward the past. In $O$ we also choose a quasi-orthonormal reference frame $\{
k^\mu, u^\mu, n_1^\mu, n_2^\mu \}$, where $u^\mu$ is the 4-velocity
of the observer and satisfies
\begin{equation}
  u^\mu u_\mu = -1
\end{equation}
and $n_1^\mu,n_2^\mu$ are two the spacelike vectors spanning the
plan orthogonal to the fiducial light ray, \emph{i.e.} to the line of
sight. They satisfy
\begin{equation}
 n_a^\mu n^b_\mu = \delta_a^b,
\quad
 n_a^\mu k_\mu = n_a^\mu u_\mu = 0
\end{equation}
with $a,b=1,2$. From this tetrad constructed in $O$, we construct a
basis at each point of the geodesic by parallel transporting it along
the fiducial geodesic.

As shown in Ref.~\cite{ub00},
$\xi^\mu$ can be decomposed as
\begin{equation}\label{decxi}
 \xi^\mu = \xi_0 k^\mu + \sum_{a=1,2}
 \xi_a n_a^\mu
\end{equation}
and one can always choose to set $\xi_0=0$. The propagation equation
of $\xi^\mu$ is obtained from the geodesic deviation equation as
\begin{equation}
 \frac{D^2}{\dd\lambda^2} \xi^\mu = R^\mu_{\;\;\nu\alpha\beta}k^\nu k^\alpha \xi^\beta,
\end{equation}
where $R^\mu_{\;\;\nu\alpha\beta}$ is the Riemann tensor and
${D}/{\dd\lambda}\equiv k^\mu\nabla_\mu$. In terms of the
decomposition (\ref{decxi}), this equation reads
\begin{equation}\label{eqn:tidal}
 \frac{\dd^2}{\dd\lambda^2}\bm{\xi} = \bm{\mathcal{R}}\bm{\xi},
\end{equation}
where $\mathcal{R}^b_{\;a} \equiv R^\mu_{\;\;\nu\alpha\beta}k^\nu
k^\alpha n_\mu^a n_b^\beta$ is known as the \emph{optical tidal
matrix}. We have used the notation $\bm{\xi}=\xi_a$ and
$\bm{\mathcal{R}}\bm{\xi} = \mathcal{R}^b_{\;\;a} \xi_b$.
Decomposing the Riemann tensor in terms of the Ricci and the Weyl
tensor, it can be rewritten as
\begin{equation}
 \mathcal{R}^b_{\;a} = - \frac{1}{2}R_{\mu\nu}k^\mu k^\nu \delta^a_b
           + C^\mu_{\;\;\nu\alpha\beta}k^\nu k^\alpha n_\mu^a n_b^\beta.
\end{equation}

The linearity of the geodesic equation implies that $\bm{\xi}$ can
be related to the initial value of $\dd\bm{\xi}/\dd\lambda$
through a linear transformation
\begin{equation}\label{eq10}
  \bm{\xi}(\lambda) = \bm{\mathcal{D}}(\lambda)\frac{\dd\bm{\xi}}{\dd\lambda}(0).
\end{equation}
Using that $\bm{\xi}(0)=0$ (bundle converging in $O$) ,
Eq.~(\ref{eqn:tidal}) gives an equation of evolution for
$\mathcal{D}^b_a$ as
\begin{equation}\label{eqn:master}
 \frac{\dd^2}{\dd\lambda^2}\bm{\mathcal{D}}
    = \bm{\mathcal{R}}\bm{\mathcal{D}}.
\end{equation}
The initial conditions in $0$ for the matrix $\bm{\mathcal{D}}$
are given by
\begin{equation}
 \bm{\mathcal{D}}(0) = 0
 \quad\mbox{and}\quad
 \bm{\mathcal{D}}^\prime(0) = \bm{\mathcal{I}},
\end{equation}
with $\bm{\mathcal{I}}$ being the $2\times 2$ identity matrix. The
matrix $\bm{\mathcal{D}}$ describes the deformation of the light
bundle induce by the spacetime geometry.

The direction of observation, $\bm{\theta}_I$, and
and the angular position of the source $\bm{\theta}_S$ are
related to the displacement field by
\begin{equation}
 \bm{\theta}_I = \frac{\dd\bm{\xi}}{\dd\lambda}(0),\qquad
 \bm{\theta}_S = \frac{\bm{\xi}(\lambda_S)}{D_A(\lambda_S)}
\end{equation}
where $\lambda_S$ is the value of the affine parameter at the source
and $D_A$ is, by definition, the angular distance of the source.
It follows that Eq.~(\ref{eq10}) takes the form
\begin{equation}
 \bm{\theta}_S = \frac{\bm{\mathcal{D}}(\lambda_S)}{D_A(\lambda_S)}\bm{\theta}_I.
\end{equation}
The deformation of the shape of background galaxies is thus
characterized by the {\it amplification matrix}
\begin{equation}
 \bm{\mathcal{A}} \equiv
   \frac{\dd \bm{\theta}_S}{\dd \bm{\theta}_I}
   =
   \frac{\bm{\mathcal{D}}(\lambda_S)}{D_A(\lambda_S)}.
\end{equation}
It can always be decomposed as
\begin{equation}\label{def:AmpliMatrix}
 \bm{\mathcal{A}}\equiv \begin{pmatrix} 1-\kappa-\gamma_1 & \gamma_2 \\
    \gamma_2 & 1-\kappa+\gamma_1 \end{pmatrix},
\end{equation}
in terms of the convergence $\kappa$ and of the shear
$\bm{\gamma}=(\gamma_1,\gamma_2)$. They can be extracted from the
amplification matrix as
\begin{equation}\label{gamkap}
 \kappa = 1 - \frac{1}{2}\,\mathrm{Tr}\,\bm{\mathcal{A}},\qquad \bm{\gamma}
 = \frac{1}{2}\begin{pmatrix} \mathcal{A}_{22}-\mathcal{A}_{11}\\
 2\mathcal{A}_{12} \end{pmatrix}.
\end{equation}
As was emphasized by many
authors~\cite{mellier,schneider92,bartelmann01} and demonstrated
experimentally by various observations~\cite{wldetect}, the shear
can be obtained from the measurement of the shape of galaxies.

All the definitions and derivations of this section do not depend on
the specific form of the metric and are thus completely general as
long as photons follow null geodesic and the geodesic deviation
equation holds.

\subsection{Background spacetime}\label{subsec2b}

We now apply this general results to the case of a background
cosmological spacetime with line element
\begin{equation}\label{metric1}
 \dd s^2 = g_{\mu\nu}\dd x^\mu \dd x^\nu =
           a^2(\eta)\bar g_{\mu\nu}\dd x^\mu \dd x^\nu
\end{equation}
where $\eta$ is the conformal time and $a$ the scale factor. We
decompose the metric as
\begin{equation}\label{metric2}
 \bar g_{\mu\nu}\dd x^\mu \dd x^\nu = -\dd\eta^2 + \gamma_{ij}\dd x^i\dd x^j
\end{equation}
where $\gamma_{ij}$ is the metric of the constant time
hypersurfaces. Since they are hypersurfaces of constant curvature
\begin{equation}\label{metric3}
 \gamma_{ij}\dd x^i\dd x^j = \dd\chi^2 + S_K^2(\chi) \dd\Omega^2
\end{equation}
where $\chi$ is the comoving radial coordinate and $\dd\Omega^2$ the
infinitesimal solid angle. The function $S_K$ is defined by
\begin{equation}\label{defsk}
  S_K(\chi) = \frac{\sin(\sqrt{K}\chi)}{\sqrt{K}}, \,
              \chi, \,
              \frac{\sinh(\sqrt{-K}\chi)}{\sqrt{-K}}
\end{equation}
respectively for $K>0, K=0, K<0$.

From Eq.~(\ref{metric3}), it is clear that $S_K$ is the comoving
angular distance so that the angular distance will be given by
\begin{equation}\label{dang}
 D_A = a(\eta)S_K(\chi)
     = a_0 \frac{S_K(\chi)}{1+z}
\end{equation}
where $a_0$ refers to the value of the scale factor today and $z=a_0/a - 1$
is the redshift. The expression of $\chi(z)$ is given by
\begin{equation}
 \chi(z) = \frac{1}{H_0a_0}\int_0^z\frac{\dd z}{E(z)}
\end{equation}
where $H_0$ is the value of the Hubble constant today and $E(z)$ is defined
by
\begin{equation}
 E(z)= H(z)/H_0 =(1+z){\cal H}/{\cal H}_0.
\end{equation}
The expression for $E(z)$ depends on the matter content of the
universe and on the theory of gravity. It is given in
Appendix~\ref{appb}.

To solve Eq.~(\ref{eqn:master}), we will use the fact that null
geodesics are left unchanged by a conformal transformation since
two conformal spaces have the same causal structure (see e.g.
Ref.~\cite{wald}). For the metric $\bar g_{\mu\nu}$,
Eq.~(\ref{eqn:master}) takes the form
\begin{equation}\label{eqn:master0}
 \frac{\dd^2}{\dd\lambda^2}\bm{\mathcal{D}}
    = -K\bm{\mathcal{D}}.
\end{equation}
The solution of this equation is trivially given by
\begin{equation}\label{sol:zeroth}
  \bm{\mathcal{D}}^{(0)}(\lambda) = S_K(\lambda)I,
\end{equation}
where $S_K$ is given by Eq.~(\ref{defsk}), so that
\begin{equation}\label{sol:Azeroth}
  \bm{\mathcal{A}}^{(0)} = \bm{\mathcal{I}}
\end{equation}
where the subscript 0 refers to the solutions in the background
spacetime.

\subsection{Perturbed spacetime}\label{subsec2c}

\subsubsection{Amplification matrix}

Let us now turn to the perturbed cosmological spacetime with a metric
\begin{eqnarray}
ds^2 &=& a^2(\eta)\left(\bar{g}_{\mu\nu} + h_{\mu\nu}\right)\dd x^\mu
\dd x^\nu.
\end{eqnarray}
Again, we work in the static conformal spacetime with metric
$\bar{g}_{\mu\nu} + h_{\mu\nu}$. As usual, we develop the tidal matrix
as
\begin{equation}
 \bm{\mathcal{D}} = \bm{\mathcal{D}}^{(0)} + \bm{\mathcal{D}}^{(1)} + \mathcal{O}
 \left(h^2\right),
\end{equation}
where $\bm{\mathcal{D}}^{(n)}$ involves terms of $n$-th order in
the metric perturbation, and consistently we solve
Eq.~(\ref{eqn:master}) order by order. Plugging the zeroth order
solution~(\ref{sol:Azeroth}), Eq.~(\ref{eqn:master}) reduces to
\begin{equation}\label{eqn:first}
   \frac{\dd}{\dd\lambda}\bm{\mathcal{D}}^{(1)}
     = \bm{\mathcal{R}}^{(1)}(\lambda)S_K(\lambda).
\end{equation}
the solution of which is explicitly given by
\begin{equation}
 \bm{\mathcal{D}}^{(1)}(\lambda) =
         \int_0^{\lambda}S_K(\lambda^\prime)
         S_K(\lambda - \lambda^\prime)
         \bm{\mathcal{R}}^{(1)}(\lambda^\prime)\dd\lambda^\prime,
\end{equation}
so that the amplification matrix is
\begin{equation}
 \bm{\mathcal{A}}^{(1)}(\lambda) =
         \int_0^{\lambda}
         \frac{S_K(\lambda^\prime)S_K(\lambda - \lambda^\prime)}{ S_K(\lambda)}
         \bm{\mathcal{R}}^{(1)}(\lambda^\prime)\dd\lambda^\prime.
\end{equation}
This expression is again very general and just assumes that
photons are propagating along null geodesics. Now, we just need to
express $\bm{\mathcal{R}}^{(1)}$ in terms of the perturbations of
the metric and go back to the original (non static) spacetime to
get the final result.

\subsubsection{Deflecting potential}

To go further, we assume that the perturbations can be
decomposed as
\begin{eqnarray}
 \frac{1}{2}h_{\mu\nu}\dd x^\mu \dd x^\nu &=&
  -\phi\dd\eta^2+ B_i\dd\eta\dd x^i\nonumber\\
  && +\left(-\psi\gamma_{ij} + \bar E_{ij}
  \right)\dd x^i\dd x^j
\end{eqnarray}
where $\phi$ and $\psi$ are the two gravitational potentials in
Newtonian gauge, $B_i$ and $\bar E_{ij}$ describe the vector and
tensor perturbations and satisfy
\begin{equation}
 \bar E_i^i = D_i\bar E^i_j = D_i B^i =0
\end{equation}
where $D_i$ is the covariant derivative associated to the spatial
metric $\gamma_{ij}$. Using that $2R_{\mu\nu\alpha\beta} =
h_{\nu\alpha,\mu\beta} + h_{\mu\beta,\nu\alpha} -
h_{\alpha\mu,\nu\beta} - h_{\nu\beta,\alpha\mu}$, we obtain that
\begin{equation}\label{eqn:Rab1}
  \mathcal{R}^{(1)}_{ab} =
       \frac{1}{2}h_{\mu\nu,\alpha\beta}k^\mu k^\nu n_a^\alpha n_b^\beta
       -\frac{1}{2}\frac{\dd}{\dd\lambda}\left(\Gamma_{\rho\beta}^\alpha
       \bar g_{\alpha\mu}k^\beta n_a^\mu n_b^\rho\right),
\end{equation}
where $\Gamma_{\rho\beta}^\alpha$ are the Christoffel symbols. If
we focus on scalar perturbations then the second term vanishes and
we end up with
\begin{equation}
 \mathcal{R}^{(1)}_{ab} = -\partial_{ab}\Phi
\end{equation}
where the {\it deflecting potential} is defined as
\begin{equation}
 \Phi = -\frac{1}{2}h_{\mu\nu}k^\mu k^\nu.
\end{equation}
In the case of a perturbed cosmological spacetime, it reduces
to
\begin{equation}
 \Phi = \phi + \psi - B_i \theta^i - \bar E_{ij} \theta^i \theta^j
\end{equation}
where $\theta^i$ is the direction of observation. This potential
includes the effect of rotation and of gravity waves. Notes that
in the case of pure scalar perturbation in general relativity and
in the absence of anisotropic stress $\phi=\psi$ and we recover
the standard result $\Phi=2\phi$.

\subsubsection{Final form}

The last step requires to go back to the original (non static)
metric to get
\begin{eqnarray}\label{Aabfin}
 \mathcal{A}_{ab} &=& I_{ab} \\
         &-&\int_0^{\chi}
         \frac{S_K(\chi^\prime)S_K(\chi - \chi^\prime)}{ S_K(\chi)}
         \partial_{ab}\Phi\left[S_K(\chi^\prime)\bm{\theta},\chi'\right]\dd\chi^\prime
         \nonumber.
\end{eqnarray}
It follows that the shear is given by by the trace of this matrix
\begin{eqnarray}\label{kappafinal}
 \kappa(\bm{\theta},\chi) &=& \frac{1}{2}\int_0^{\chi}
         \frac{S_K(\chi^\prime)S_K(\chi - \chi^\prime)}{ S_K(\chi)}
         \Delta_2\Phi\,\dd\chi^\prime,
\end{eqnarray}
with $\Delta_2\equiv \partial_a\partial^a$. The deflecting
potential is evaluated along the line of sight as $\Phi =
\Phi\left[S_K(\chi^\prime)\bm{\theta},\chi'\right]$.

In this derivation, the source has been assumed to be at a given
redshift, it generalizes easily to a more general redshift
distribution, $p_\chi(\chi)\dd\chi$, as
\begin{equation}
 \kappa(\bm{\theta}) = \int
 p_\chi(\chi)\kappa(\bm{\theta},\chi)\dd\chi.
\end{equation}
In conclusion, the shear takes the form
\begin{eqnarray}\label{kappa1}
 \kappa(\bm{\theta}) &=& \frac{1}{2}\int_0^{\chi_{\rm H}}
         g(\chi)S_K(\chi)
         \Delta_2\Phi\left[S_K(\chi)\bm{\theta},\chi\right]\dd\chi.
\end{eqnarray}
where $g$ is defined as
\begin{eqnarray}\label{kappa2}
 g(\chi) &=& \int_\chi^{\chi_{\rm H}}
         p(\chi')
         \frac{S_K(\chi'- \chi)}{ S_K(\chi')}.
\end{eqnarray}

\subsection{Comments}

Again, let us stress that this construction does not assume the
validity of general relativity. It is based only on the validity
of the geodesic deviation equation. This is the case in particular
for all metric theory of gravity such as scalar-tensor theories.

We also emphasize that using a conformal transformation to solve
the problem in a static spacetime has greatly simplified the
derivation.

In the late universe, when we are interested by gravitational
lensing effects, the matter content is mainly dominated by
pressureless matter and dark energy. It follows that we can define
an {\it effective density field}, $\deff$, by the relation
\begin{equation}\label{defdeff}
 \Delta{\Phi} = {3}H_0^2\Omega_{m,0}\frac{\deff}{(a/a_0)}
\end{equation}
where $\Omega_{m,0}$ is the value of the matter density parameter
today. The subsecript ``$m$" refers to the matter fields,
including at least two components, dark matter (cdm) and baryonic
matter (b). If we express the 3-dimensional Laplacian as $\Delta_2
+\partial^2_z$ where $\partial_z$ is the derivative along the line
of sight, then the convergence (\ref{kappa1}) can be expressed in
terms of the effective density contrast as
\begin{equation}\label{kappa3}
 \kappa(\bm{\theta}) = \frac{3}{2}H_0^2\Omega_{m,0}\int_0^{\chi_{\rm H}}
 g(\chi)S_K(\chi) \frac{\deff\left[S_K(\chi)\bm{\theta},\chi\right]}{(a/a_0)}\dd\chi.
\end{equation}

%----------------------------------------------------------------------------------------
\section{Weak lensing observables}\label{sec5}

In the former section, we have derived the expression of the
convergence in terms of the deflecting potential. In cosmology, we
are interested in the statistical properties of the convergence.

This section is devoted to the observable quantities to be
compared with weak lensing data. \S~\ref{subsec3a} describes the
computation of the power spectrum and \S~\ref{subsec3c} is de
voted to the computation of various 2-point statistics, such as
the shear and the variance of the aperture mass. This requires to
smooth the shear field. We then describe our numerical
implementation in \S~\ref{subsec3b}.

\subsection{Power spectra}\label{subsec3a0}

The convergence is a function on the 2-sphere and it can be
expanded in a 2-dimensional Fourier transform as
\begin{equation}
 \kappa(\bm{\theta}) = \int  \hat\kappa(\bm{\ell})
 \hbox{e}^{i\bm{\ell}\cdot\bm{\theta}}\,\frac{\dd^2\bm{\ell}}{2\pi}.
\end{equation}
From the coefficients $\hat\kappa(\bm{\ell})$, one can define the
power spectrum of the shear as
\begin{equation}
 \left<\hat\kappa(\bm{\ell})\hat\kappa^*(\bm{\ell}')\right> = P_\kappa(\ell)
 \delta^{(2)}(\bm{\ell}-\bm{\ell}').
\end{equation}
The shear components can be decomposed alike. It is convenient to
use a complex notation for the shear, $\bm{\gamma}\equiv\gamma_1
+i\gamma_2$. Since the shear and the convergence derive from the
same potential, they are not independent and it can be shown that
the convergence can be obtained from the shear as
\begin{equation}
 \hat\kappa(\bm{\ell}) =
 \hat{\bm{\mathcal{K}}}^*(\bm{\ell})\hat{\bm{\gamma}}(\bm{\ell})/\pi
\end{equation}
for all $\bm{\ell}=(\ell_1,\ell_2)\not=\bm{0}$. The kernel
function $\hat{\bm{\mathcal{K}}}^*$ is defined as
\begin{equation}
 \hat{\bm{\mathcal{K}}} = \pi(\ell_1^2 - \ell_2^2 +
 2i\ell_1\ell_2)/\ell^2.
\end{equation}
Interestingly, this implies $|\hat\kappa|^2 =
\hat{\bm{\gamma}}\hat{\bm{\gamma}}^*$ so that the shear and
convergence have same power spectrum
\begin{equation}
  P_\gamma(\ell)=P_\kappa(\ell).
\end{equation}
Note that the shear can then be obtained as
\begin{equation}\label{masssh}
 \kappa(\bm{\theta}) = \kappa_0
 +\frac{1}{\pi}\int \mathcal{K}^*(\bm{\theta}-\bm{\theta}')
 \,\bm{\gamma}(\bm{\theta'})\dd^2\bm{\theta}'.
\end{equation}
The integration constant $\kappa_0$ is related to any constant
uniform mass distribution that will contribute to the convergence
but not to the shear.

Analogously, all 3-dimensional fields can be developed in Fourier
modes as
\begin{equation}\label{f3D}
 \phi(\bm{r},\eta) = \int  \hat\phi(\bm{k},\eta)
 \hbox{e}^{i\bm{k}\cdot\bm{r}}\,\frac{\dd^3\bm{r}}{(2\pi)^{3/2}}
\end{equation}
and are associated to a power spectrum
\begin{equation}
 \langle\hat\phi(\bm{k},\eta)\hat\phi(\bm{k}',\eta)\rangle = P_\phi(k,\eta)
 \delta^{(3)}(\bm{k}-\bm{k}').
\end{equation}

\subsection{Shear power spectrum}\label{subsec3a}

As Eqs.~(\ref{kappa1}-\ref{kappa2}) show, the shear is obtained as
a weighted projection of the deflecting potential and takes the
general form
\begin{equation}\label{eqn:kq}
 \kappa = \int
 q(\chi)\deff\left[S_K(\chi)\bm{\theta},\chi\right]\dd\chi.
\end{equation}
This implies that if $\deff$ is a homogeneous and isotropic
Gaussian random field, so will $\kappa$. Its angular correlation
function, $\xi_\kappa \equiv \langle
\kappa(\bm{\varphi}+\bm{\theta}) \kappa(\bm{\varphi})\rangle$, can
be expressed as
\begin{eqnarray}
 \xi_\kappa(\theta)&=&\int\dd\chi\dd\chi'q(\chi)q(\chi')\nonumber\\
 &&\times\langle
 \deff\left[S_K(\chi)\bm{\varphi},\chi\right]
 \deff\left[S_K(\chi)(\bm{\varphi}+\bm{\theta}),\chi\right]
 \rangle.
\end{eqnarray}
The correlation function that appears in this integral can be
computed if $\deff$ is expanded in Fourier modes~(\ref{f3D}) to
give
\begin{eqnarray}
  &&\int
  \hbox{e}^{-i \bm{k}_\perp.\bm{\varphi}f_K(\chi)}\hbox{e}^{-i{k}_3\chi}
  \hbox{e}^{-i \bm{k'}_\perp\cdot(\bm{\varphi}+\bm{\theta})f_K(\chi')}\hbox{e}^{-ik'_3\chi'}
  \nonumber\\
  &&\qquad\times\langle\deff(\bm{k},\chi)\deff(\bm{k}',\chi')\rangle
  \frac{\dd^3\bm{k}\dd^3\bm{k'}}{(2\pi)^3}.
\end{eqnarray}
On small angular scales ($\theta\ll1$), $k_\perp^2\gg k_3^2$ and
the power is mostly carried by $\bm{k}_\perp$. We thus approximate
$$
\langle\deff(\bm{k},\chi)\deff(\bm{k}',\chi')\rangle\simeq
\langle\deff(\bm{k}_\perp,\chi)\deff(\bm{k}'_\perp,\chi')\rangle
\delta^{(1)}(k_3+k_3').
$$
The integration over $k_3$ gives a factor
$2\pi\delta^{(1)}(\chi-\chi')$ so that after integration on
$\chi'$ we end up with the expression
\begin{eqnarray}
 \xi_\kappa(\theta) = \int\dd\chi q^2(\chi)
 \frac{\dd^2\bm{k}_\perp}{(2\pi)^2}P_\eff(k_\perp,\chi)
 \hbox{e}^{iS_K(\chi)\bm{k}_\perp\cdot\bm{\theta}}
\end{eqnarray}
We conclude, that the 2-dimensional power spectrum of the
convergence is related to the 3-dimensional power spectrum of
$\deff$ by
\begin{eqnarray}\label{PkappaGEN}
 P_\kappa(\ell) = \frac{9H_0^4}{4}\Omega_\mat^0\int
 \left[\frac{g(\chi)}{a(\chi)}\right]^2
 P_\eff\left[\frac{\ell}{S_K(\chi)},\chi\right]\dd\chi.
\end{eqnarray}

If we filter the shear field by a window function,
$U(\theta',\theta)$ of angular radius $\theta$ as
\begin{equation}
 \kappa(\theta) = \int\dd^2\bm{\theta}'\kappa(\bm{\theta}')U(\theta',\theta).
\end{equation}
then the variance of the filtered convergence is given by
\begin{equation}
 \langle\kappa^2\rangle = 2\pi\int_0^\infty \dd\ell
 \ell P_\kappa(\ell)
 \left(\int \theta'U(\theta',\theta) J_0(\ell\theta) \dd\theta\right)^2
\end{equation}
where $J_0$ is a Bessel function.

\subsection{2-point statistics of the shear field}\label{subsec3c}

The shear field $\bm{\gamma}$ has two components that can be
decomposed in various ways. From a reference point, such as the
center of the filter, one can define a radial and tangential shear
as
\begin{equation}
 \gamma_{\rm r} =\rm{Re}(\bm{\varphi}^*\bm{\gamma}),\qquad
 \gamma_{\rm t} =\rm{Im}(\bm{\varphi}^*\bm{\gamma})
\end{equation}
where $\bm{\varphi}$ represents the unit vector pointing from the
reference point to the point where $\bm{\gamma}$ is defined. From
these components we can define the correlation functions
\begin{equation}
 \xi_{\rm r}(\theta) = \langle\gamma_{\rm r}\gamma_{\rm r}\rangle,\qquad
 \xi_{\rm t} = \langle\gamma_{\rm t}\gamma_{\rm t}\rangle
\end{equation}
and by symmetry the correlation $\langle\gamma_{\rm t}\gamma_{\rm
r}\rangle$ strictly vanishes. We can also combine this functions
as
\begin{equation}
 \xi_\pm(\theta) = \xi_{\rm t} \pm \xi_{\rm r}.
\end{equation}
Developing the shear in Fourier modes, one gets
\begin{eqnarray}
 \xi_+(\theta) &=& \int_0^\infty\frac{\dd\ell}{2\pi}\ell P_\kappa(\ell)J_0(\ell\theta)\\
 \xi_-(\theta) &=& \int_0^\infty\frac{\dd\ell}{2\pi}\ell
 P_\kappa(\ell)J_4(\ell\theta).
\end{eqnarray}
Since the shear and convergence have same power spectra, we also
deduce that
\begin{equation}
 \langle\bm{\gamma}^*\cdot\bm{\gamma}\rangle(\theta) =
 \int_0^\infty\frac{\dd\ell}{2\pi}\ell P_\kappa(\ell)
 \left[\frac{2J_1(\ell\theta)}{\ell\theta}\right]^2.
\end{equation}

An interesting statistics arises when one uses a compensated
filter, that is
\begin{equation}\label{compense}
 \int\theta'\,U(\theta',\theta)\dd\theta = 0.
\end{equation}
With such a filter, any constant mass density, related to the
integration constant $\kappa_0$ in Eq.~(\ref{masssh}) will not
bias the statistics. It is usual to define the {\it aperture mass}
as
\begin{equation}
 M_{\rm ap} = \int
 \dd^2\bm{\theta}'\,U(\theta',\theta)\kappa(\bm{\theta}').
\end{equation}
Interestingly, it can be expressed in terms of the tangential
shear as
\begin{equation}
 M_{\rm ap} = \int
 \dd^2\bm{\theta}'\,Q(\theta',\theta)\gamma_{\rm t}(\bm{\theta}')
\end{equation}
with the filter $Q$ defined as
\begin{equation}
 Q(\theta',\theta) = \frac{2}{\theta^2}\int
 \dd\theta'\,\theta' U(\theta',\theta)
 -U(\theta',\theta).
\end{equation}
A widely used family of filter that satisfies the condition
(\ref{compense}) is
\begin{equation}
 U(\theta',\theta) = \frac{3}{\pi\theta^2}\left[1-\left(\frac{\theta'}{\theta}\right)^2\right]
 \left[1-3\left(\frac{\theta'}{\theta}\right)^2\right]
\end{equation}
for which the aperture mass has a variance given by
\begin{equation}
 \langle M_{\rm ap}^2 \rangle(\theta) = \frac{288}{\pi}
 \int_0^\infty \dd\ell\ell P_\kappa(\ell)
 \left[\frac{J_4(\ell\theta)}{\ell^2\theta^2}\right]^2.
\end{equation}
All these 2-point statistics derive from the same power spectrum
are thus not independent.

\subsection{Numerical integration}\label{subsec3b}

%------------------------------------------------------------
\begin{center}
\begin{figure*}[ht]
\unitlength=1cm
\begin{picture}(14,14)
 \thicklines
 \put(-1.5,0) {\framebox(7.5,9.5){}}
 \put(-1.5,10.75){\framebox(7.5,1.25){}}
 \put(7.5,0){\framebox(7.5,7){}}
 \put(7.5,8){\framebox(7.5,1.5){}}
 \put(7.5,10.75){\framebox(7.5,1.25){}}
 \put(6,11.35){\vector(1,0){1.5}}
 \put(11.25,10.75){\vector(0,-1){1.25}}
 \put(2.25,10.75){\vector(0,-1){1.25}}
 \put(11.25,8){\vector(0,-1){1}}
 \put(-1,11.5){CMB code:}
 \put(0,11){input: $\Omega$, $P(k),\ldots$}
 \put(7.8,11.5){Background: $\chi(z)$}
 \put(7.8,11){Perturbations: $\delta(\bm{k},z)$, $\varphi(\bm{k},z)$,
            $\phi(\bm{k},z)$, $\psi(\bm{k},z),\ldots$}
 \put(7.8,9){Linear to nonlinear mapping}
 \put(11,8.5){$\lbrace\bm{k}_{NL},P_{NL}\rbrace$}
 \put(-0.8,9){\framebox{
 \psfig{file=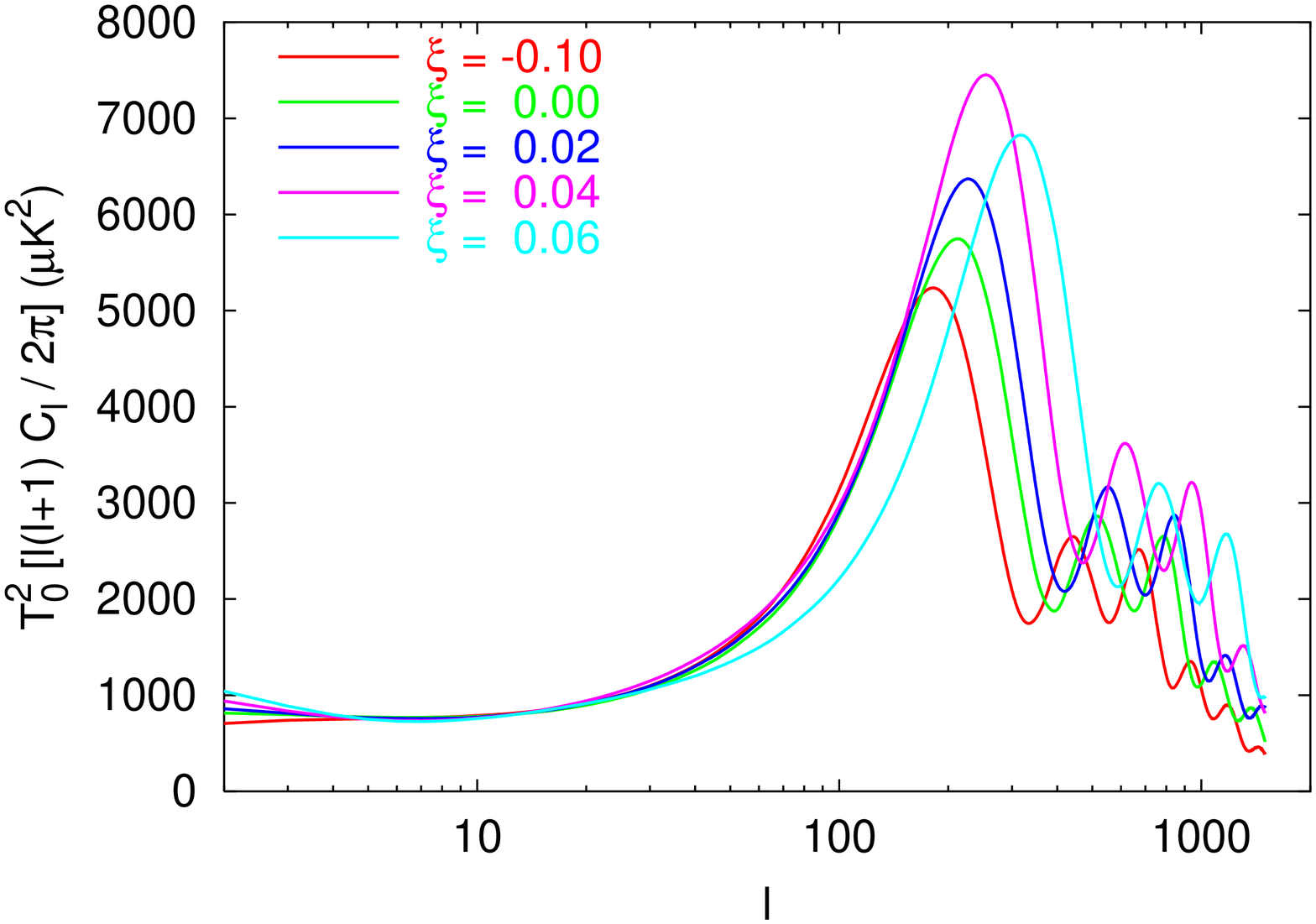,angle=270,width=6cm}}}
 \put(-0.8,4.5){\framebox{
 \psfig{file=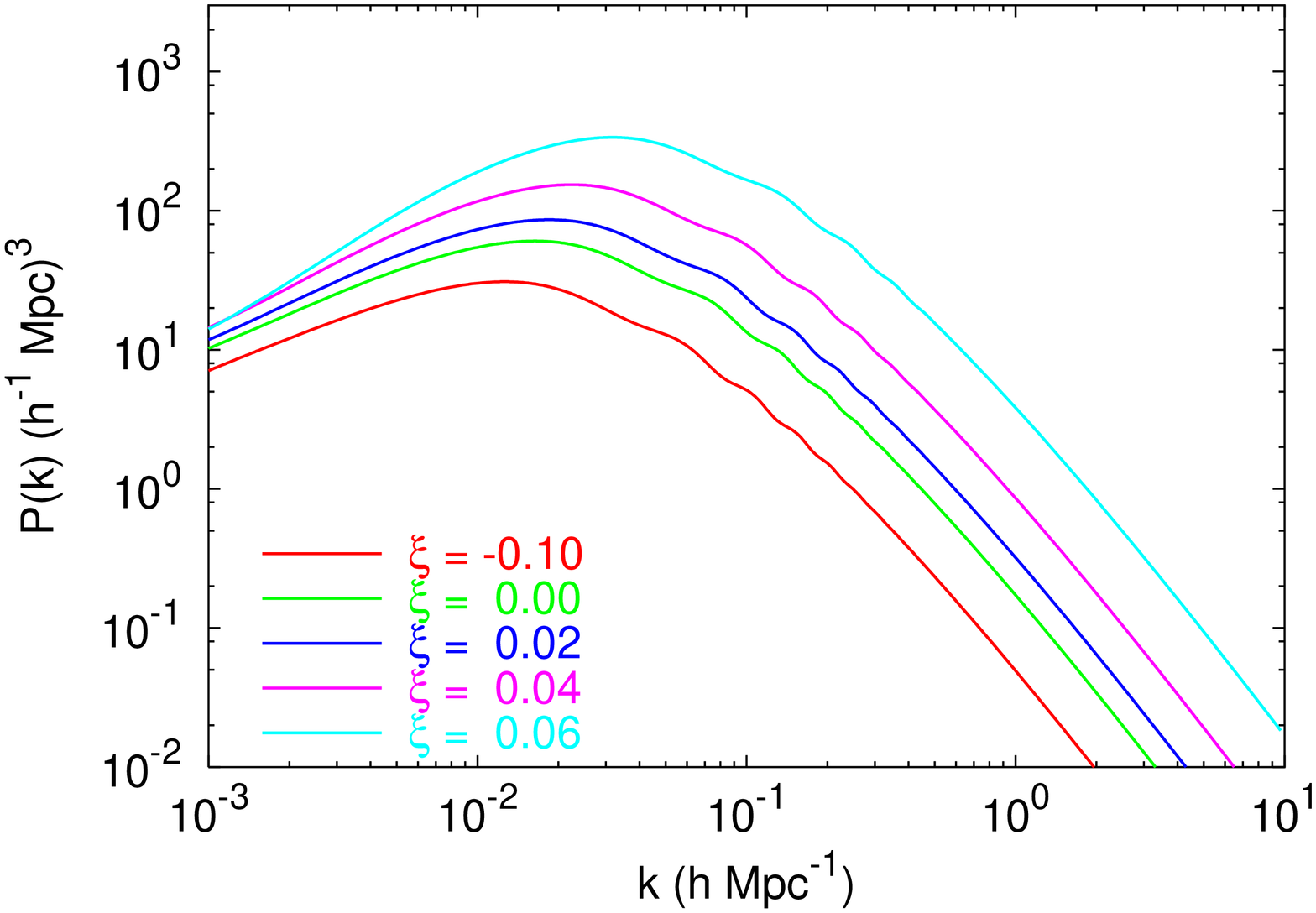,angle=270,width=6cm}}}
 \put(7.8,2.75){\framebox{\epsfxsize=6.5cm
 \epsffile{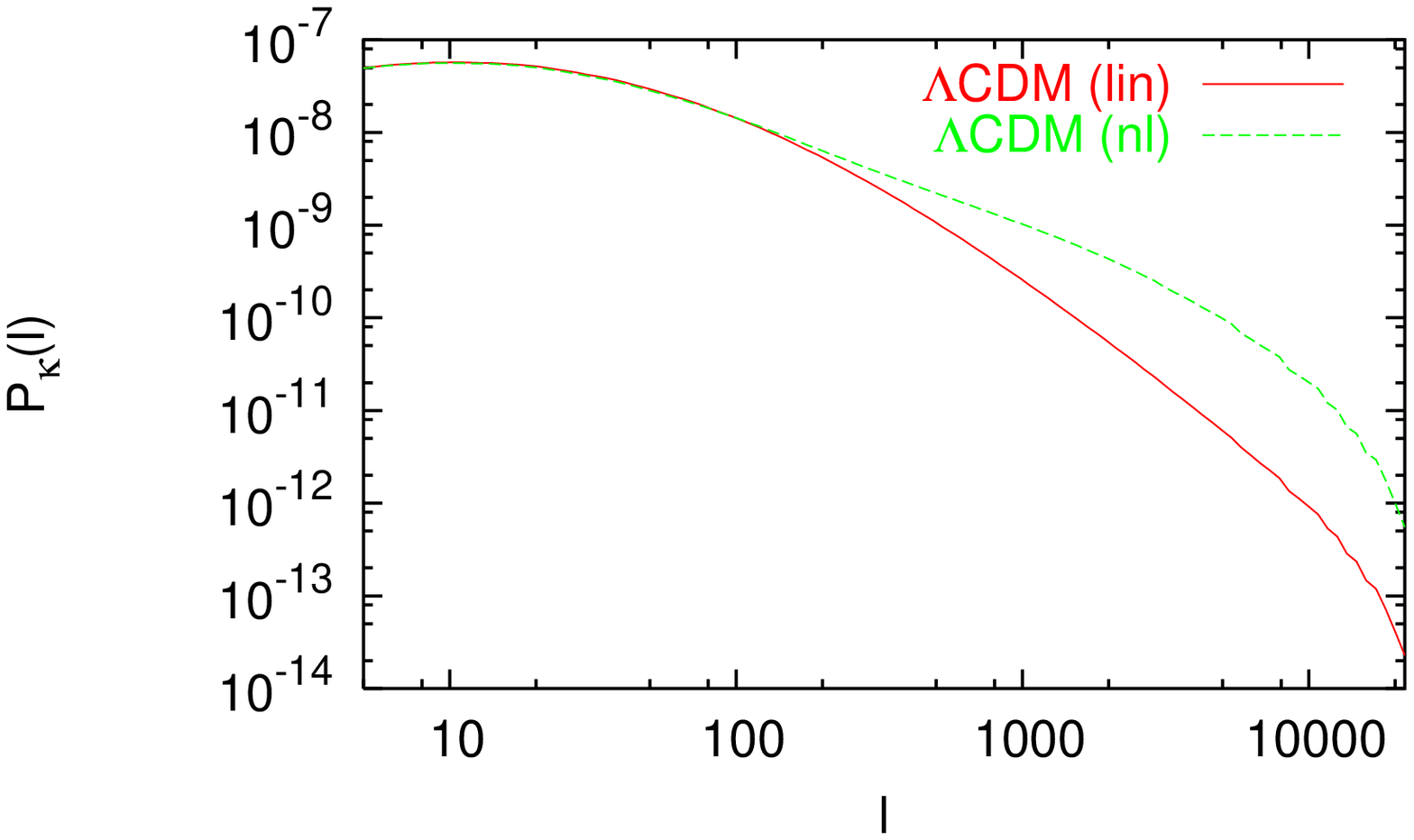}}}
  \put(11.25,2.65){\vector(0,-1){1.45}}
  \put(9.5,0.5){$\langle\gamma^2\rangle,\,\langle M_{\rm ap}^2\rangle,\xi_-,\xi_+\,\ldots$}
\end{picture}
\caption{General description of the computation of the lensing
  observables. The input are generated by a CMB code that deals with
  scalar-tensor theories and extended quintessence~\cite{ru02}.
  Once the linear to non-linear mapping has been applied, we can compute
  the shear power spectrum and then all the lensing observables with the same
  normalization as used for the CMB observables.}
\label{figmethod}
\end{figure*}
\end{center}
%--------------------------------------------------------------

The computation of the shear power spectrum as well as the 2-point
statistics have been implemented in a numerical code that can be
used both with general relativity and scalar-tensor theories. It
is based on the Boltzmann code described in Ref.~\cite{ru02},
implemented to study the evolution of the background and of the
linear perturbations for a general scalar-tensor theory specified
either in Jordan or Einstein frame (see Appendix~\ref{appb}). In
particular, this code allows the study of the magnitude-redshift
relation and of the CMB angular power spectrum. It integrates the
Einstein and fluid perturbation equations forward in time (toward
the future) and sets the initial conditions on the scalar field by
a shooting method in order for the dark energy density parameter
and the Newton constant to agree with their value today (see
Ref.~\cite{ru02} for details).

The CMB code gives access to the value of the perturbation
variables, among which the two gravitational potential $\phi$ and
$\psi$ and the density perturbation, as a function of the wave
numbers $k$ and the redshift $z$. Therefore we do not need, as is
usually done in the Newtonian regime, to decompose the density
perturbation in an initial random field and a growth factor. In
such a case, the lensing observables depend on the normalization
and shape of the transfer function~\cite{peebles93}, that is on
$\sigma_8$ and $\Gamma$. In our approach, all observables are CMB
normalized at $\ell=111$, according to Ref.~\cite{spergel}, and we
need not introduce these parameters. The CMB code also gives the
value of $\chi(z)$, from which we can derive the angular
distances. This implies that the lensing plug-in code (see
Fig.~\ref{figmethod}) we developed, does not need to solve any
evolution equations. It only integrates the deflecting potential
along the lines of sight accounting for the source distribution,
according to the two-dimensional projection described in
Sec.~\ref{subsec3a0} and deal with the linear to non-linear
mapping (see \S~\ref{lnlst}).

We are interested in angular scales ranging from 1 arcmin to 2
degrees and we considers multipoles in the range [$\ell_{\rm
min}=2$,$\ell_{\rm max}=22000$].  We thus have to consider
comoving wave numbers in the range [$k_{\rm min},k_{\rm max}$]
with
\begin{equation}\label{kmin}
 k_{\rm min}\sim\ell_{\rm min}/\chi(z_{\rm max})
 \sim 10^{-3}h\,\mathrm{Mpc}^{-1}
\end{equation}
and we set the cut-off $k_{\rm max}\sim\ell_{\rm max}/\chi(z_{\rm
min})$ to 10 $h$ Mpc$^{-1}$. Since some of the scales are in the
non-linear regime, we carry out the linear to non-linear mapping
applying the procedure of Ref.~\cite{smithetal}. It has to be
stressed that these mappings have been calibrated on numerical
simulations assuming that gravity was described by general
relativity and $\Lambda$CDM cosmological models. The validity of
this hypothesis will call for further checking, but it is to be
expected that the scalar field does not affect too much the
clustering of the matter during the non-linear phase and that its
main effect is through its contribution to the expansion of the
universe, as we will discuss later.

We assume a redshift sources distribution parameterized as
$$
 p(z)=\Gamma^{-1}\left(\frac{1+\alpha}{\beta}\right)
 \frac{\beta}{z_s}\left(\frac{z}{z_s}\right)^\alpha
 \exp\left[-(z/z_s)^\beta\right]
$$
with $(z_s,\alpha,\beta)=(0.8,2,1.5)$ consistently with a limited
magnitude $I_{AB}=24.5$ (for details concerning this choice, see
Ref.~\cite{sourcedistr}).

\subsection{Non-linear power spectrum}

The computation of weak lensing observables requires to determine the
power spectrum in the non-linear regime.

\begin{figure}[h]
 \centerline{\epsfig{figure=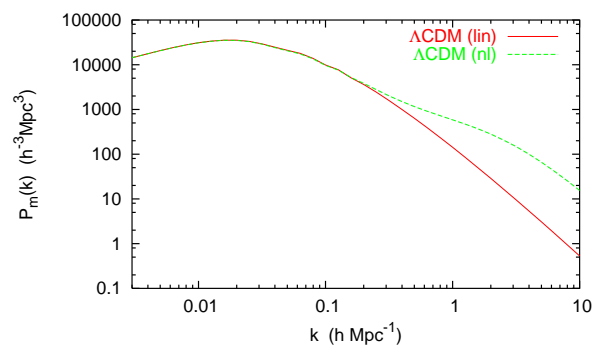,width=8cm}}
 \caption{Matter power spectrum $P_m(k)$ for a flat fiducial $\Lambda$CDM
 model defined by $\Omega_\Lambda=0.7$, $\Omega_b h^2=0.019$, $h=0.72$.
 Linear (solid) and non-linear (dashed) regimes are presented.}
 \label{fig2b}
\end{figure}

For $\Lambda$CDM models, various mappings have been proposed in
the literature. Assuming stable clustering, it was
argued~\cite{hamilton,peacockd} that the effects non-linear
evolution can be described by a mapping between the linear and
non-linear power spectra involving a universal function, $f_{\rm
nl}$. Introducing $\Delta$ as
\begin{equation}\label{nl1}
 \Delta^2(k) \equiv 4\pi k^3 P(k),
\end{equation}
the non-linear power spectrum is obtained by
\begin{equation}\label{nl2}
 \Delta_{\rm nl}^2\left(k_{\rm nl}\right) =
        f_{\rm nl}\left[\Delta^2(k)\right]
\end{equation}
where the wavenumber, $k_{\rm nl}$, is related to the linear
wavenumber, $k$, by
\begin{equation}\label{nl3}
 k^3 = \left[1+\Delta_{\rm nl}^2\left(k_{\rm nl}\right)\right]^{-1}
 k_{\rm nl}^3.
\end{equation}
The function $f_{\rm nl}$ is determined by $N$-body simulations
and it depends on the value of the cosmological parameters. It has
also been shown~\cite{peacockd} that at large values of its
argument this function behaves as
\begin{equation}
 f_{\rm nl}(x) \sim \left(\frac{D}{a}\right)^{-3}x^{3/2},
\end{equation}
$D$ being the linear growth factor [see Eq.~(\ref{decompo})].
Because of this simpler asymptotic analytic form,  we will use the
relations (\ref{nl1}-\ref{nl3}) for general analytic arguments.
Numerically, we have however implemented the more realistic
mapping described in Ref.~\cite{smithetal}. Fig.~\ref{fig2b}
depicts the linear and nonlinear power spectra of the density
perturbation computed for the fiducial $\Lambda$CDM model.

As we have already stressed, all the mappings have to be
calibrated on $N$-body simulations. No full $N$-body simulations
for quintessence models and have been performed so far (see
however Ref.~\cite{Nbodyquint} where $N$-Body simulations with a
modified expansion rate to take into account quintessence have
been investigated). However, as it has been shown in
Ref.~\cite{specQCDM}, the shape of the linear power spectrum for
sub-Hubble modes, but not its absolute amplitude, of quintessence
models is very similar to the one of a pure $\Lambda$CDM. Hence we
assume that these mapping also apply to quintessence models. In
particular, it was argued~\cite{lnlquint} that the mapping
(\ref{nl1}-\ref{nl3}) was reasonably accurate for effective
quintessence models, at least at low redshift. In particular, we
assume that the non-linear regime always reaches a stable
clustering regime. Notice that various mappings lead to results
that agrees only at a 5-10\% level (see e.g. Ref.~\cite{ludoben}).

As far as scalar-tensor theories are concerned, as for
quintessence, there is no $N$-body simulations to calibrate the
mapping. We have to assume that the mapping~\cite{smithetal}
calibrated on pure $\Lambda$CDM still hold. It can be argued that
we do not expect the change of the theory of gravity to
drastically affect this mapping as long as strong field effects,
such as spontaneous scalarization~\cite{def}, appear. Even though,
we emphasize that the mapping procedure has to be adapted in the
case of scalar-tensor theories, as it will be discussed in see
\S~\ref{lnlst}.

%----------------------------------------------------------------------------------------
\section{Weak lensing in general relativity}\label{sec3}

\begin{figure*}[ht]
 \centerline{\epsfig{figure=LCDM_Pk.ps,width=6cm}
              \epsfig{figure=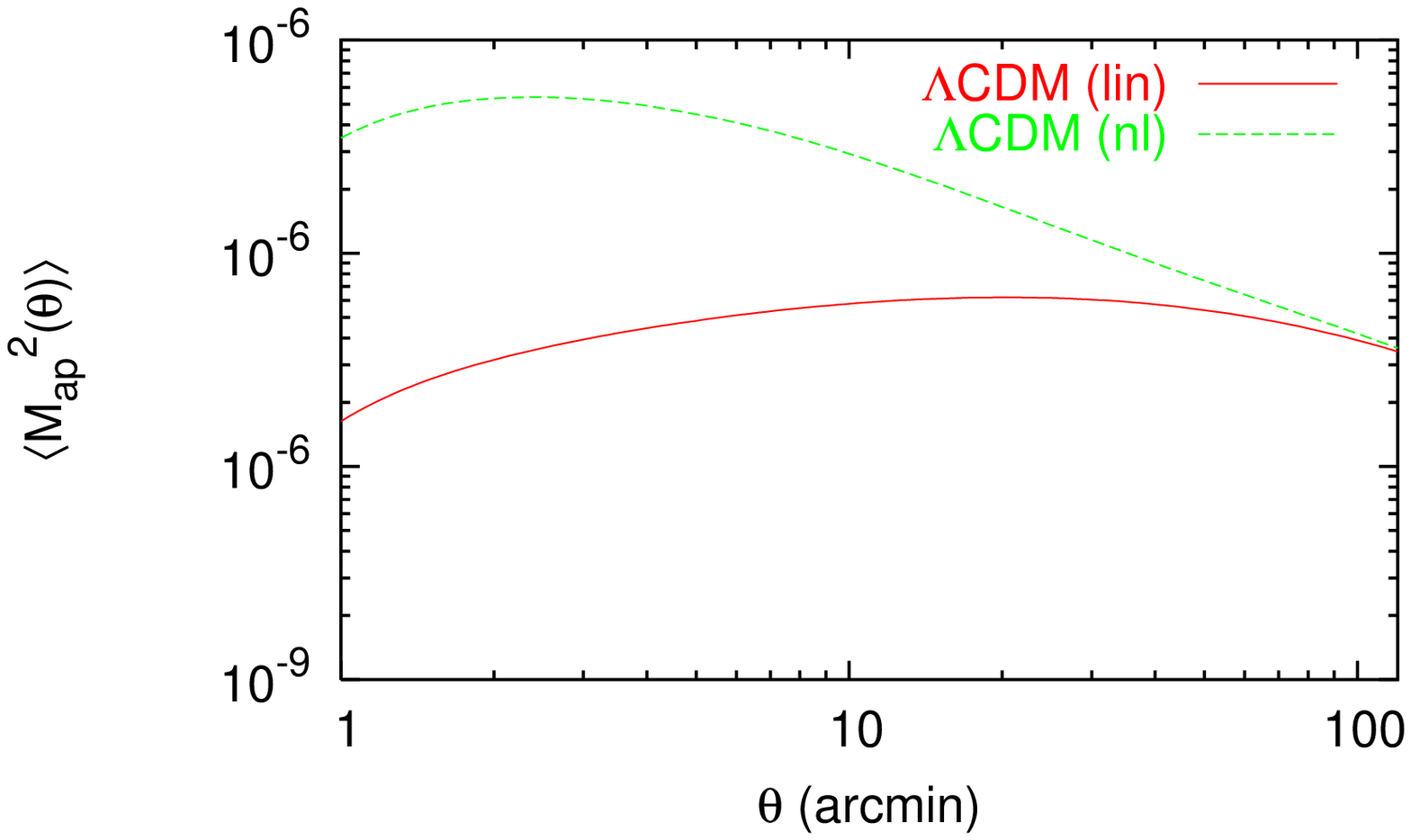,width=6cm}
              \epsfig{figure=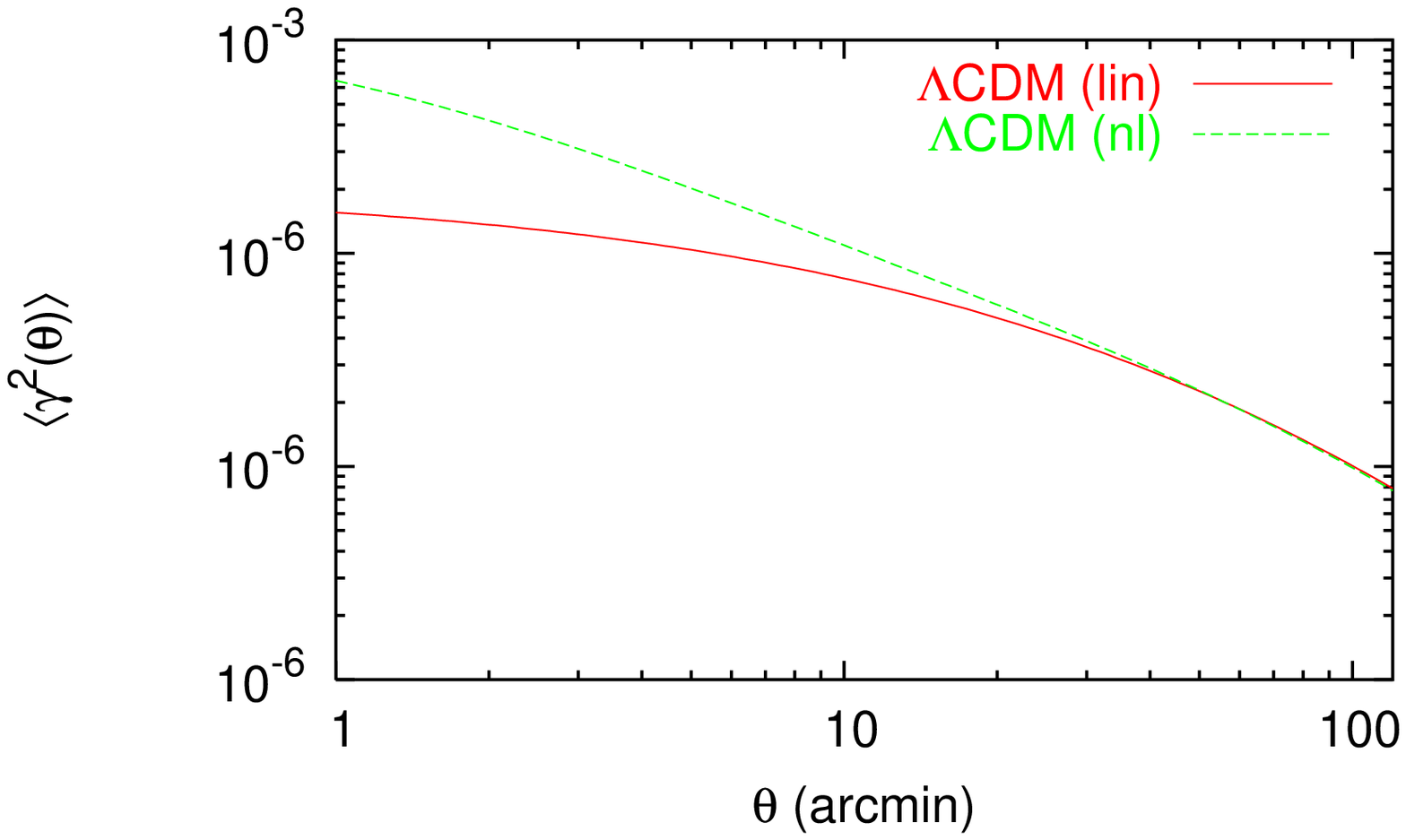,width=6cm}}
  \caption{(left) Convergence power spectrum $P_\kappa (\ell)$ as a function
  of the multipole $\ell$ and the 2-point statistics of the shear
  field: (middle) the aperture mass variance and (right) the shear
  variance in the case of the fiducial $\Lambda$CDM model
  (\ref{refLCDM}).}
  \label{fig2}
\end{figure*}

\subsection{Generalities}

When gravity is described by general relativity, then at late time
matter dominates and the anisotropic stresses are negligible so
that the two gravitational potentials are equal
\begin{equation}
 \phi=\psi
\end{equation}
consequently the deflecting potential is simply given by
\begin{equation}
\Phi=2\phi.
\end{equation}
On sub-Hubble scales, the gravitational potential is related to
the matter density perturbation by the Poisson equation which
implies that
\begin{equation}
 \deff = \delta_m,
\end{equation}
where $\delta_m=\delta\rho_m/\rho_m$ is the matter density
contrast. Let us stress that on large scales, one must take into
account the radiation anisotropic stress which will induce a
departure from this equality (see Fig.~\ref{figautre}). Starting
from an initial time, $a_i$, where the modes of interest are
sub-Hubble, one can decompose the density field as
\begin{equation}\label{decompo}
\delta_m(\bm{k},a) = D(a)\delta(\bm{k},a_i) = D(a)\delta_i
\end{equation}
where $D(a)$ is the growth factor. It follows that $P_\eff(k,a) =
D^2(a) P_\delta(k)$. Its equation of evolution, in the Newtonian
regime, is given by
\begin{equation}
 \ddot D + \HH \dot D -4\pi G \rho_m a^2D = 0.
\end{equation}
It can be rewritten by using the redshift as variable as
\begin{equation}
 D'' + \left(\frac{H'}{H} - \frac{1}{1+z}\right)D' -
 \frac{3}{2}\Omega_{m,0} (1+z) D = 0,
\end{equation}
setting $a_0=1$ and where a prime refers to a derivative with
respect to $z$. This equation has two solutions, a decaying mode,
$D \propto H$, and a growing mode
\begin{equation}
 D(a) = \frac{5}{2}\frac{H(a)}{H_0}\Omega_{m,0}\int_0^a \frac{\dd
 \tilde{a}}{\left[\tilde{a}E(\tilde{a})\right]^3}.
\end{equation}
The convergence power spectrum takes the simplified form
\begin{equation}\label{PkappaRG}
 P_\kappa(\ell) = \frac{9H_0^4}{4}\Omega_{m,0}\int
 \left[\frac{g(\chi)}{a(\chi)}\right]^2
 P_\delta\left[\frac{\ell}{S_K(\chi)}\right]D^2(\chi)\dd\chi.
\end{equation}

\subsection{$\Lambda$CDM}

The power spectrum (\ref{PkappaRG}) depends on the cosmological
parameters through the growth function and the angular distances and
on the normalization of the power spectrum. As emphasized above, we do
not perform such a splitting and use the complete expression
(\ref{PkappaGEN}) so that our results depend on the cosmological
parameters and the primordial spectrum are CMB-normalized.

As a reference model, we choose a flat $\Lambda$CDM with
\begin{equation}\label{refLCDM}
 \Omega_{\Lambda,0} = 0.7,\,\,
 \Omega_{m,0} = 0.3,\,\,
 \Omega_{b,0} = 0.037,\,\,
 h = 0.72.
\end{equation}

Figure~\ref{fig2} depicts the convergence power spectrum, the
aperture mass and the shear variance for a $\Lambda$CDM model.
This model will be our reference model and we will try to quantify
the deviation from its prediction on various models.

\begin{figure*}[htb]
 \centerline{\epsfig{figure=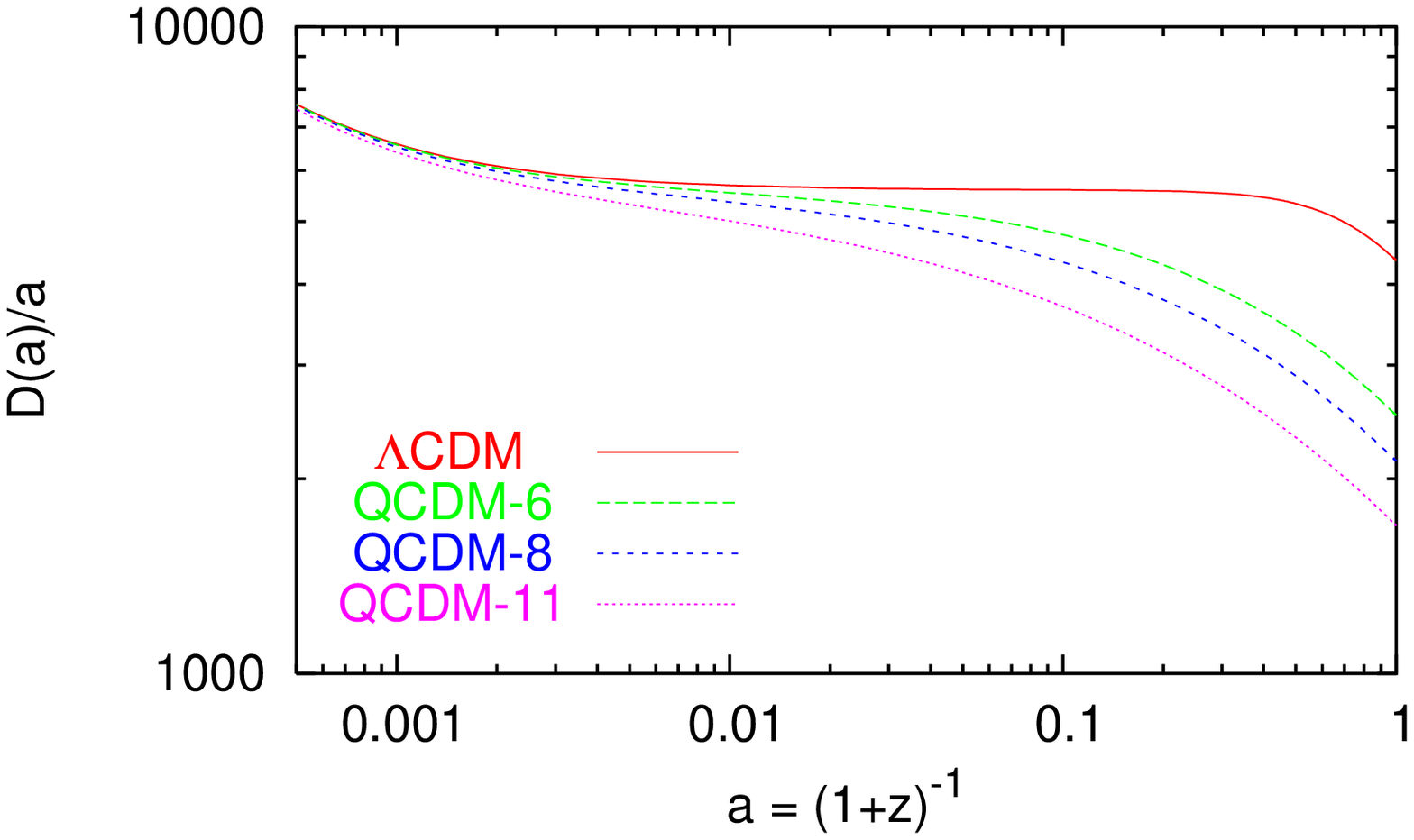,width=6cm}
             \epsfig{figure=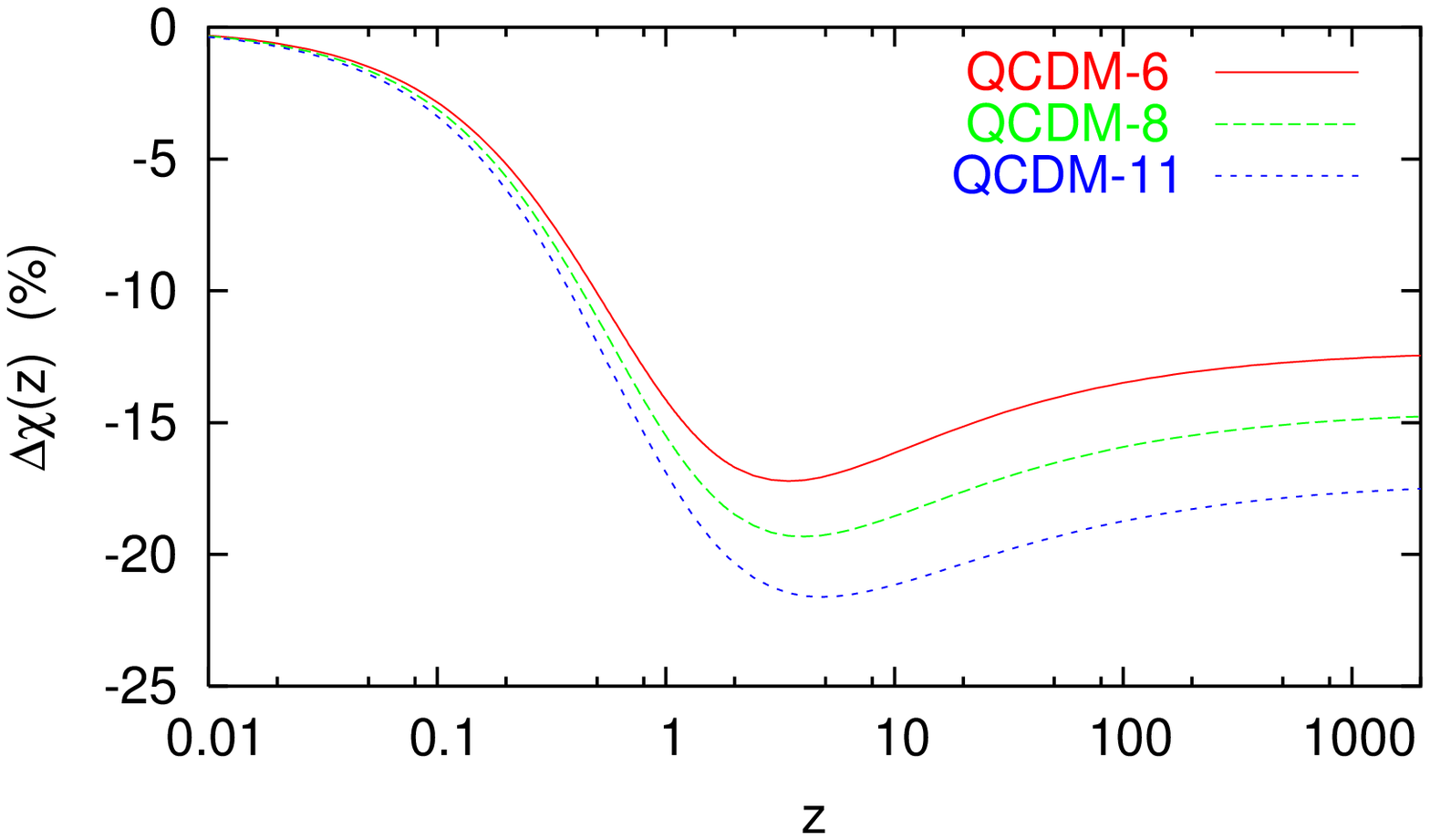,width=6cm}
             \epsfig{figure=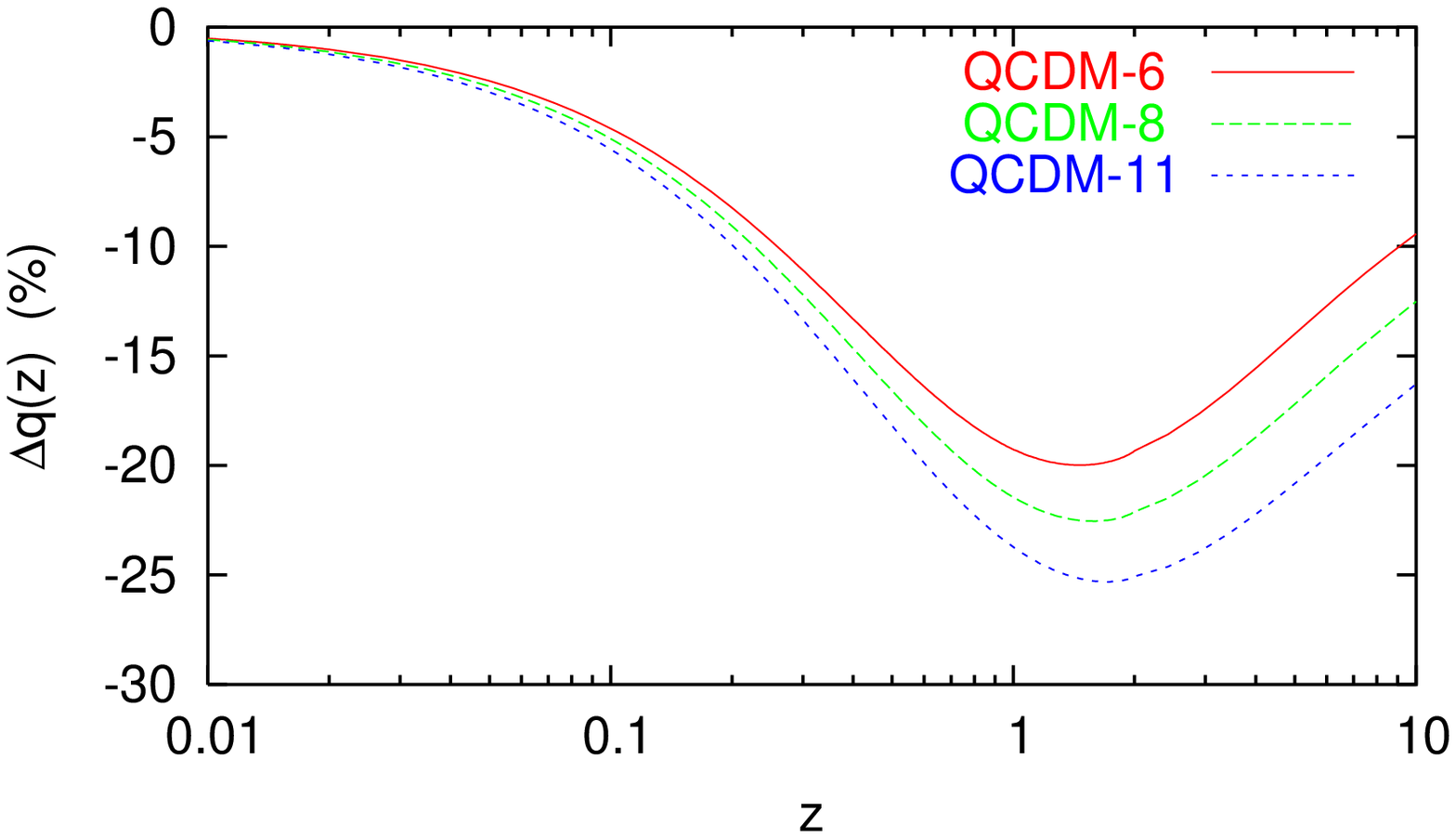,width=6cm}}
 \caption{Comparison between the fiducial $\Lambda$CDM model
 (\ref{refLCDM}) and various quintessence models. (left) the
 growth factor normalized to its value in an Einstein-de Sitter
 universe, $D(a)/a$, as a function of the scale factor $a$,
 normalized at high redshift; (middle) Relative deviation
 from $\Lambda$CDM on the comoving radial distance, $\chi(z)$,
 and (right) on the geometrical factor, $q(z)$.
 In all plots, we consider three QCDM models with potential
 (\ref{eqn:RP}) with $m = 6, 8, 11$ from top to bottom.}
 \label{fig3bc}
\end{figure*}

\subsection{Quintessence: effect of the potential}\label{qrg2}

As a first generalization to the previous $\Lambda$CDM model we
consider a class of quintessence models (QCDM) with runaway
potentials of the form
\begin{equation}\label{eqn:RP}
 V(\varphi) = M^4\varphi^{-m}.
\end{equation}
In the slow-roll regime, the quintessence field acts as a
repulsive matter component that replaces the cosmological
constant. Gravity is still described by general relativity and the
main effects on the lensing observables arise from the
modification of the Friedman equations, and thus of the angular
distances and of the growth factor of the density field.

To estimate the amplitude of the effects, let us follow
Ref.~\cite{berben} and assume that the sources are located at a
redshift $z_s$ so that the source distribution is simply given by
$p(\chi)=\delta(\chi-\chi_s)$ and thus $g(\chi) =
S_K(\chi_s-\chi)/S_K(\chi_s)$. The function
\begin{equation}
 \mathcal{W}(\chi,\chi_s) \equiv \frac{S_K(\chi)S_K(\chi_s-\chi)}{S_K(\chi_s)a}
\end{equation}
is peaked around $\chi_s/2$ so that it can be approximated by
\begin{equation}
 \mathcal{W}^2 \simeq \mathcal{W}_{1/2}^2\delta(\chi-\chi_{1/2})
\end{equation}
with $\chi_{1/2}\simeq\chi_s/2$ and
\begin{equation}\label{eq:weff}
 \mathcal{W}_{1/2}^2 = \int_0^{\chi_s}\mathcal{W}^2(\chi,\chi_s)\dd\chi.
\end{equation}
Plugging this approximation into Eq.~(\ref{PkappaGEN}) allows to
estimate the shear power spectrum as
\begin{equation}
 P_\kappa(\ell)\sim\frac{9}{4}H_0^2\Omega_{m,0}
 \left[\frac{\mathcal{W}_{1/2}}{S_K(\chi_{1/2})}\right]^2
 P\left(\frac{\ell}{S_K(\chi_{1/2})},\chi_{1/2}\right).
\end{equation}
Assuming that the initial power spectrum takes the form
\begin{equation}
 P_\delta(k) = A k^n,
\end{equation}
we get that in the linear regime $P_{\rm lin} = D^2 A k^n$ and
$P_{\rm nl}\sim a^3Ak^{3(3+n)/2}$ so that
Eqs.~(\ref{nl1}-\ref{nl3}) imply that
\begin{equation}
 k \sim \left(k_{\rm nl}^2/Aa^2\right)^{1/(5+n)}.
\end{equation}
As long as we are considering modes in the linear regime, the
effect of the quintessence field is mainly encoded in the growth
function. We get
\begin{eqnarray}\label{spectreQlin}
 \left.\frac{P_\kappa^Q}{P_\kappa^\Lambda}\right|_{\rm lin} &\sim&
 \left(\frac{\mathcal{W}_{1/2}^Q}{\mathcal{W}_{1/2}^\Lambda}\right)^2
 \left(\frac{S^\Lambda_K(\chi_{1/2}^\Lambda)}{S^Q_K(\chi_{1/2}^Q)}\right)^{2+n}
 \nonumber\\
 &&\qquad\qquad
 \left(\frac{D^Q_{1/2}/D^Q_0}{D^\Lambda_{1/2}/D^\Lambda_0}\right)^2
 \frac{P_0^Q}{P_0^\Lambda},
\end{eqnarray}
where $D_{1/2}$ and $D_0$ are the values of the growth factor at
$\chi=\chi_{1/2}$ and $\chi_0$ respectively and  $P_0$ i the value
of the matter power spectrum today, evaluated at $k = 1
h\,\mathrm{Mpc}^{-1}$.

In the non-linear regime, the matter power spectra are not
distorted in the same way because the same value of $k_{\rm nl}$
does not correspond to the same $k$. It follows that
\begin{eqnarray}\label{spectreQnl}
 \left.\frac{P_\kappa^Q}{P_\kappa^\Lambda}\right|_{\rm nl}  &\sim&
 \left(\frac{\mathcal{W}_{1/2}^Q}{\mathcal{W}_{1/2}^\Lambda}\right)^2
 \left(\frac{S^\Lambda_K(\chi_{1/2}^\Lambda)}{S^Q_K(\chi_{1/2}^Q)}\right)^{2}
 \nonumber\\ &&\qquad\qquad
 \times\left(\frac{z_{1/2}^\Lambda}{z_{1/2}^Q}\right)^3
 \left(\frac{P_0^Q}{P_0^\Lambda}\right)^{3/(5+n)}
\end{eqnarray}
if the spectral index is the same in both models. The relations
(\ref{spectreQlin}) and (\ref{spectreQnl}) show that on small scales
the shape of the matter power spectrum is modified in quintessence
models.

\begin{table}[htb]
\caption{\label{tab:Qbkg}\underline{Quintessence models (QCDM)}:
background effects. (Upper table) Maximum relative deviation on the
comoving radial distance, $\chi (z)$, geometrical factor, $q(z)$, and
linear growth factor, $D(z)$, from the fiducial $\Lambda$CDM model for
three models with inverse power law potential,
Eq.~(\protect{\ref{eqn:RP}}). (Lower table) Quantities used to
estimate the amplitude of the effects on the weak lensing observable,
according to Eq.~(\protect{\ref{spectreQlin}}). The amplitude of the
matter power spectrum $P_0$ is evaluated at $k = 1\, h\,
\mathrm{Mpc}^{-1}$. For the $\Lambda$CDM model $z_{1/2} \simeq 0.51$,
for all the quintessence models $z_{1/2} \simeq 0.48$.}
\begin{ruledtabular}
\begin{tabular}{lccc}
$m$ &$\Delta\chi(z)\;(\%)$&$\Delta q(z)\;(\%)$&$\Delta
D(z)\;(\%)$\footnotemark[1] \\ \hline $6$&$-17$ at $(z=3.4)$&$-20$ at
$(z=1.4)$&$-42$ \\ $8$&$-19$ at $(z=3.9)$&$-23$ at $(z=1.6)$&$-51$ \\
$11$&$-21$ at $(z=4.7)$&$-26$ at $(z=1.7)$&$-61$ \\
\end{tabular}
\begin{tabular}{lcccc}
$m$
&$\chi_{1/2}/\chi^\Lambda_{1/2}$&$\mathcal{W}_{1/2}^2/(\mathcal{W}_{1/2}^2)^\Lambda$&$D_{1/2}/D_0$&$P_0/P^\Lambda_0$
\\ \hline $6$&$0.853$&$0.611$&$1.052$&$0.275$ \\
$8$&$0.844$&$0.580$&$1.056$&$0.189$ \\
$11$&$0.835$&$0.549$&$1.060$&$0.113$ \\
\end{tabular}
\end{ruledtabular}
\footnotetext[1]{Evaluated at $z=0$.}
\end{table}

\begin{table*}[ht]
\caption{\underline{Quintessence models (QCDM)}: absolute values
 of the convergence power spectrum, $P_\kappa(\ell)$, aperture mass
 variance, $\langle M_{ap}^2(\theta)\rangle$, and shear variance,
 $\langle\gamma^2(\theta)\rangle$ at two angular scales respectively
 in the linear and non-linear regimes. For the small scales we give
 both the value in the linear and non-linear regime (values within
 parenthesis). Each model is labelled by
 $\mathrm{QCDM}m$, where $m$ defines the inverse power law
 potential, Eq.~(\ref{eqn:RP}).}
\begin{ruledtabular}
\begin{tabular}{lcccccc}
model &\multicolumn{2}{c}{$P_\kappa(\ell)$}&\multicolumn{2}{c}{$\langle M_{ap}^2(\theta)\rangle$}&\multicolumn{2}{c}{$\langle\gamma^2(\theta)\rangle$}\\
 &$\ell=180$\footnotemark[1]&$\ell=7200$\footnotemark[1]&$\theta=2^\circ$&$\theta=3^\prime$&$\theta=2^\circ$&$\theta=3^\prime$ \\
\hline
$\Lambda\mathrm{CDM}$&$7\times 10^{-9}$&$2\times 10^{-12}\;(4\times 10^{-11})$&$3\times 10^{-6}$&$4\times 10^{-6}\;(5\times 10^{-5})$&$8\times 10^{-6}$&$1\times 10^{-4}\;(3\times 10^{-4})$  \\
$\mathrm{QCDM}6$&$1\times 10^{-9}$&$4\times 10^{-13}\;(6\times 10^{-12})$&$7\times 10^{-7}$&$6\times 10^{-7}\;(6\times 10^{-6})$&$2\times 10^{-6}$&$2\times 10^{-5}\;(4\times 10^{-5})$  \\
$\mathrm{QCDM}8$&$9\times 10^{-10}$&$2\times 10^{-13}\;(3\times 10^{-12})$&$5\times 10^{-7}$&$4\times 10^{-7}\;(3\times 10^{-6})$&$1\times 10^{-6}$&$1\times 10^{-5}\;(2\times 10^{-5})$  \\
$\mathrm{QCDM}11$&$4\times 10^{-10}$&$1\times 10^{-13}\;(1\times 10^{-12})$&$3\times 10^{-7}$&$2\times 10^{-7}\;(1\times 10^{-6})$&$7\times 10^{-7}$&$9\times 10^{-6}\;(1\times 10^{-5})$ \\
\end{tabular}
\end{ruledtabular}
\footnotetext[1]{The multipoles $\ell=180, 7200$ correspond to the angle
 $\theta = 2^\circ, 3^\prime$.}\label{tab:XIQCDMwl}
\end{table*}

\begin{figure}[htb]
 \centerline{\epsfig{figure=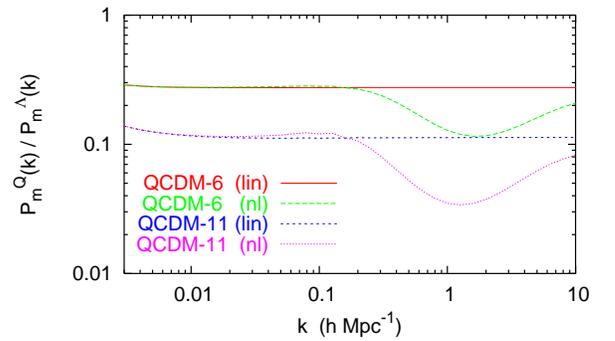,width=8cm}}
 \caption{The ratio of the three dimensional power spectrum of the
 matter perturbation $P_m(k)$ of two quintessence models to the
 reference $\Lambda$CDM model~(\ref{refLCDM}) for $m=6$ and $m=11$.}
 \label{fig2e}
\end{figure}

\begin{figure*}[htb]
 \centerline{\epsfig{figure=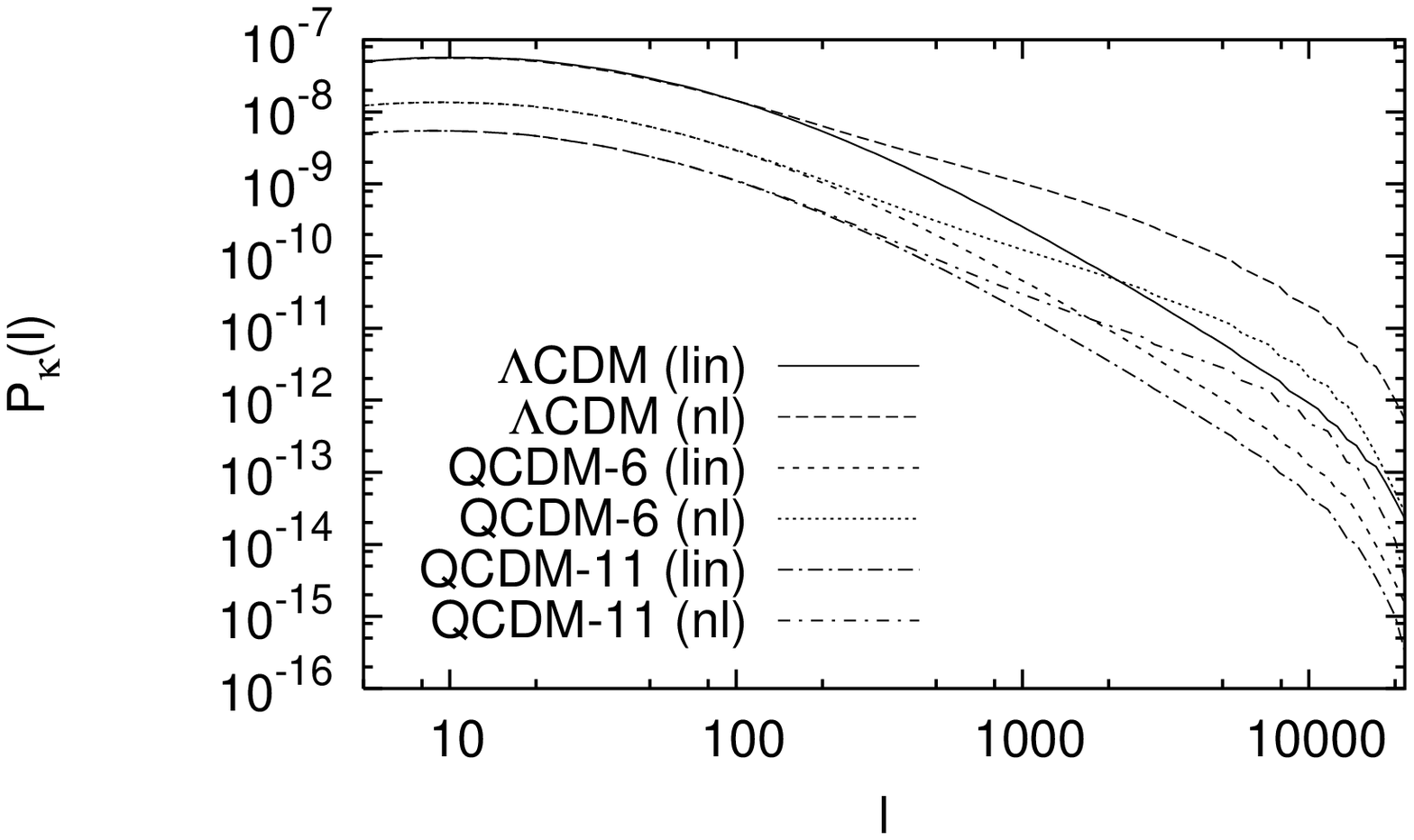,width=6cm}
             \epsfig{figure=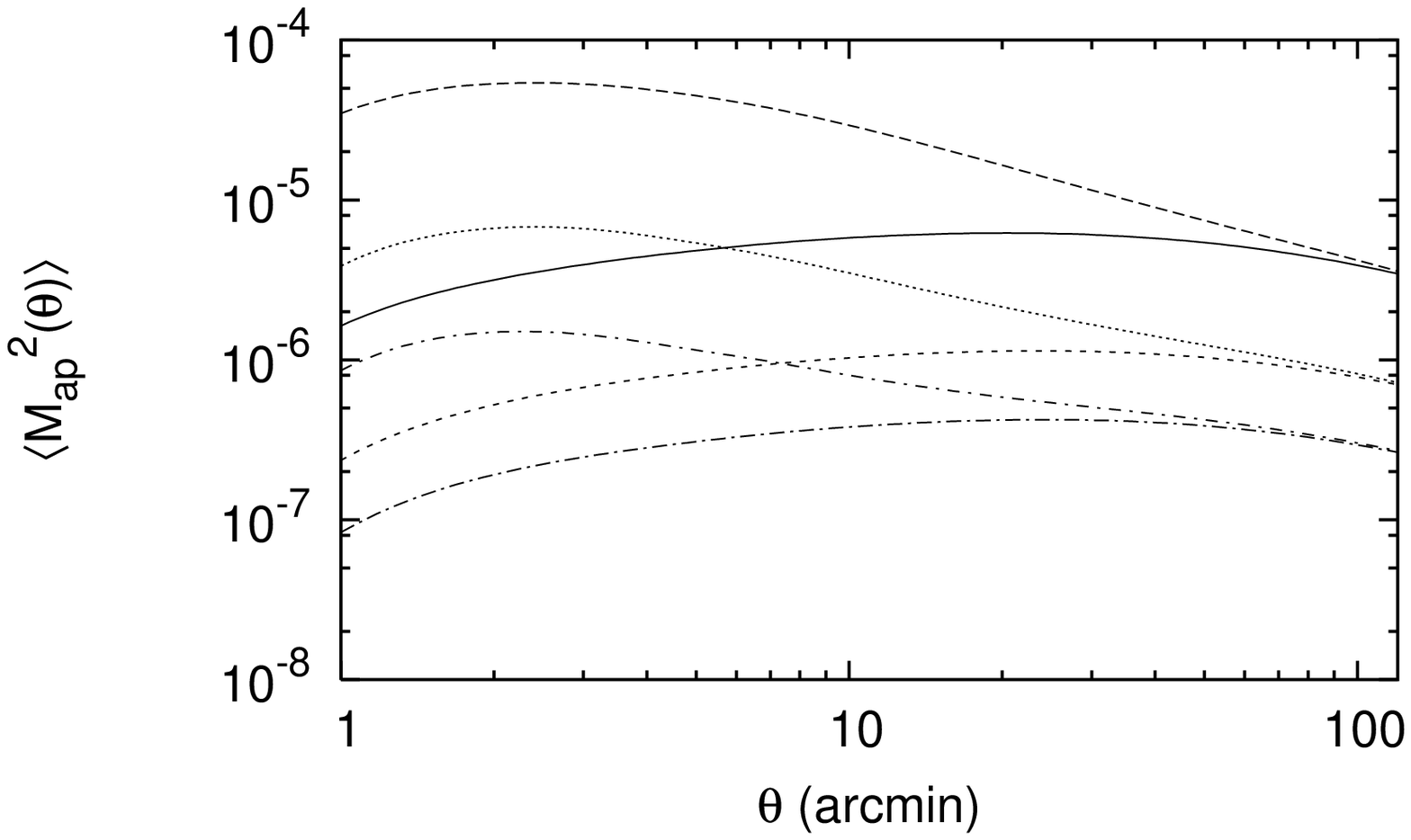,width=6cm}
             \epsfig{figure=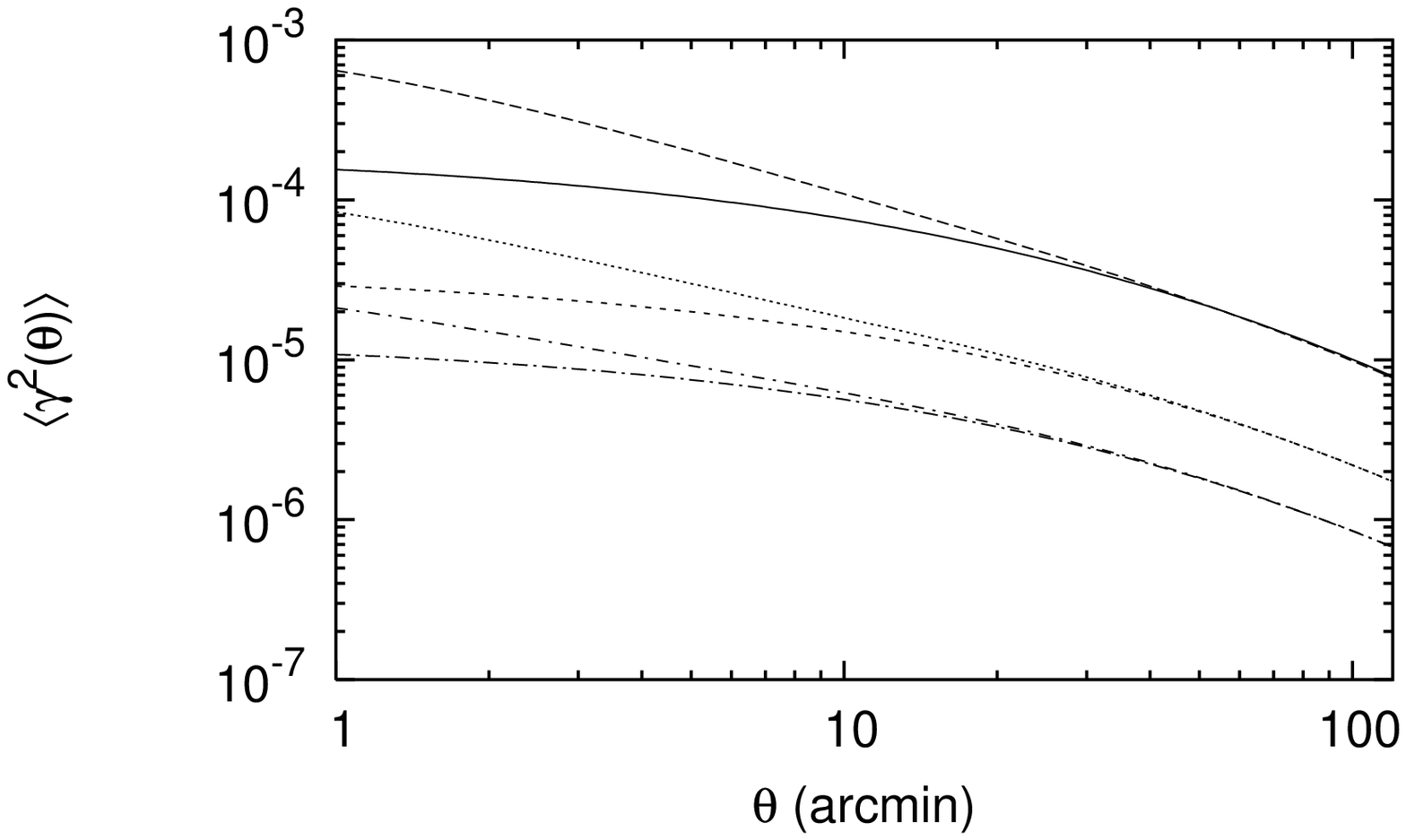,width=6cm}}
 \caption{(left) Convergence power spectrum $P_\kappa (\ell)$ as a function
  of the multipole $\ell$ and the 2-point statistics of the shear
  field: (middle) the aperture mass variance and (right) the shear
  variance for the reference $\Lambda$CDM model~(\ref{refLCDM})
 (solid/long-dashed lines) and two quintessence models
 ($m = 6$, short-dashed/dotted line; $m = 11$,
 long-dashed-dotted/short-dashed-dotted line).}
 \label{fig2d}
\end{figure*}

As can be seen on Fig.~\ref{fig3bc}, matter perturbations grow more
slowly than in a $\Lambda$CDM model. The CMB normalization implies
that at lower redshift the amplitude of density perturbations are
smaller in quintessence models. Thus, a given scale enters the
non-linear regime later than in a $\Lambda$CDM.  This time delay
accounts for the sharp modification of the spectrum at small
wavenumbers. Table~\ref{tab:Qbkg} gives the order of magnitude of the
expected effects on the background quantities, the growth factor and
the lensing observables. For instance let us concentrate of a
quintessence model with $m=6$. Table~\ref{tab:XIQCDMwl} shows that the
convergence power spectrum is 80\% to 85\% smaller in quintessence
than in $\Lambda$CDM. Now, assume we normalize the power spectra on
linear scales at $z=0$, it will have implied to multiply the linear
power spectrum by $\sim7$ and the non-linear one by $\sim\sqrt{7}$.
Thus, we would have found that the non-linear regime will differ
roughly by 50\% from the $\Lambda$CDM. This conclusions are similar to
the ones of Ref.~\cite{berben} and show that the change of spectrum
between the linear and non-linear part of the spectrum sets strong
constraints on the time evolution of the dark energy.

%----------------------------------------------------------------------------------------
\section{Weak lensing in scalar-tensor theories}\label{sec4}

We now turn to the case where gravity is not described by general
relativity but by a scalar-tensor theory.

\subsection{Newtonian regime}

Before we discuss explicit models, we can try to evaluate and
discuss the expected effects on lensing observables. For that
purpose, let us first look at the perturbation equations in the
Newtonian regime. We consider modes with wavelengths smaller than
the Hubble length and also assume that the scalar field is light.

In the matter era and on sub-Hubble scales, the equation
(\ref{B7}) of Appendix~\ref{appb} reduces to
\begin{equation}\label{steq1}
 \psi - \phi = \frac{F_\varphi}{F} \delta\varphi,
\end{equation}
so that non-minimal coupling induces an extra-contribution to the
anisotropic stress, while Eq.~(\ref{B9}) reduces to a generalized
Poisson equation
\begin{equation}\label{steq2}
 F\Delta\phi = 4\pi G_*\rho a^2 \delta_m -\frac{F_\varphi}{2}\Delta\delta\varphi.
\end{equation}
The evolution equation (\ref{B5}) of the matter field takes the
form
\begin{equation}\label{steq3}
 \dot\delta_m = -\Delta V + 3\dot\psi
\end{equation}
and the Euler equation~(\ref{B6}) to
\begin{equation}\label{steq4}
 \dot V + \HH V = - \phi.
\end{equation}
This set has to be completed by the Klein-Gordon equation for the
scalar field evolution
\begin{equation}\label{steq5}
 \left(\Delta - U_{\varphi\varphi}a^2\right)\delta\varphi = F_\varphi\Delta(\phi-2\psi).
\end{equation}

\subsection{Effects and amplitudes}

Various effects are expected. As in the case of quintessence, the
Friedmann equation is modified so that the geometrical function
$q$ as well as the growth of density fluctuations are modified.

Besides, there are two specific effects which do not appear in
pure quintessence models due to the modification of Einstein
equations:
\begin{enumerate}
 \item The Poisson equation does not take its standard Newtonian form so
 that
 \begin{equation}
 \deff\not=\delta_m.
 \end{equation}
 \item The two gravitational potentials are not equal anymore so
 that
 \begin{equation}
 \Phi\not=2\phi.
 \end{equation}
By extension to the post-Newtonian formalism, we can define the
parameter $\gamma^\ppn$ in the cosmological context by
\begin{equation}
 \psi = \gamma^\ppn \phi
\end{equation}
so that $\Phi = [1 + \gamma^\ppn(z)]\phi$.
\end{enumerate}

\subsubsection{Linear regime}

The two equations of evolution (\ref{steq3}-\ref{steq4}) can be
combined to get the evolution of the density field
\begin{equation}
 \ddot \delta_m + 2\HH\delta_m - \Delta\phi = 3a^2\left(\dot\psi
 a^{-2}\right)^..
\end{equation}
When $U_{\varphi\varphi}$ is much smaller than the wavelength of
the modes we consider then  we can combine Eq.~(\ref{steq1}) and
Eq.~(\ref{steq5}) to get the fluctuation of the scalar field as
\begin{equation}\label{dphi}
 \delta\varphi \simeq -\frac{FF_\varphi}{F + 2 F_\varphi^2}\phi.
\end{equation}
In particular, this implies that
\begin{equation}
 \gamma^\ppn(z) - 1 = \frac{F_\varphi^2}{F + 2F_\varphi^2}
\end{equation}
similarly, as obtained in \S~\ref{IIB} in a non-cosmological
context. Inserting Eq.~(\ref{dphi}) in the generalized Poisson
equation (\ref{steq2}), one gets
\begin{equation}\label{poissoneff}
 \Delta\phi \simeq 4\pi G_\cav\rho a^2 \delta_m
\end{equation}
where $G_\cav$ is defined by Eq.~(\ref{Gcav}) and is indeed time
dependent. Using the Eq.~(\ref{defdeff}), this equation implies that
\begin{equation}\label{deffdm}
 \deff \simeq
 \frac{1}{2}\frac{G_\cav(z)}{G_\eff(0)}\left[1+\gamma^\ppn(z)\right]\delta_m
  \simeq \frac{F_0}{F}\left[1+\eta(\varphi)\right]\delta_m
\end{equation}
where $\eta(\varphi)\equiv 2F_\varphi^2/(2F+3F_\varphi^2)$. Today
$\eta(\varphi_0)$ is expected to be smaller than a few $10^{-3}$.

The density evolution follows an equation that is similar to the
pure Newtonian case but where the gravitational constant has to be
replaced by its time dependent value
\begin{equation}\label{DdeA}
 \ddot \delta_m + \HH\dot \delta_m -4\pi G_\cav \rho a^2\delta_m = 0.
\end{equation}
If we decompose the density field as in Eq.~(\ref{decompo}), we
obtain that the effective growth factor is
\begin{equation}
 D_\eff \simeq \frac{F_0}{F}\left[1+\eta\right]^2 D(a)
\end{equation}
where $D$ is solution of Eq.~(\ref{DdeA}). Written in terms of the
redshift, it takes the form
\begin{equation}
 D'' + \left(\frac{H'}{H} - \frac{1}{1+z}\right)D' -
 \frac{3}{2}\Omega_{m,0} (1+z)\frac{G_\cav(z)}{G_\eff(0)} D = 0,
\end{equation}
where $\Omega_{m,0}$ is defined by Eq.~(\ref{a_defOm}). It follows
that the shear power spectrum takes the form
\begin{eqnarray}\label{PkappaSTlr}
P_\kappa(\ell) &=& \frac{9H_0^4}{4}\Omega_{m,0}\int
 \left[\frac{g(\chi)}{a(\chi)}\right]^2\left(\frac{F_0}{F}\right)^2\left[1+\eta\right]^2
 \nonumber\\
 &&\qquad\qquad
 \times D^2(\chi)P_\delta\left[\frac{\ell}{S_K(\chi)},\chi\right]\dd\chi.
\end{eqnarray}

It follows from this expression that we can estimate $P_\kappa$ by
solving the background equations solely. In particular, because of
the form (\ref{poissoneff}) of the Poisson equation, we do not
expect effects as the ones described in Ref.~\cite{ub00} on small
scales (see however Fig.~\ref{figautre}).

\subsubsection{Remark on the linear to non-linear
mapping}\label{lnlst}

The linear to non-linear mapping procedure has to be extended when
working in scalar-tensor theory, in particular because we will
have to deal with the scalar field perturbations. As
Eq.~(\ref{PkappaGEN}) shows, we must determine $P_\eff^{\rm nl}$.
For that purpose, we use the definition~(\ref{defdeff}) to define
$\deff$ and we decompose it as
\begin{equation}
 \deff = \delta_m + \delta_X
\end{equation}
where $\delta_X$ contains the contribution of the scalar field
perturbation and its derivative.

Assuming that the scalar field does not cluster, as it is the
case in quintessence, $\delta_X$ will not enter the non-linear
regime so that we can assume that
\begin{equation}
 \delta_X^{\rm nl} = \delta_X^{\rm lin}.
\end{equation}
In the Newtonian linear regime, Eq.~(\ref{defdeff}) implies that
\begin{equation}
 \deff^{\rm lin} = \frac{a/a_0}{3H_0^2\Omega_{m,0}}\Delta\Phi^{\rm
 lin}.
\end{equation}
The output of our CMB code gives access to $\Phi^\mathrm{lin}$ and
$\delta^{\rm lin}_m$ from which we can deduce $\deff^\mathrm{lin}$
and thus $\delta_X^\mathrm{lin}$.  With this ansatz, we get the
effective perturbation in the non-linear regime
\begin{equation}
 \deff^{\rm nl} = \delta_m^{\rm nl} + \delta_X^{\rm lin}.
\end{equation}
It follows that
\begin{equation}
 P_\eff^{\rm nl}(k,z) = P_m^{\rm nl}(k,z) + P_X^{\rm lin}(k,z) + 2\sqrt{P_m^{\rm nl}(k,z)P_X^{\rm
 lin}(k,z)}.
\end{equation}

When we are deeply in the Newtonian regime, but still in
the linear regime, Eq.~(\ref{deffdm}) implies that
$\delta_X\rightarrow (F_0/F -1)\delta_m$ so that
$$
 P_\eff^{\rm lin}(k,z) \longrightarrow
 \left(\frac{F_0}{F}\right)^2P_m^{\rm lin}(k,z),
$$
neglecting the contribution in $\eta$. In fact, notice that the ratio
$P^{\rm lin}_\eff/P^{\rm lin}_m$ evaluated today is always grater than
one, because of the non vanishing anisotropic stress of the radiation
and non-minimally coupled scalar fields. In the non-linear regime,
$$
 P_\eff^{\rm nl}(k,z) \longrightarrow \left[\sqrt{P_m^{\rm nl}(k,z)} +
 \left(\frac{F_0}{F}-1 \right)\sqrt{P_m^{\rm lin}(k,z)}\right]^2
$$
so that at $z=0$, $P_\eff^{\rm nl}(k)\rightarrow P_m^{\rm nl}(k)$, as
expected from our assumption that the scalar field does not
cluster.

Hence, our ansatz amounts for an interpolation between the
super-Hubble regime where the contribution of the scalar field
perturbation cannot be neglected and the Newtonian regime where
the effective Poisson equation (\ref{poissoneff}) hold. Note that
it assumes that the scalar field does not enter the non-linear
regime. In fact the mapping procedure of Ref.~\cite{smithetal}
uses a halo model and it was shown~\cite{vdb} that in a spherical
collapse, the scalar field does indeed not cluster (see also
Ref.~\cite{carlo,amen1}). So, we can hope this interpolation to be
justified but, at the time being, we have no possibility to check
it.

\subsubsection{Effect on the shear power spectrum}\label{qrg3}

Following the approximation of Section~\ref{qrg2}, we can compare two
models which differ only by the theory of gravity (i.e. same
quintessence potential or $\Lambda$CDM models).

In the linear regime, we get
\begin{eqnarray}\label{ordregrandeurL}
 \left.\frac{P_\kappa^{ST}}{P_\kappa^{RG}}\right|_{\rm lin} &\sim&
 \left(\frac{\mathcal{W}_{1/2}^{ST}}{\mathcal{W}_{1/2}^{RG}}\right)^2
 \left(\frac{S^\Lambda_K(\chi_{1/2}^{RG})}{S^{ST}_K(\chi_{1/2}^{ST})}\right)^{2+n}
 \left(\frac{F_0}{F(z_{1/2})}\right)^2
 \nonumber\\
 &&\qquad\qquad
 \times\left(\frac{D^{ST}_{1/2}/D^{ST}_0}{D^{RG}_{1/2}/D^{RG}_0}\right)^2
 \frac{P_0^{ST}}{P_0^{RG}}
\end{eqnarray}
while in the non-linear regime we obtain
\begin{eqnarray}
 \left.\frac{P_\kappa^{ST}}{P_\kappa^{RG}}\right|_{\rm nl}  &\sim&
 \left(\frac{\mathcal{W}_{_{1/2}}^{ST}}{\mathcal{W}_{_{1/2}}^{RG}}\right)^2
 \left(\frac{S^{RG}_K(\chi_{_{1/2}}^{RG})}{S^{ST}_K(\chi_{_{1/2}}^{ST})}\right)^{2}
 \left(\frac{F_0}{F(z_{_{1/2}})}\right)^2
 \nonumber\\
 &&\qquad\qquad
 \times\left(\frac{z_{_{1/2}}^{RG}}{z_{_{1/2}}^{ST}}\right)^3
 \left(\frac{P_0^{ST}}{P_0^{RG}}\right)^{3/(5+n)}.
\end{eqnarray}

\begin{figure}[htb]
 \centerline{\epsfig{figure=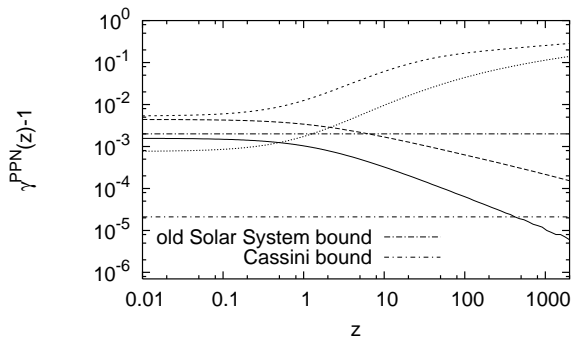,width=8cm}}
 \caption{Deviation from general relativity parameterized by
 $\gamma_{\rm PPN}(z)-1$ for several non-minimally coupled models with
 Ratra-Peebles potential, Eq.~(\protect{\ref{eqn:RP}}), as a function
 of redshift $z$. Solid line (long dashed line): non-minimal coupling,
 $\xi = +0.001$ and $m = 6 \;(m = 11)$ [potential defined in Jordan
 frame]. Short dashed line (dotted line): exponential Damour-Nordvedt
 type coupling, with $\beta = 4, B = 0.5$ and $m = 6 \;(m = 11)$
 [potential defined in Einstein frame]. Horizontal lines: level of the
 upper bounds measured by gravitational experiment on Solar System
 scales ($z\sim 0$).}  \label{fig7}
\end{figure}

The main contribution to the modification of the shear power
spectrum, and consequently on the 2-points statistics, is expected
to arise from the evolution of the matter power spectrum. Because
of the normalization to the CMB, i.e. at high redshift, its
amplitude evaluated today accounts for integrated effect over a
wide redshift range. Therefore it leaves trace on the shear power
spectrum, even if the computation of the shear power spectrum
involves an integration just up to the source redshift $z_s$.
Instead, as for quantities depending on the redshift range
$[0,z_s]$, eventually evaluated at $z_{1/2}$, their contribution
can become relevant only if $z_{1/2}$ is sufficiently high. In
particular, for $z_{1/2}\simeq 0.5$ the deviation from general
relativity described by $F_0/F_{1/2}$ is negligible (see
Fig.~\ref{fig7}, hence the peculiar effects of scalar-tensor
theories are ultimately encoded in the amplitude of the matter
power spectrum evaluated today. However notice that, accounting
for sources at higher redshift, where the deviations from general
relativity could become significant, and looking at lenses at low
redshift, one should detect deviation of order 10\% on
weak-lensing observables, peculiar of a scalar-tensor theory.

%---------------------------------------------------------------------------

\section{Investigation of two explicit models}\label{sec6}

We consider two test models to discuss in more details the
amplitude of scalar-tensor theories on lensing observations.
Indeed, scalar-tensor theories have two free functions that need
to be specified, which open a parameter space much larger than a
standard $\Lambda$CDM. Any quintessence model is specified by its
potential $U(\varphi)$. To embed it in scalar-tensor theories, one
has two possibilities: (1) assume that the potential in Jordan
frame, $U(\varphi)$ is the quintessence potential and choose a
coupling function $F(\varphi)$ or (2) assume that the potential in
Einstein frame, $V(\varphi_*)$, is the quintessence potential and
choose a coupling function $\alpha(\varphi_*)$. The second
possibility is probably more secure from a theoretical point of
view since it amounts to specify the property of the true spin-0
degree of freedom of the theory.

We will however investigate the two possibilities. We first
consider in \S~\ref{subsec6a} the case of a non-minimally coupled
scalar field and in \S~\ref{subsec6b} the case of scalar-tensor
models that incorporate attraction toward general relativity.

\begin{figure*}[htb]
 \centerline{\epsfig{figure=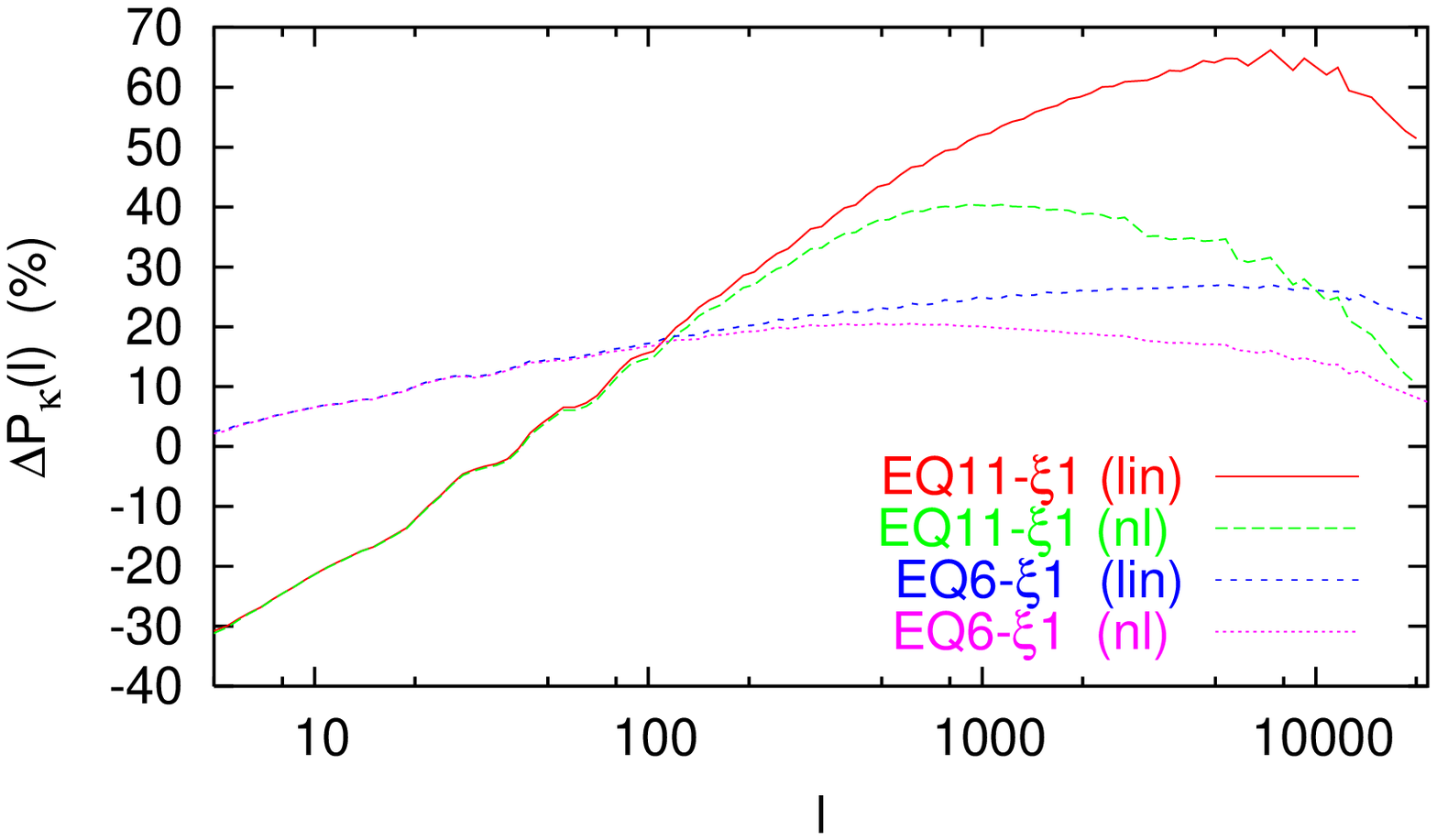,width=6cm}
             \epsfig{figure=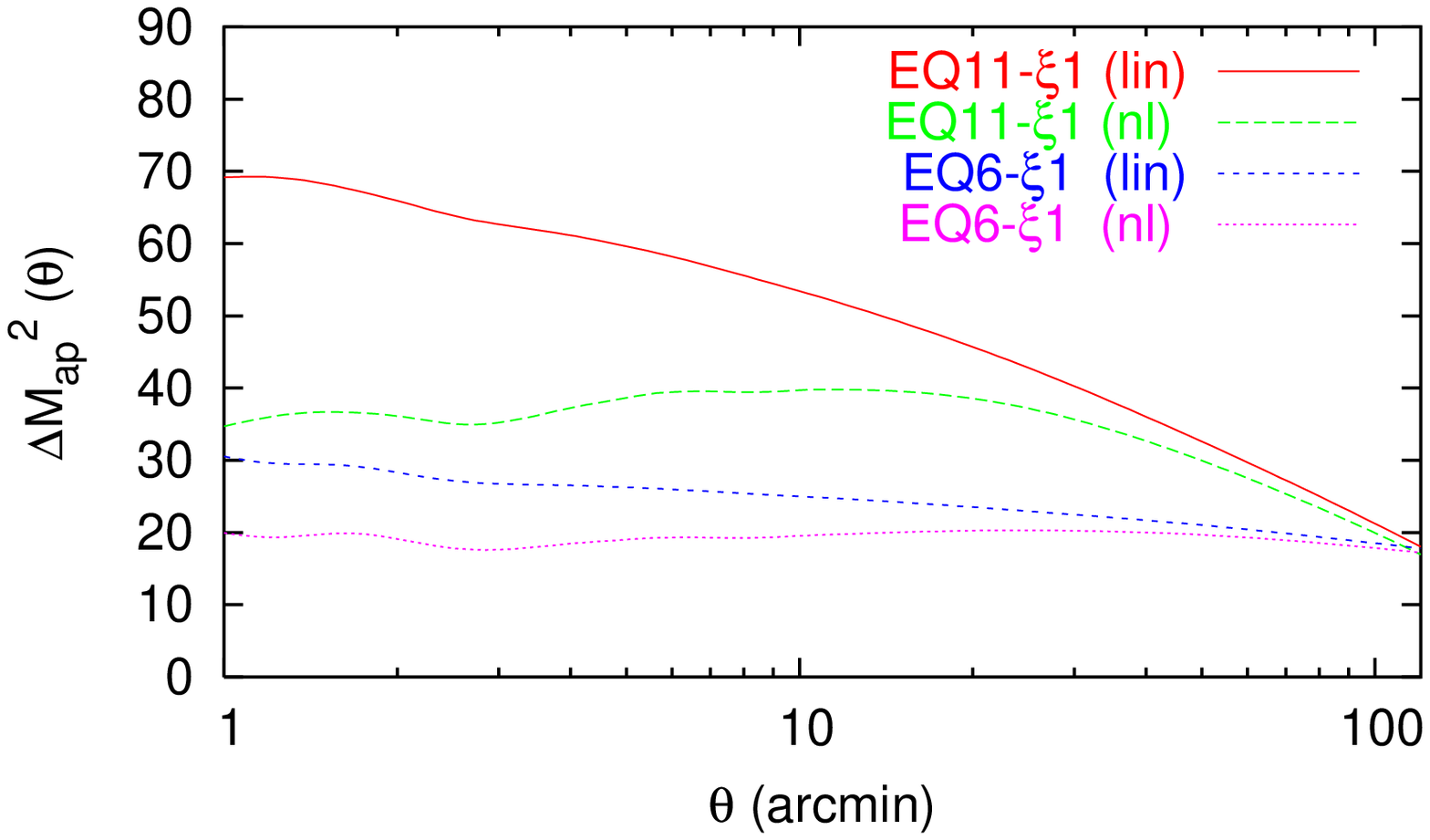,width=6cm}
             \epsfig{figure=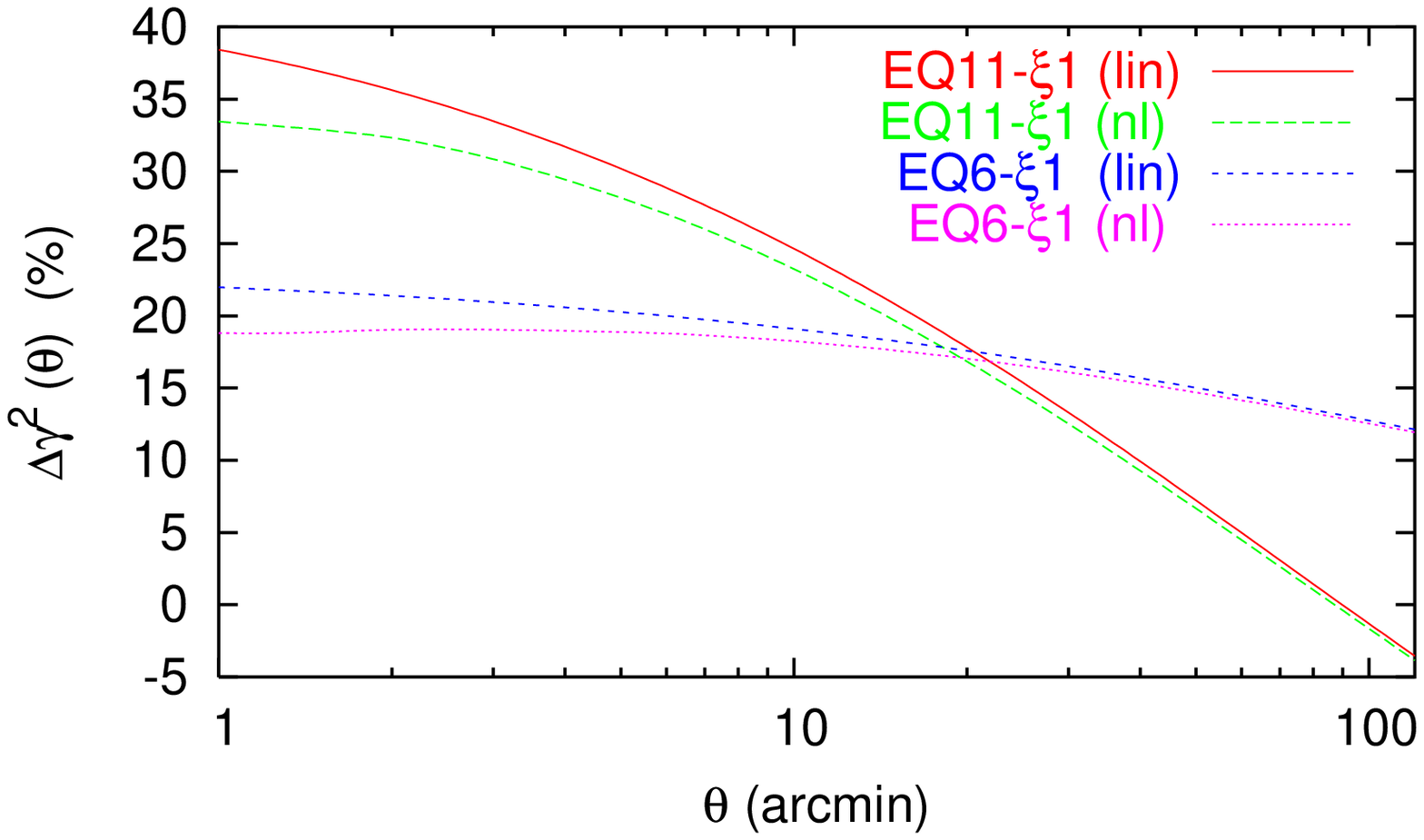,width=6cm}}
 \caption{Comparison between an extended quintessence model with
 $\xi=10^{-2}$ and $m=6$ and a quintessence model with same
 potential. (left) convergence power spectrum
 $([P_\kappa]_{\rm{EQ}-\xi}/[P_\kappa]_{\rm QCDM} -1)$,
 (middle) aperture mass variance and (right) and shear variance.
  Solid (long dashed): linear (non-linear) regime for Ratra-Peebles potential with $m = 6$.
  Short dashed (dotted): linear (non-linear) regime for Ratra-Peebles potential with $m =
  11$.}
 \label{fig8}
\end{figure*}

%---------------------------------------------------
\subsection{Non-minimally coupled case}\label{subsec6a}

We consider the simplest model we can think of for which the
coupling of the scalar field to gravity is described by the
function
\begin{equation}
 F(\varphi) = 1 + \xi\varphi^2.
\end{equation}
The constraint (\ref{systsolcon}) implies that
\begin{equation}\label{xic1}
 \left|\xi\varphi_0\right| < 0.55\times10^{-3}.
\end{equation}
Generically~\cite{ru02}, $\varphi_0\sim\mathcal{O}(1)$ today so
that this constraints implies that
\begin{equation}\label{xic2}
 \left|\xi\right|\lesssim 0.5\times10^{-3}.
\end{equation}
In this class of models, all deviations from general relativity
will scale like $\xi$ so that we expect $\delta P_\kappa/P_\kappa
\sim\mathcal{O}(\xi)$, that is very small effects.

\subsubsection{$\xi\Lambda$CDM: effect of the coupling}

To quantify the effect of this coupling, we first assume that
$\varphi$ has no potential and that there exists a cosmological
constant.

At this point, we should stress that there is no unique way of
defining a cosmological constant in scalar-tensor theories. Either
one adds this constant in Jordan frame, in which case the
associated energy density is constant. But, from
Eq.~(\ref{jf_to_ef3}), this induces a potential for the true
spin-0 degrees of freedom of the theory. On the other hand, one
could argue that imposing a constant potential in Einstein frame
lets the scalar field massless hence generalizing a constant
potential. But, it will not correspond to a constant energy
density in the physical (Jordan) frame. Let us also mention that
adding a cosmological constant in Jordan frame modifies the
potential as $U\rightarrow U + \Lambda$, corresponding to the
change $V\rightarrow V + \Lambda A^4/2$. It follows that the
spin-0 degree of freedom can remain massless only if
$$
 V''(\varphi_*) + 2\Lambda A^4(\varphi_*)\left[4\alpha^2+\beta\right] = 0.
$$

We have considered various models with $U(\varphi)=\Lambda$ and
with the constraint $\Omega_\Lambda = 0.7$ for various value of
$\xi$, typically ranging from $-0.005$ to $0.1$. None of them
exhibit any departure from the standard $\Lambda$CDM reference
model~(\ref{refLCDM}), as expected from our general arguments.

\subsubsection{Extended quintessence}

We now take into account both effects of a non-minimal coupling
and of a runaway potential in Jordan frame. We consider models with
 $m=6,\, 11$ and $\xi = 10^{-2}, 10^{-3}, 10^{-4}, -5\times 10^{-3}$.
We will refer to these models by the label $\mathrm{EQ}m-\xi n$,
where $m$ stands for the exponent of the runaway potential and $n$
for the value of the coupling ($n = 1,\dots , 4$ for
$\xi=10^{-2}\ldots -5\times10^{-3}$, respectively).

As expected, the main contribution to the deviation on the shear
power spectrum and consequently on the 2-points statistics come
from the amplitude of the matter power spectrum evaluated today.
The relative differences occurring in the (comoving) angular
distance, in the window function and in the growth factor at
$z=z_{1/2} \simeq 0.5$ are negligible. Notably, the evolution of
the coupling function $F(\varphi)$ is negligible. Anyway, the
scaling in $\xi$ is obvious (see Tables~\ref{tab:EQbkg}
and~\ref{tab:EQpert}).

As can be shown on Fig.~\ref{fig8} we obtain an effect of order
20\% for $\xi=0.01$, a value excluded by Solar System constraints,
see Fig.~\ref{fig7}. For $\xi = 10^{-3}$, our results agree with
those of Ref.~\cite{acqua04}. They argue that, taking into account
the different normalization of the scalar field compared to our
work,
$$
 \delta P_\kappa/P_\kappa \sim -16\xi\varphi_0^2,
$$
hence leading to a $\sim2\%$ effect on the shear power spectrum once the
constraints (\ref{xic1}-\ref{xic2}) are taken into account. In
general, since the effect scales like $\xi$, we can conclude that for
values compatible with Solar System constraints, the amplitude of the
effects will be smaller than a percent. Such a low value makes these
models, in practice, indistinguishable from their corresponding
quintessence models.

\subsection{Attraction toward GR}\label{subsec6b}

\subsubsection{Mechanism}

Another interesting class of models is the one in which the
scalar-tensor theory is attracted toward general relativity today.
This feature is better described in Einstein frame. The Klein
Gordon equation (cf. Eq.~\ref{KG_EF}) can be rewritten in terms of
the new variable
\begin{equation}
 p = \ln a_*
\end{equation}
as
\begin{eqnarray}
 && \frac{2[1+v(\varphi_*)]}{3-\varphi_*^{'2}}\varphi_*''
 +[1-w+2v(\varphi_*)]\varphi_*' =\nonumber\\
 &&\qquad\qquad\quad -\alpha(\varphi_*)(1-3w) - v(\varphi_*)
 \frac{\dd \ln V}{\dd\varphi_*}
\end{eqnarray}
where a dash temporarily refers to a derivative with respect to
$p$, $w$ refers to the equation of state of all matter fields but
$\varphi_*$ and where we have introduced the reduced potential
\begin{equation}
 v(\varphi_*) = \frac{V(\varphi_*)}{4\pi G_*\rho_*}.
\end{equation}
The coupling function $A(\varphi_*)$ is decomposed as
\begin{equation}
 A(\varphi_*) = \hbox{e}^{a(\varphi_*)}
\end{equation}
so that $\alpha = \dd a/\dd\varphi_*$.

The mechanism of attraction is then well illustrated by the
original model~\cite{dn1} in which the scalar field potential is
flat and where $a(\varphi_*)$ is quadratic. Setting
$a(\varphi_*)=\beta\varphi_*^2/2$, the Klein-Gordon equation takes
the form of the equation of motion of a particle with
velocity-dependent inertial mass,
$m(\varphi_*)=2/(3-\varphi_*^{'2})$, subject to a damping force
$-(1-w)\varphi_*'$ in a potential, $(1-3w)a(\varphi_*)$,
\begin{equation}
 \frac{2}{3-\varphi_*'{}^2}\varphi_*''
 +(1-w)\varphi_*' = -\beta(1-3w)\varphi_*.
\end{equation}
The positivity of the energy implies that $m(\varphi_*)>0$. During
radiation era $w=1/3$ and the field is decoupled from the
potential and will tend to a constant value, $\varphi_{*,R}$,
whatever its initial velocity. During the matter era, the
evolution of $\varphi_*$ is the one of a damped oscillator that
starts with a vanishing initial velocity. $\varphi_*$ will thus
moved toward the minimum of $a(\varphi_*)$ where $\alpha=0$, if
$\beta>0$. That is the scalar-tensor theory will become infinitely
close to general relativity.

\subsubsection{Massless scalar field models}

We first investigate a class of models with a vanishing potential
and a quadratic coupling defined in the Einstein frame, as in the original Ref.~\cite{dn1}. The coupling
function takes the form
\begin{equation}\label{aquad}
 a(\varphi_*) = a_m + \frac{1}{2}\beta(\varphi_* - \varphi_m)^2
\end{equation}
where $\varphi_m$ is the value of the field at minimum. The
functions $\alpha$ and $\beta$ defined in Eqns.~(\ref{eqalpha})
and~(\ref{eqbeta}) are thus given by
\begin{equation}
 \alpha(\varphi_*) = \beta(\varphi_* - \varphi_m),\qquad
 \beta(\varphi_*) = \beta.
\end{equation}
The coupling (\ref{aquad}) can be rewritten as
\begin{equation}
a(\varphi_*) = a_0 + \alpha_0(\varphi_*-\varphi^*_0) +
\frac{1}{2}\beta(\varphi_* - \varphi_0)^2
\end{equation}
where $\varphi_0$ is the value of $\varphi_*$ today. Without loss
of generality, a redefinition of units allows to reduce to this
form
\begin{equation}
 a(\varphi_*) = \frac{1}{2}\beta\varphi_*^2.
\end{equation}
The post-Newtonian constraints (\ref{boundgamma}-\ref{boundold})
imply that
\begin{equation}\label{bounddn}
 \alpha_0^2 <10^{-5}, \quad
 -8.5\times10^{-4} < \alpha_0^2(1+\beta_0) < 1.5\times 10^{-4}.
\end{equation}
A detailed analysis~\cite{dp} of the primordial nucleosynthesis in
this particular model also give constraints on $(\alpha_0,\beta)$.

During the matter era, assuming $\varphi_*'{}^{2}\ll 3$, the
Klein-Gordon equation simplifies to
\begin{equation}
\frac{2}{3}\varphi_*'' + \varphi_*' + \beta\varphi_*=0.
\end{equation}
According to the value of $\beta$, it has two different kinds of
solutions. For $\beta<3/8$, that is for a small curvature of the
coupling,
\begin{equation}
 \varphi_* (z_*) = A_+ (1+z_*)^{3(1+r)/4} + A_- (1+z_*)^{3(1-r)/4}
\end{equation}
where we have set $r=\sqrt{1-8\beta/3}$. The solution compatible
with the bound (\ref{bounddn}) is thus given by
\begin{equation}
 \varphi_* (z_*) = \varphi_0 (1+z_*)^{3(1-r)/4}.
\end{equation}

For $\beta>3/8$, we have damped oscillations around the minimum of
$a(\varphi_*)$
\begin{eqnarray}
 \varphi_* (z_*) &=& (1+z_*)^{3/4}
     \left\lbrace
     A\cos\left[\frac{3}{4}r\ln(1+z_*)\right]\right.\nonumber\\
     &&\qquad\left.+ B\sin\left[\frac{3}{4}r\ln(1+z_*)\right]\right\rbrace.
\end{eqnarray}
In this situation, $G_\cav$ may be oscillating during the matter
era. In the radiation era, the solution for the scalar field can
be shown~\cite{dn1} to behave as
\begin{equation}
 \varphi_* = \pm\sqrt{3}\,\mathrm{arctanh}\sqrt{1-A(1+z)^{-2}} + B.
\end{equation}

\begin{figure}[htb]
 \centerline{\epsfig{figure=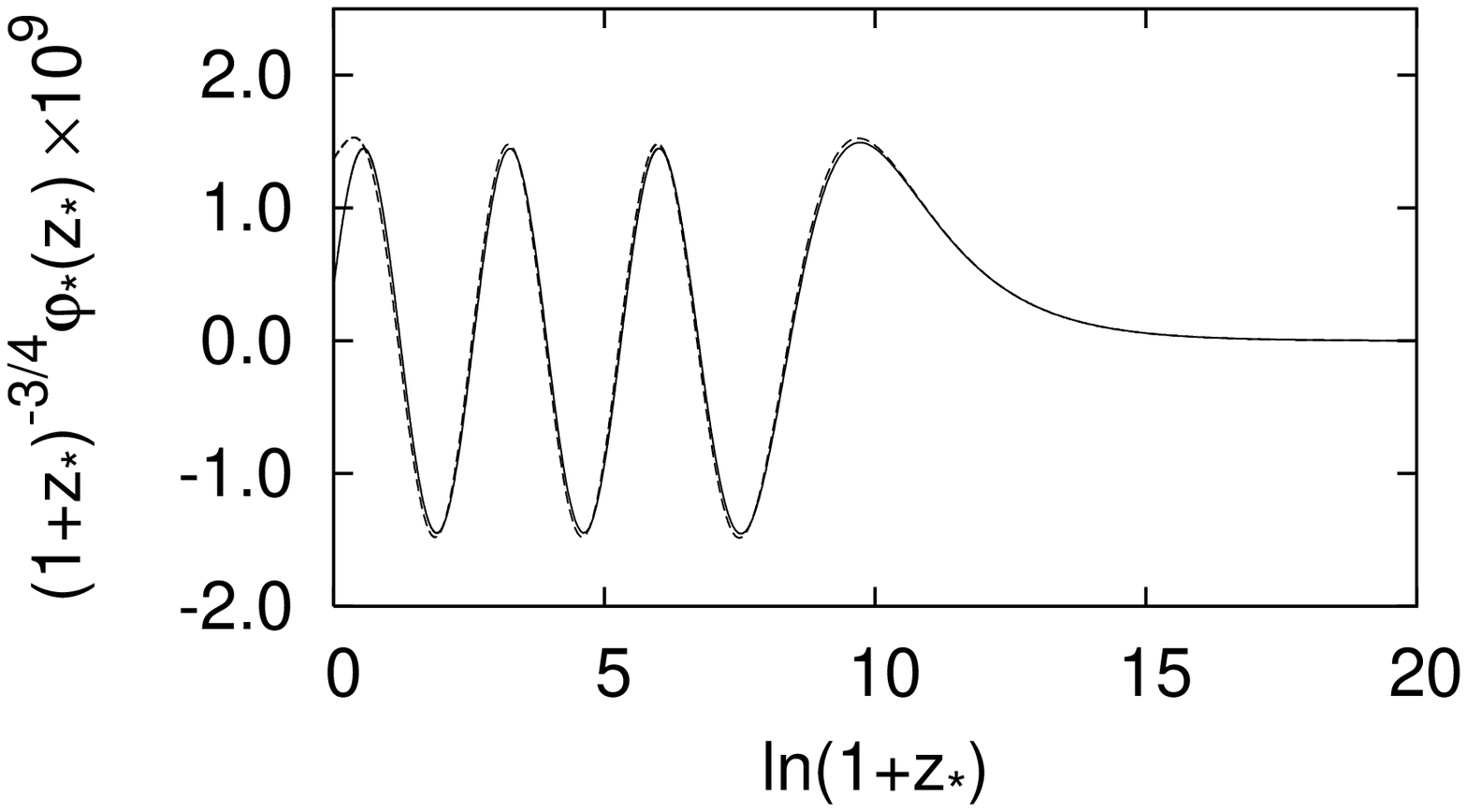,width=4.1cm}
             \epsfig{figure=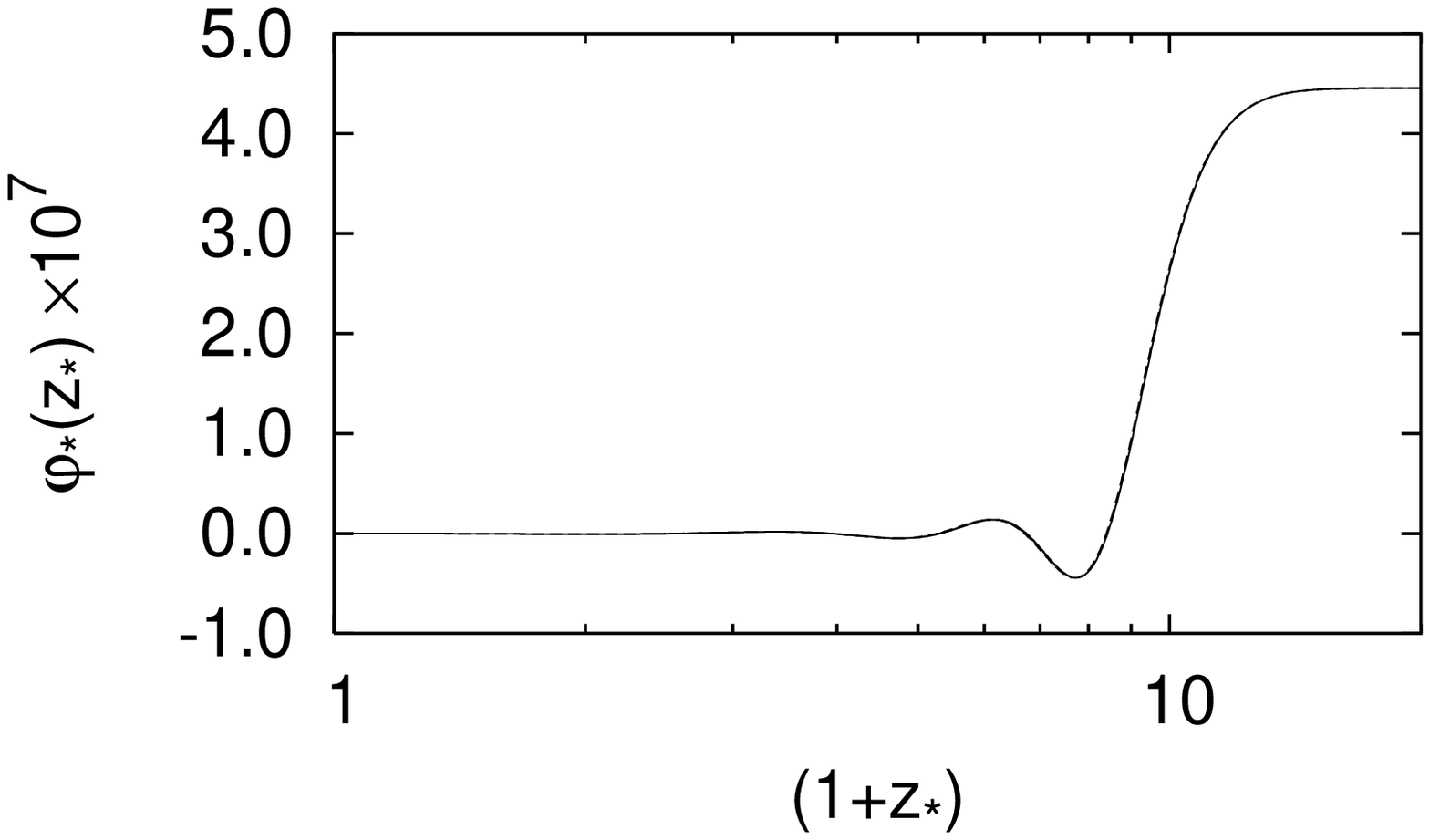,width=4.1cm}}
 \caption{Evolution of the scalar field in Einstein-frame in a model
 of massless dilaton with a quadratic coupling ($\alpha_0=10^{-4}$, $\beta=10^2$). We compare the case of a vanishing
 potential (solid) and a constant potential (dash). The solutions differ only recently.
 The period of the oscillation is compatible with the analytic result, as can be
 seen on the left plot where $\varphi_*$ has been divided by the damping
 dactor $(1+z_*)^{3/4}$.}
 \label{figoscill}
\end{figure}
In figure~\ref{figoscill}, we show that the numerically computed
evolution of the scalar field fits very well the previous analytic
solutions. Interestingly, the models considered in
Fig.~\ref{figoscill} are compatible with nucleosynthesis and Solar
System bounds. We can check that the oscillation of the scalar
field do not leave any imprint on both matter and CMB angular
power spectra.

\begin{figure}[htb]
 \centerline{\epsfig{figure=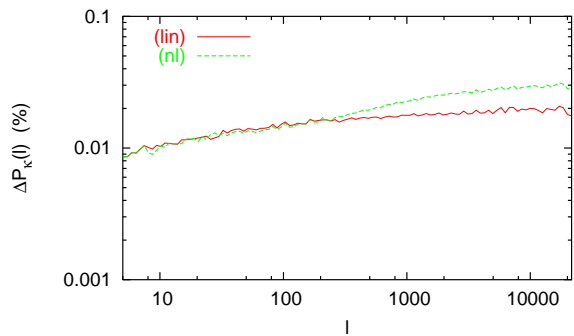,width=8cm}}
 \caption{Relative deviation $(P_\kappa/[P_\kappa]_{\rm SCDM} -1)$
 from a SCDM model ($\Lambda=0$) on the convergence power spectrum of the massless
 scalar fields models, with quadratic coupling
 with $(\alpha_0^2,\beta)=(10^{-4},0.1)$, as a function of the multipole $\ell$.
 Linear (solid) and non-linear (dashed) regime are presented. The field is attracted
 rapidly toward the minimum of the coupling function so that the model does not
 differ from a SCDM by more than 0.1\%.}
 \label{fig8bis}
\end{figure}

Figure~\ref{fig8bis} depicts the shear power spectrum for a model
with a massless dilaton with quadratic coupling (\ref{aquad}) with
$(\alpha_0^2,\beta)=(10^{-4},0.1)$. The scalar field is attracted
toward the minimum of the coupling very efficiently so that the
model can hardly be differentiated from the analogous model in
general relativity.  We emphasize this class of models, which have
also been constrained by BBN~\cite{dp}, do not account for the
late acceleration of the universe. Thus, we have compared them to
a standard cold dark matter model with $\Lambda=0$.

Note that adding a constant potential, that is a cosmological
constant defined in Einstein frame, lets the dilaton massless and
does not affect its dynamics apart from at very small redshift
(see Fig.~\ref{figoscill}).

As for non-minimal quadratic model, this class of model will not
leave any significant signature on lensing observables.

\subsubsection{Extended quintessence}

In the previous models, the scalar field accounts for a new
interaction but not for the acceleration of the universe that was
driven by a pure cosmological constant.

In the context of quintessence, extended quintessence models have
been widely considered. In particular, it was realized~\cite{bp}
that scalar-tensor quintessence models with attraction toward
general relativity can be constructed. This is the case for
instance when quintessence is constructed as a runaway
dilaton~\cite{runaway1,runaway2}. In this case, the coupling
function typically reduces to
\begin{equation}
 \alpha(\varphi_*) = -B \hbox{e}^{-\beta\varphi_*}
\end{equation}
and the potential takes the form
\begin{equation}\label{eqn:RP_EF}
 V(\varphi_*) = M_*^4\varphi_*^{-m}.
\end{equation}
During radiation era, the coupling is not effective and the scalar
field evolution will get its standard attractor solution. In the
matter era, the field will start by slow rolling. If we assume it
is always slow-rolling and that it explains the acceleration of
the universe today, then the Klein-Gordon equation reduces to
\begin{equation}
 \varphi_*' \simeq B \hbox{e}^{-\beta\varphi_*}
\end{equation}
whose solution is well approximated by
\begin{equation}
 \hbox{e}^{\beta\varphi_*} =  \hbox{e}^{\beta\varphi_0}
 + B\beta\ln(1+z_*).
\end{equation}
It follows that
\begin{equation}
 \alpha(z_*) = \frac{\alpha_0}{1-\beta\alpha_0\ln(1+z_*)}
\end{equation}
where $\alpha_0= -B\hbox{e}^{-\beta\varphi_0}$ and
$\beta_0=-\beta\alpha_0$. Besides, as in quintessence, the value
of $\varphi_0$, is obtained from the constraint on
$\Omega_\Lambda$. Note also that no clear bounds from BBN but the
general ones~(\ref{bbnbound}) have been inferred. Another
constraint on $\varphi'_0$ arises from the bounds on the time
variation of the gravitational constant.

\begin{figure*}[ht]
 \centerline{\epsfig{figure=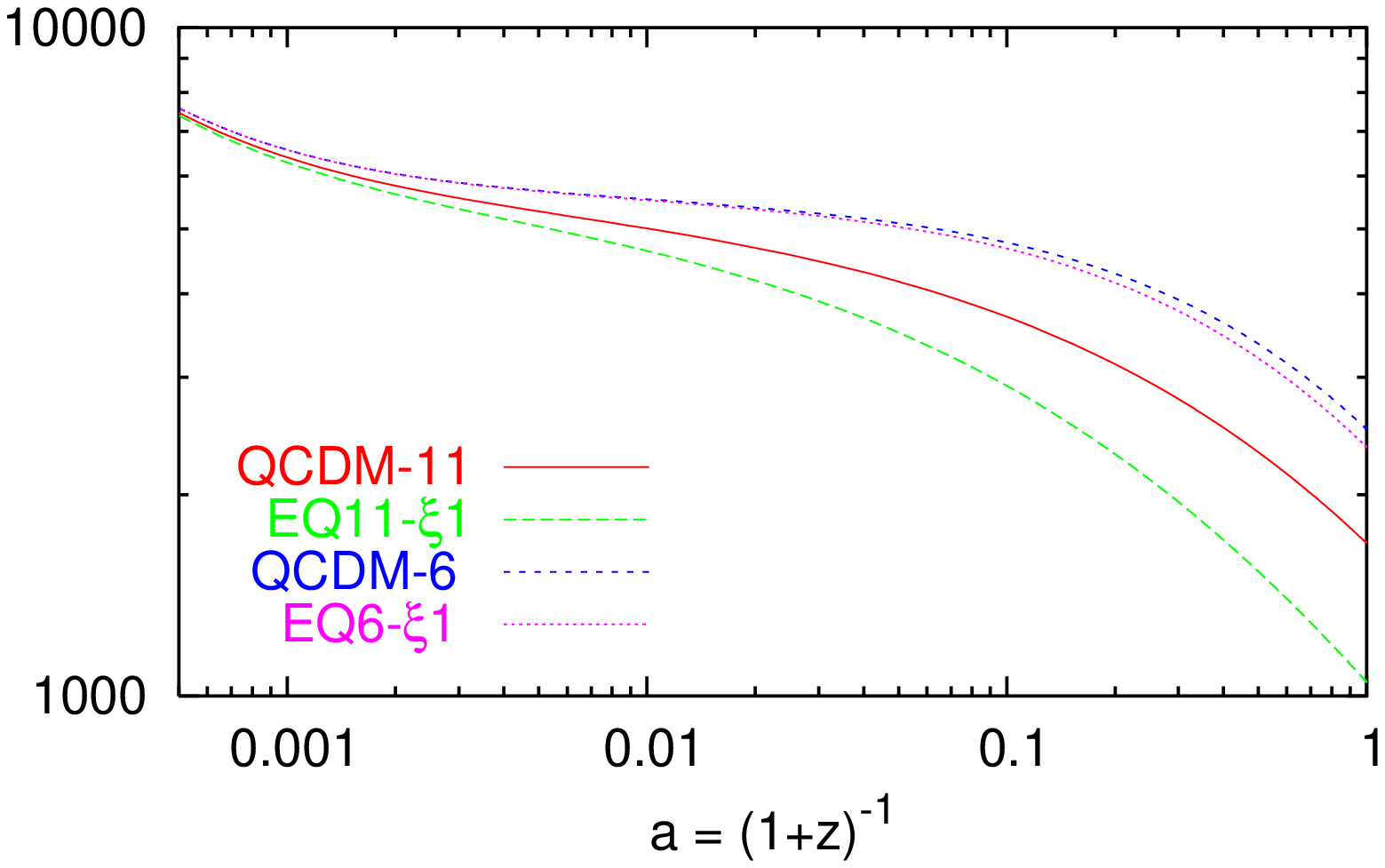,width=8cm}
             \epsfig{figure=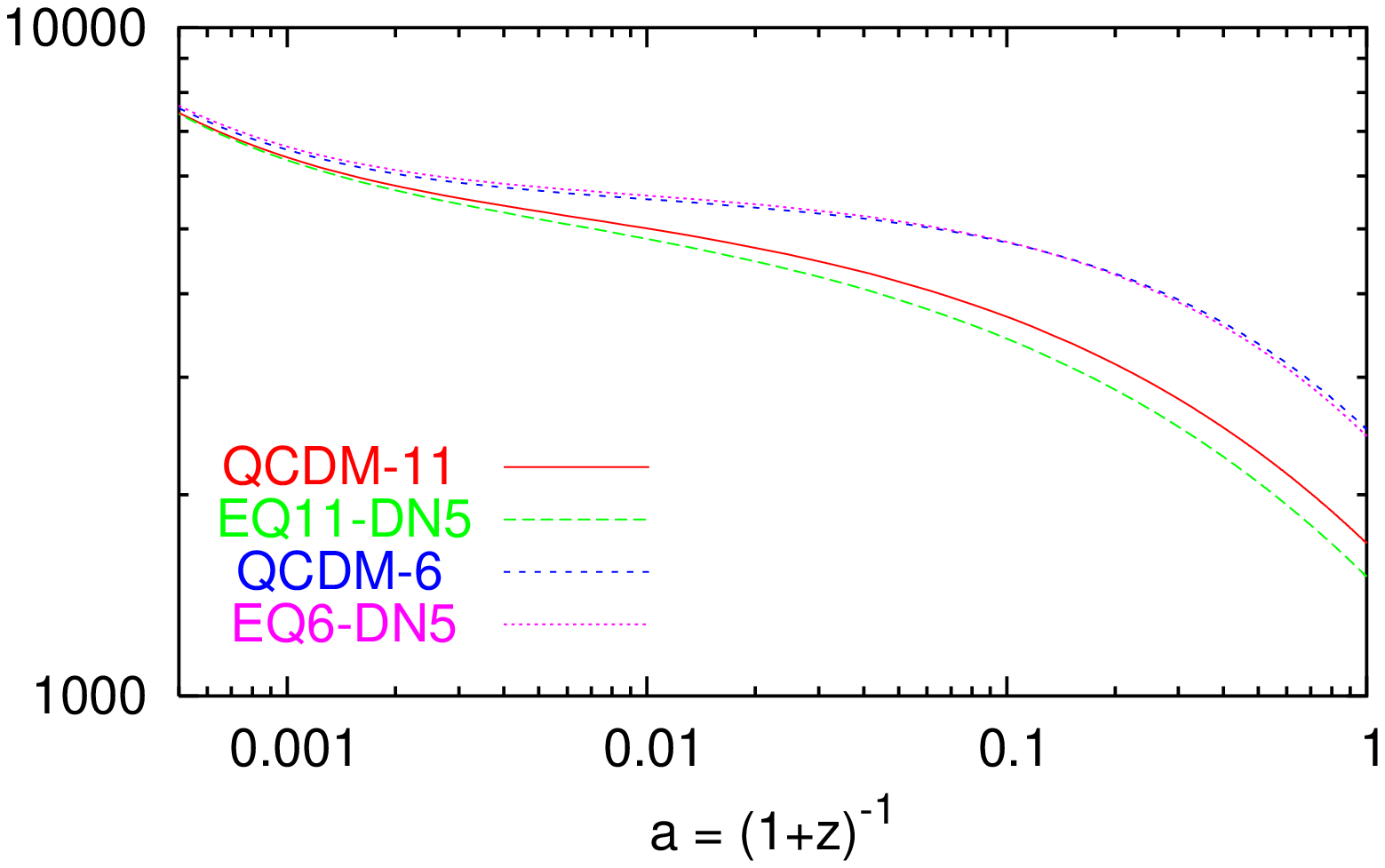,width=8cm}}
 \caption{Ratio $D(a)/a$ as a function of the scale factor $a$,
 normalized at high redshift, for extended quintessence models
 with quadratic ($\xi = +0.01$, left panel) and exponential
 coupling ($\beta = 4, B = 0.1$, right panel) and for the corresponding
 minimally coupled models. Notice that the potential are defined in different frame.}
 \label{fig3hi}
\end{figure*}

\begin{figure*}[htb]
 \centerline{\epsfig{figure=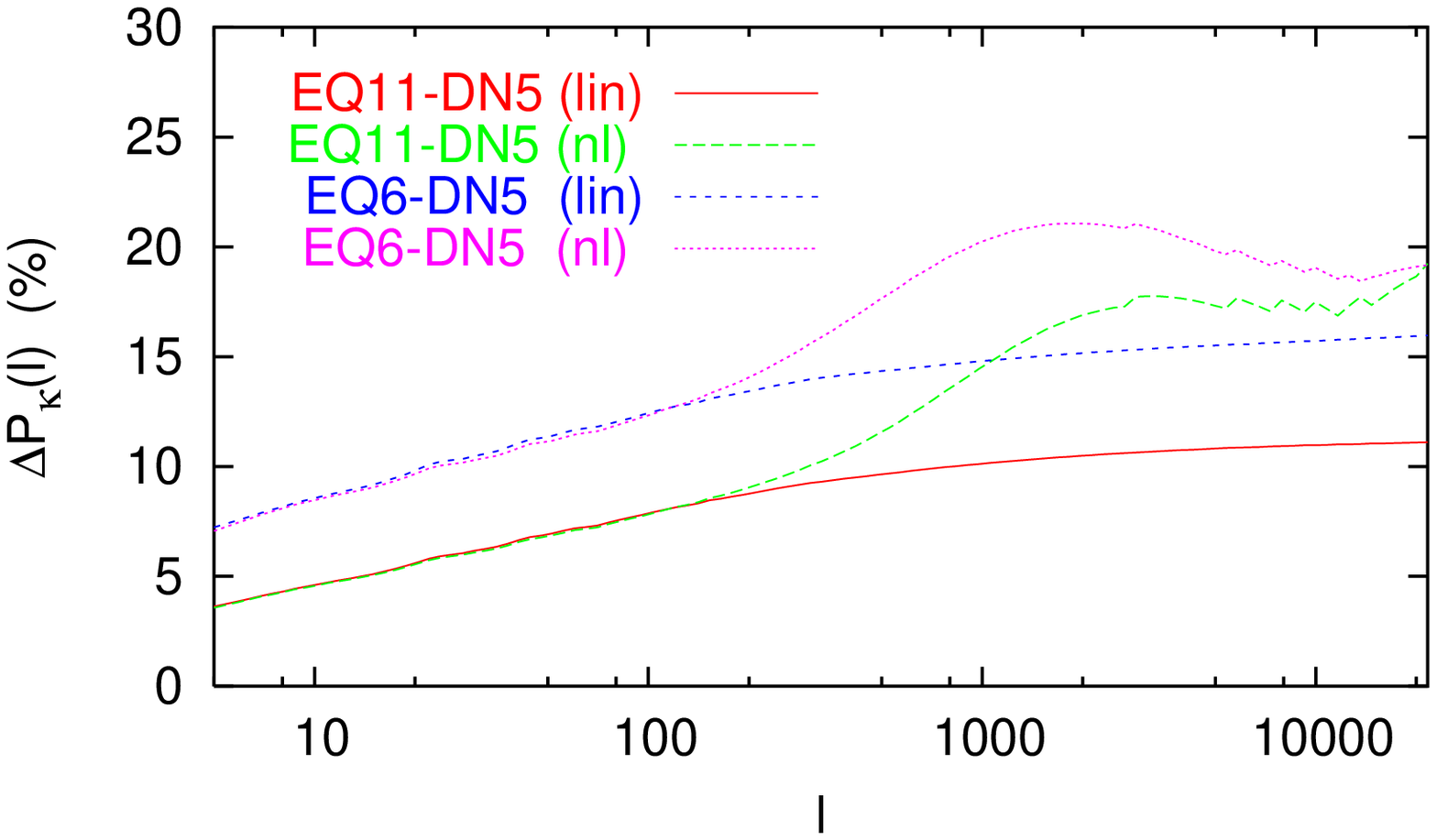,width=6cm}
             \epsfig{figure=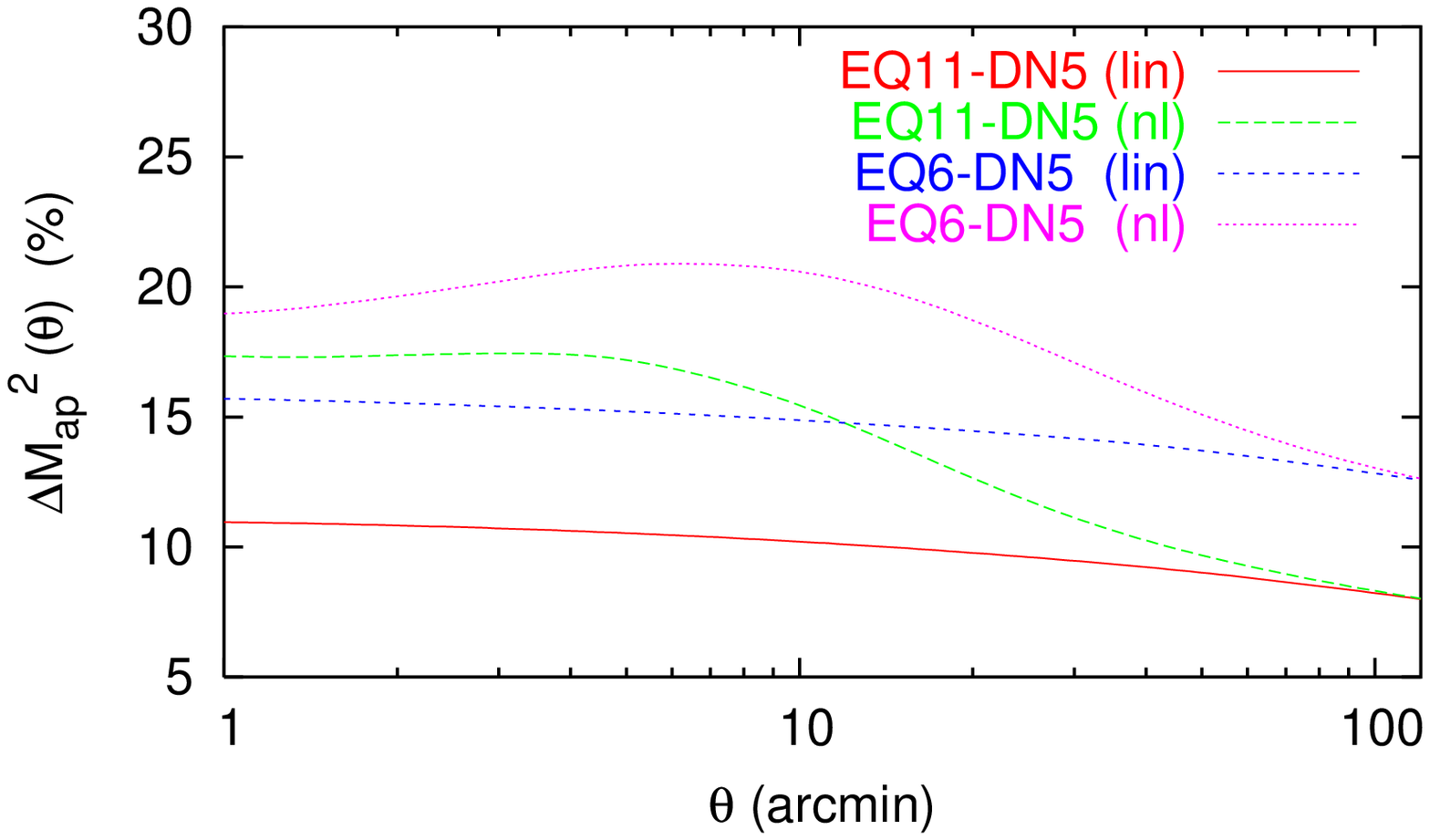,width=6cm}
             \epsfig{figure=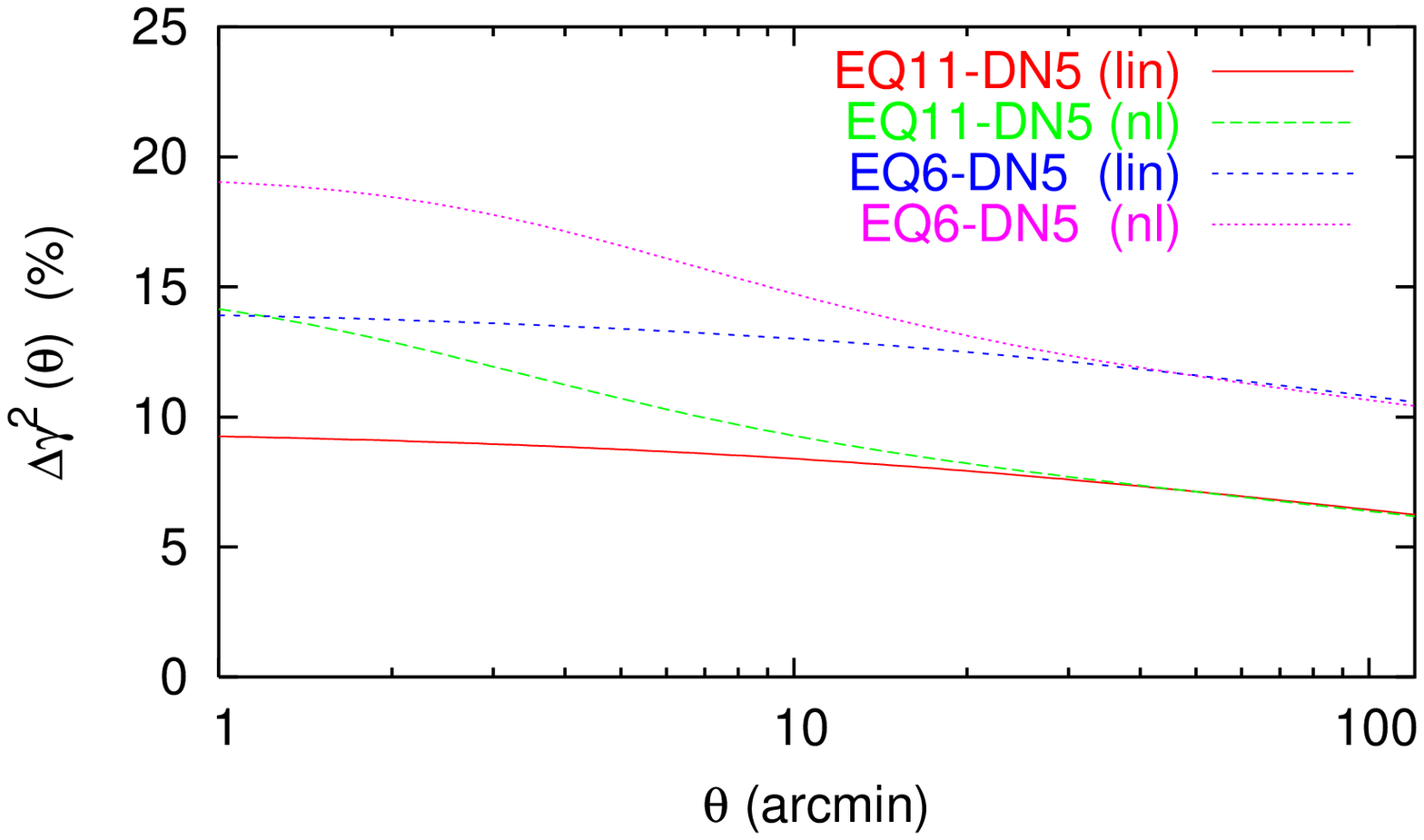,width=6cm}}
 \caption{Comparison between an runaway dilaton model with
 $\beta = 4, B = 0.1$ and $m=6,11$ and a quintessence model with same
 potential. (left) convergence power spectrum
 $([P_\kappa]_{\rm{EQ}-\xi}/[P_\kappa]_{\rm QCDM} -1)$,
 (middle) aperture mass variance and (right) and shear variance.
 Solid (long dashed): linear (non-linear) regime for Ratra-Peebles potential with $m = 6$.
  Short dashed (dotted): linear (non-linear) regime for Ratra-Peebles potential with $m =  11$.}
 \label{fig9}
\end{figure*}

\begin{table*}\label{tab:EQbkg}
\caption{\underline{Extended quintessence models}: background effects with respect
to the corresponding minimally coupled (QCDM) models (according to
\S~\ref{qrg3}). The models are labeled by $\mathrm{EQ}m-\xi
/\mathrm{DN}\, n$, the number $m$ defining the inverse power law
potential, Eq.~(\protect{\ref{eqn:RP}}), $\xi /\mathrm{DN}$
standing for the (quadratic) non-minimal coupling in Jordan frame
or for the exponential Damour-Nordvedt type coupling in Einstein
frame, and $n$ singling out the specific model, namely the
couplings parameters (see text). For the models $\mathrm{EQ}6-\xi
4$ and $\mathrm{EQ}11-\xi 1$it is $z_{1/2} \simeq 0.49$ and
$z_{1/2} \simeq 0.47$, respectively; otherwise it is $z_{1/2}
\simeq 0.48$.}
\begin{ruledtabular}
\begin{tabular}{lcccccc}
model &$\chi_{1/2}/\chi^Q_{1/2}$&$\mathcal{W}_{1/2}^2/(\mathcal{W}_{1/2}^2)^Q$&$(D_{1/2}/D_0)/(\cdots)^Q$&$P_0/P^Q_0$&$F_0/F_{1/2}$ \\
\hline
$\mathrm{EQ}6-\xi 1$&$1.021$&$1.031$&$1.006$&$1.114$&$1.0387$ \\
$\mathrm{EQ}6-\xi 2$&$1.000$&$0.994$&$1.000$&$1.008$&$1.0032$ \\
$\mathrm{EQ}6-\xi 3$&$1.000$&$0.999$&$1.000$&$1.005$&$1.0003$ \\
$\mathrm{EQ}6-\xi 4$&$1.031$&$0.011$&$0.989$&$0.982$&$0.0985$ \\
$\mathrm{EQ}6-\mathrm{DN}4$&$1.005$&$1.055$&$1.020$&$2.101$&$1.0081$ \\
$\mathrm{EQ}6-\mathrm{DN}5$&$1.000$&$1.002$&$1.004$&$1.135$&$1.0015$ \\
$\mathrm{EQ}11-\xi 1$&$1.005$&$1.066$&$1.044$&$1.211$&$1.0807$ \\
$\mathrm{EQ}11-\xi 2$&$1.000$&$0.989$&$1.002$&$1.006$&$1.0060$ \\
$\mathrm{EQ}11-\xi 3$&$1.000$&$0.999$&$1.000$&$1.000$&$1.0006$ \\
$\mathrm{EQ}11-\xi 4$&$1.029$&$1.116$&$0.980$&$0.994$&$0.0975$ \\
$\mathrm{EQ}11-\mathrm{DN}4$&$1.000$&$1.003$&$1.027$&$1.727$&$1.0026$ \\
$\mathrm{EQ}11-\mathrm{DN}5$&$1.000$&$1.000$&$1.005$&$1.099$&$1.0004$ \\
\end{tabular}
\end{ruledtabular}
\end{table*}

\begin{table*}
\label{tab:EQpert} \caption{\underline{Extended quintessence models}: maximum
deviation on the convergence power spectrum, $\Delta
P_\kappa(\ell)$, aperture mass variance, $\Delta \langle
M_{ap}^2(\theta)\rangle$,  and shear variance, $\Delta
\langle\gamma^2(\theta)\rangle$, from the corresponding minimally
coupled models (QCDM). Every weak lensing observable is evaluated
at two angular scales and the effects of the linear and non-linear
regime (values within parenthesis) are accounted for. In second
and third column, deviation from general relativity parametrized
by $\gamma^\mathrm{PPN}(z)-1$, Eq.~(\ref{gppn}), is quoted. Model
labeled as in table~(\ref{tab:EQbkg}).}
\begin{ruledtabular}
\begin{tabular}{l|cc|cccccc}
model &\multicolumn{2}{c}{$(\gamma^\mathrm{PPN}-1)\times 10^{-3}$}&\multicolumn{2}{c}{$\Delta P_\kappa(\ell)\; (\%)$}&\multicolumn{2}{c}{$\Delta \langle M_{ap}^2(\theta)\rangle\; (\%)$}&\multicolumn{2}{c}{$\Delta \langle\gamma^2(\theta)\rangle\; (\%)$}\\
 &$z=0$\footnotemark[1]&$z=1$&$\ell=180$\footnotemark[2]&$\ell=7200$\footnotemark[2]&$\theta=2^\circ$&$\theta=3^\prime$&$\theta=2^\circ$&$\theta=3^\prime$ \\
\hline
$\mathrm{EQ}6-\xi 1$&$140$&$105$&$20$&$27\;(16)$&$18$&$27\;(18)$&$12$&$19\;(21)$  \\
$\mathrm{EQ}6-\xi 2$&$1.6$&$1.0$&$1.8$&$2.2\;(0.8)$&$\lesssim2$&$\sim 2\;(1)$&$1$&$\lesssim 2\;(\lesssim 1)$  \\
$\mathrm{EQ}6-\xi 3$&$1.5\times 10^{-2}$&$0$&$0.7$&$0.7\;(0.7)$&$0.6$&$\lesssim 0.7\;(\lesssim0.7)$&$<0.6$&$\lesssim 0.7\;(0.8)$ \\
$\mathrm{EQ}6-\xi 4$&$31$&$20$&$4$&$8\;(2)$&$4$&$\lesssim 8\;(\lesssim 2)$&$4\;(2)$&$6\;(2)$ \\
$\mathrm{EQ}6-\mathrm{DN}4$&$5.3$&$125$&$107$&$133\;(165)$&$98$&$130\;(176)$&$79$&$110\;(154)$ \\
$\mathrm{EQ}6-\mathrm{DN}5$&$0.27$&$0.61$&$14$&$\lesssim 16\;(19)$&$13$&$16\;(19)$&$11$&$14\;(19)$ \\
$\mathrm{EQ}11-\xi 1$&$312$&$293$&$27$&$32\;(66)$&$18$&$63\;(35)$&$4$&$33\;(31)$ \\
$\mathrm{EQ}11-\xi 2$&$4.4$&$3.3$&$1.8$&$3.2\;(0.3)$&$\lesssim 2$&$3\;(\lesssim 1)$&$0$&$\lesssim 2\;(\lesssim1)$  \\
$\mathrm{EQ}11-\xi 3$&$4.2\times 10^{-2}$&$0$&$0.1$&$0.3\;(-0.1)$&$0.1$&$\lesssim 0.2\;(-0.1)$&$<0.1$&$\lesssim 0.2\;(0)$ \\
$\mathrm{EQ}11-\xi 4$&$67$&$51$&$4\;(18)$&$11\;(15)$&$5\;(20)$&$11\;(14)$&$0\;(26)$&$7\;(17)$ \\
$\mathrm{EQ}11-\mathrm{DN}4$&$0.32$&$0.83$&$59$&$77\;(129)$&$54$&$75\;(133)$&$41$&$61\;(88)$ \\
$\mathrm{EQ}11-\mathrm{DN}5$&$1.6\times 10^{-2}$&$4.9\times 10^{-2}$&$9$&$11\;(17)$&$8$&$11\;(17)$&$6$&$9\;(14)$ \\
\end{tabular}
\end{ruledtabular}
\footnotetext[1]{The ``old'' Solar System bound is
$|\gamma^\mathrm{PPN}-1|\lesssim 2\times 10^{-3}$. The ``new'' Cassini
bound is $\gamma^\mathrm{PPN}-1 = (2.1 \pm 2.3)\times 10^{-5}$.}
\footnotetext[2]{$\ell=180, 7200$ correspond to the angle $\theta
= 2^\circ, 3^\prime$ respectively.}
\end{table*}

To illustrate this scenario, we consider four flat quintessence
models with $\Omega_{\varphi,0}=0.7$, runaway potential
(\ref{eqn:RP_EF}) with $m=6, 11$ and
\begin{enumerate}
 \item $B=0.5$, $\beta = 4$, labelled by DN4; \item $B=0.1$, $\beta =
 4$, labelled by DN5.
\end{enumerate}
The two choices of the coupling parameters are chosen to satisfy
the ``old'' Solar System and ``new'' Cassini constraints,
respectively. These models are labelled by
$\mathrm{EQ}m-\mathrm{DN}n$, where $m$ defines the inverse power
law potential and $n = 4, 5$. We have to stress that the
deviations on BBN constraints for such a class of models need to
be studied.

We compare each model with its related quintessence model.
Figure~\ref{fig3hi} depicts the effect of the coupling on the
growth of density perturbations. As can be shown on
Fig.~\ref{fig9}, this implies effects of more than 10\% on the
shear power spectrum and thus on all the 2-point statistics. On
the scales of interest, we can safely use the relations quoted in
\S~\ref{qrg3}. Tables~\ref{tab:EQbkg} and~\ref{tab:EQpert} show
that the amplitude of the matter power spectrum evaluated today is
the main source of deviation on $P_\kappa$. Indeed this amplitude
takes into account the whole history of the modes since the CMB,
and in particular in eras where large deviations from general
relativity are now possible, see Fig.~\ref{fig7}. Interestingly,
in the shallow universe, deviations from general relativity are
small and effect on the background quantity around $z_{1/2}\simeq
0.5$ are of the order of a couple of percents. The comparison of
the number presented in tables~\ref{tab:EQbkg}
and~\ref{tab:EQpert} with the estimate (\ref{ordregrandeurL}) in
the linear regime agrees within a factor of a few percent.

For instance, the model labelled by $\mathrm{EQ}11-\mathrm{DN}5$
fits the Solar System (Cassini) constraints and lead to a
deviation of order 10\% from the corresponding minimally coupled
(QCDM) model on the shear power spectrum.

\begin{figure*}[htb]
 \centerline{\epsfig{figure=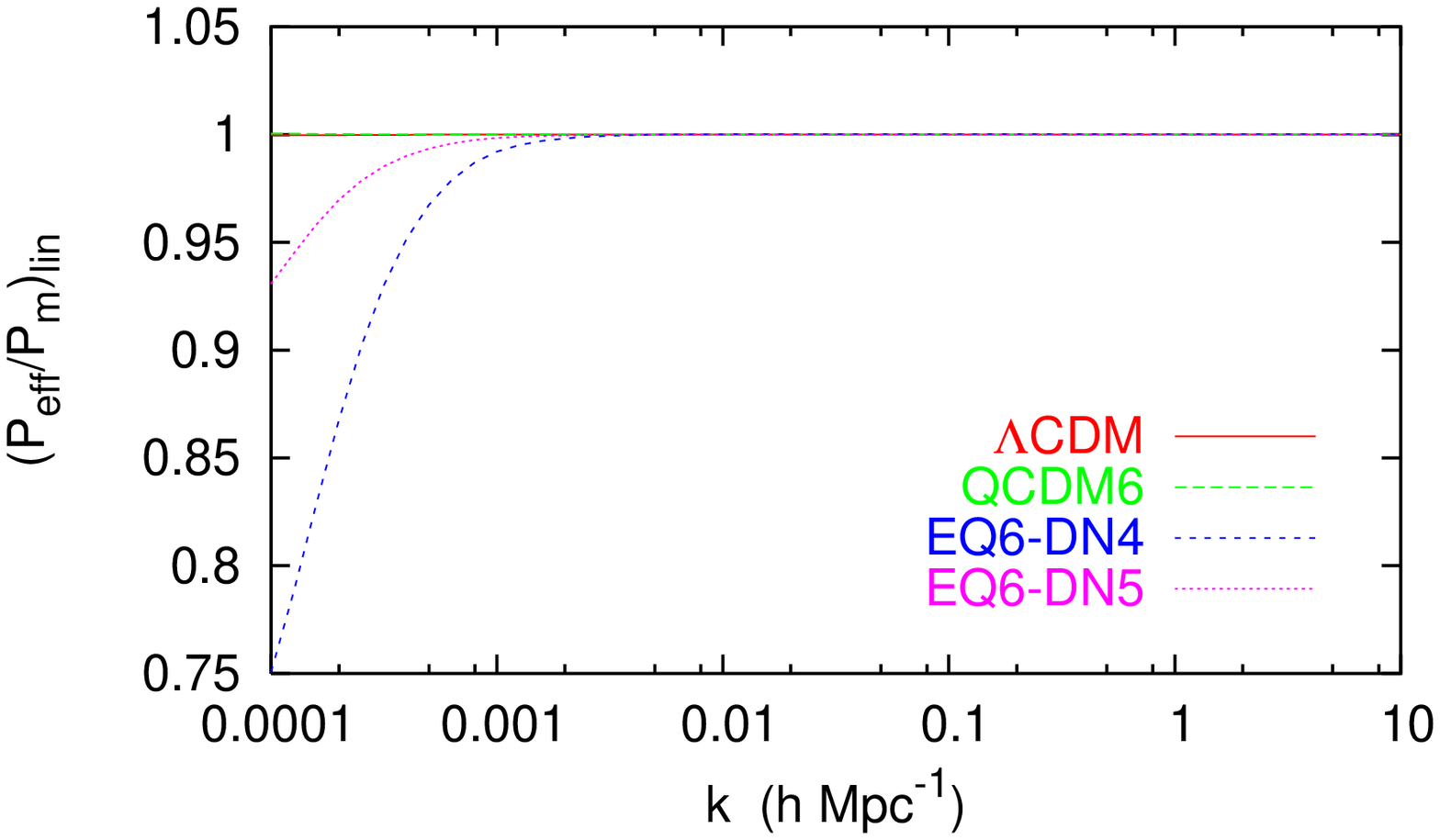,width=6cm}
              \epsfig{figure=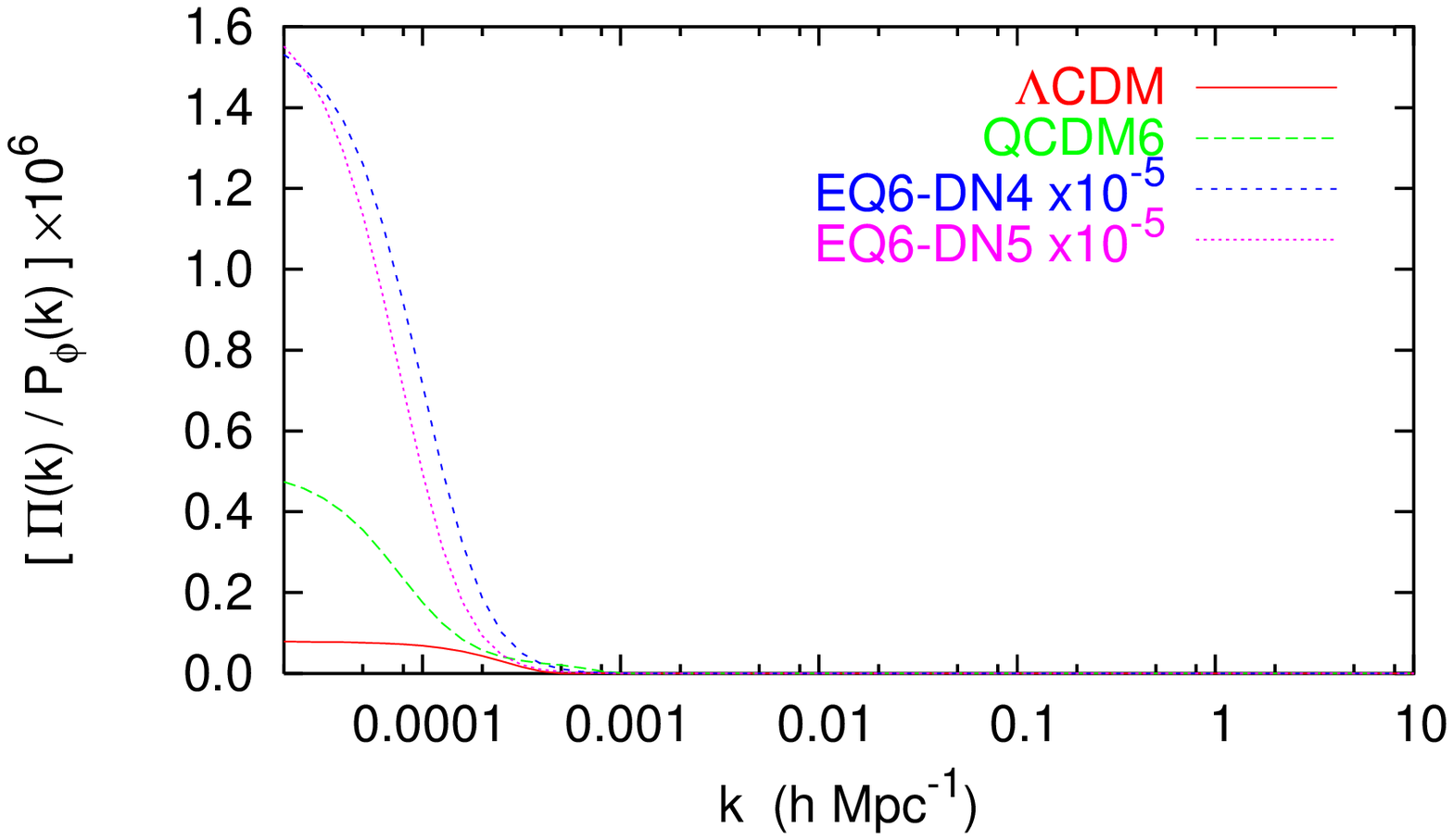,width=6cm}
              \epsfig{figure=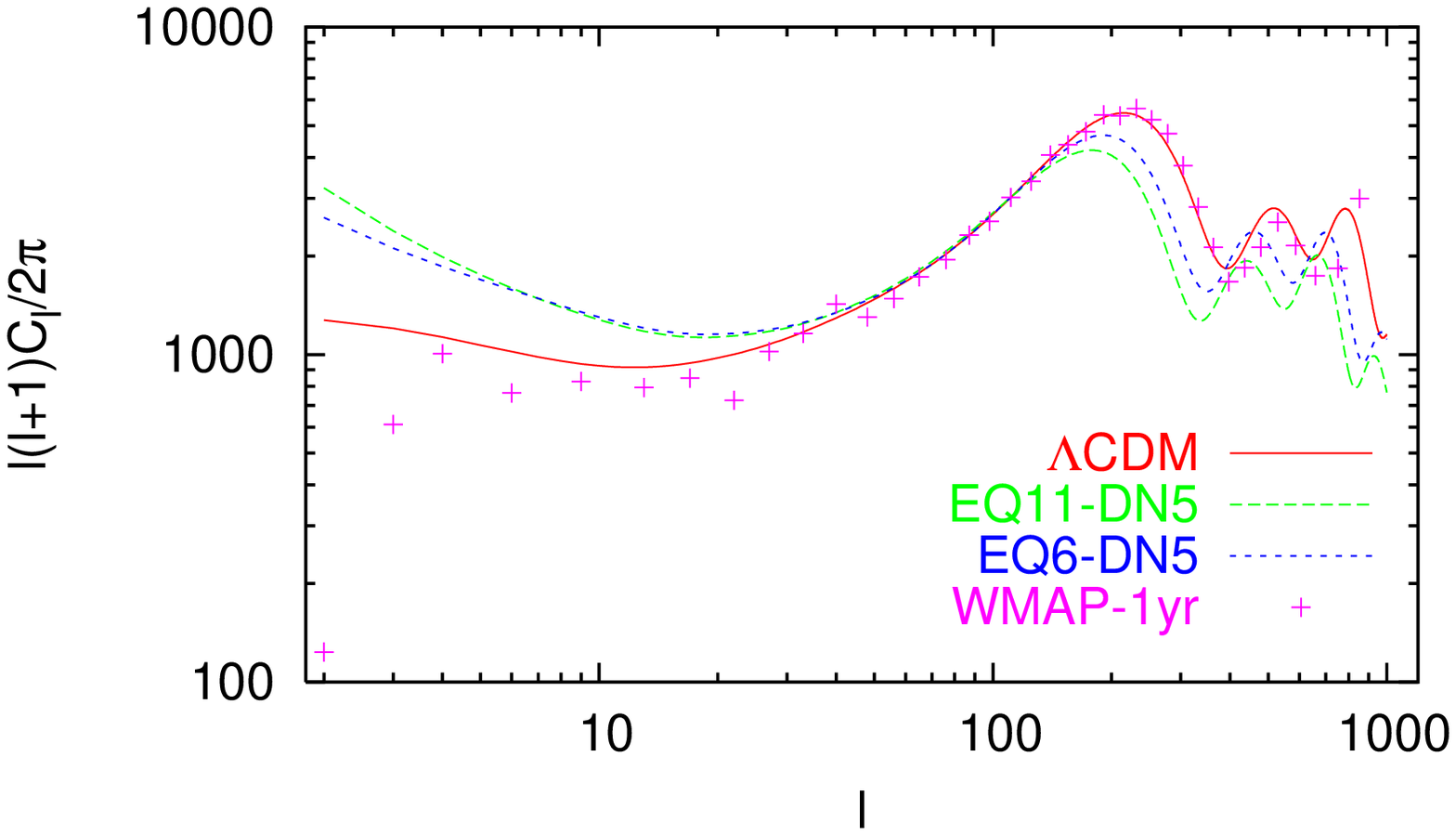,width=6cm}}
 \caption{Other possible signatures of the runaway dilaton
 models. (left) The spectrum of the density perturbation $P_m$
 differs from the power spectrum of the metric perturbation
 because of the modification of the Poisson equation. The lensing
 is sensitive to the deflecting potential $\Phi$ and thus enable to
 measure $\deff$ (Eq.~\ref{defdeff}). (middle) The anisotropic
 stress has two contributions (see Eq;~\ref{B7}) that are important only on
 large scales, the ansiotropic
 stress of the radiation and the contribution of the scalar field.
 The latter is dominant by a large factor in scalar-tensor theories.
 (right) The WMAP data (cross) compared to the runaway dilaton
 model prediction assuming an optical depth of $\tau = 0.16$.
 for $m=11$ (dash) and $m=6$ (dot) compared to a $\Lambda$CDM.}
 \label{figautre}
\end{figure*}

Besides, there are other signatures of these models described in
Fig.~\ref{figautre}. First, the Poisson equation differs from its
standard expression since the scalar field, which is a dark
component contributes to the total energy density. This implies
that $\deff\not=\delta_m$. Indeed on small scales we have found
that they are proportional [see Eq.~(\ref{poissoneff})] and that
the proportionality factor is $k$-independent. This is not the
case on large scales anymore. This implies that the shape of $P_m$
and $P_\eff$ will differ on small $k$. Lensing is sensitive to
$P_\eff$ while galaxy catalog may determine $P_m$. This effect is
of order 10\% and can be hoped to be detected, as first pointed
out in Ref.~\cite{ubpoisson}. Second, in this model the CMB
angular power spectrum will be modified: the integrated
Sachs-Wolfe effect will be amplified and the amplitude of the
secondary peaks will be smaller.

Note however than on large angular scales, $P_\eff\not=P_m$ even
for a pure $\Lambda$CDM model (Fig.~\ref{figautre}). The reason is
that on these scales, one cannot neglect the anisotropic stress of
the radiation (see Eq.~\ref{B7}). Even if the contribution of the
scalar field involves a deviation from the usual Poisson equation,
it is not clear that this effect is not blurred by the effect of
the radiation.

%----------------------------------------------------------------------------------------
\section{Discussion and conclusions}\label{sec8}

We have investigated the imprint of quintessence models, both in
general relativity and scalar-tensor theories, on weak lensing
observations.

For that purpose, we have derived the lensing quantities in a way
that does not assume general relativity. Then, we have developed a
numerical extension to our CMB code~\cite{ru02} that computes the
shear and convergence statistics. This allows us to ``CMB
normalize'' all spectra.

We have investigated various models and reach different
conclusions:
\begin{enumerate}
 \item Concerning quintessence models, density perturbations grow more
 slowly than in a $\Lambda$CDM model. This imply some difference of
 about 10\%-20\% in the linear regime that can be amplified to more
 than 50\% in the non-linear regime, affecting both the shape and
 amplitude of the spectra. This conclusion was reached in various
 previous investigations.  \item For scalar-tensor gravity we have
 shown that, given the constraints in the Solar System, a non-minimal
 quadratic coupling in Jordan frame will not change the prediction by
 more than 1\%.  \item On the other hand, runaway dilaton models that
 incorporate attraction toward general relativity can lead, for the
 same scalar field potential, to change of order 10\% in the
 predictions. Besides the effect on the amplitude, there exists a
 differential effect which modifies the shape of the spectra.  This
 opens some hope to be able track such a coupling.
\end{enumerate}
To illustrate the effect on the growth of density perturbations, one
can look at the value of $\sigma_8$, see
table~\ref{tab:XIQCDM_sigma8}.  Let us also note that the redshift
dependence of the source distribution plays an important role.

There are however some hypothesis and limitations of our analysis that
have to be stressed. First, while investigating the non-linear regime,
we have adopted universal mappings. These mappings are calibrated on
$N$-body simulations for $\Lambda$CDM models. We have argued that
these mappings should hold for quintessence models and in
scalar-tensor theories, as long as we do not enter the strong field
regime. The verification of this hypothesis will require devoted
numerical simulation but we do not expect the order of magnitude of
the effects discussed here to change drastically.

\begin{figure}[htb]
 \centerline{\epsfig{figure=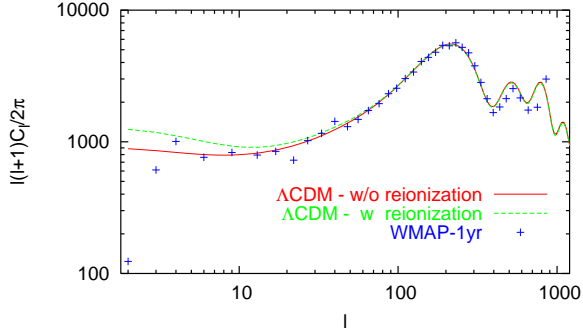,width=8cm}}
 \caption{CMB angular power spectra: the WMAP data (cross) compared to two
 models differing only by their optical depth $\tau = 0$ (solid) and $\tau=0.16$ (dash).
 This shows the influence of the reionization when one chooses to normalize
 to the CMB temperature anisotropies.}
 \label{figCMB}
\end{figure}

A second point to be mentioned is related to the possibility that our
universe has been reionized. Reionization affects the global
normalization of the CMB anisotropies and thus our normalization of
the density parameters. As an example of its effect,
table~\ref{tab:XIQCDM_sigma8} compares the results for vanishing
optical depth (upper) and for an optical depth
$\tau=0.16$. Generically it changes the value of $\sigma_8$ by 10\% to
20\%. But, it will not change the shape of the power spectra. The
effect of the reionization on the normalization can easily been
understood by looking at the CMB angular power spectrum
(Fig.~\ref{figCMB}).

Let us also emphasize an effect that may be of importance while
interpreting lensing data on large scale, as could be obtained by
a wide field imager. The radiation anisotropic stress of the
radiation implies, even if gravity is described by General
Relativity, that the two gravitational potentials are not equal so
that the value of the deflecting potential is not equal to twice
the gravitational potential.

Weak lensing observables combine the effects of the background
properties and the growth of the perturbations and extend from the
linear to non-linear regime. These two regimes are complementary
mainly because of the sensitivity on the time at which the modes
enter the non-linear regime. In conclusion weak lensing survey
appear to be a key observation of the shallow universe to
investigate the nature of the dark energy.

\begin{table}
\caption{\label{tab:XIQCDM_sigma8}Root mean square of the
variance of the matter density contrast on scale of $8 h^{-1}$Mpc
today, $\sigma_8$, assuming vanishing optical depth (upper) or accounting for the reionization with optical depth $\tau =
0.16$ (lower).}
\begin{ruledtabular}
\begin{tabular}{lcccc}
coupling&$m=0$&$m=6$&$m=8$&$m=11$ \\
\hline
no       &$0.90$ &$0.50$&$0.41$&$0.32$  \\
$\xi 1$  &$0.95$ &$0.57$&-&$0.40$ \\
$\xi 2$  &$0.95$ &$0.50$&-&$0.32$ \\
$\xi 3$  & -     &$0.50$&-&$0.32$ \\
$\xi 4$  &$0.95$ &$0.48$&-&$0.31$ \\
DN1      &$0.83$ &-&-&- \\
DN2      &$0.83$ &-&-&- \\
DN3      &$0.83$ &-&-&- \\
DN4      &-      &$0.71$&-&$0.40$ \\
DN5      &-      &$0.53$&-&$0.33$ \\
\hline
no&$1.11$&$0.58$&$0.48$&$0.37$  \\
$\xi 1$&$1.11$&$0.67$&-&$0.47$ \\
$\xi 2$&$1.11$&$0.58$&-&$0.37$ \\
$\xi 3$&-&$0.58$&-&$0.37$ \\
$\xi 4$&$1.11$&$0.56$&-&$0.36$ \\
DN1&$0.96$&-&-&- \\
DN2&$0.96$&-&-&- \\
DN3&$0.96$&-&-&- \\
DN4&-&$0.82$&-&$0.46$ \\
DN5&-&$0.62$&-&$0.39$ \\
\end{tabular}
\end{ruledtabular}
\end{table}

%------------------------------------------------------------------------------

\vskip0.5cm \noindent{\bf Acknowledgements:} CS is partially supported
by the Marie Curie programme ``Improving the Human Research Potential
and the Socio-economic Knowledge Base'' under contract
n.~HMPT-CT-2000-00132. We thank K. Benabed, F.  Bernardeau,
G. Esposito-Far\`ese, Y. Mellier, S. Prunet, I. Tereno, L. van Waerbeke
for discussions. JPU dedicates this work to Tamara.

%%%%%%%%%%%%%%%%%%%%%%%%%%%%%%%%%%%%%%%%%%%%%%%%%%%%%%%%%%%%%%%%%%%%%%%%%%%%%%%%%%%%%%%%%
\appendix

\section{Background equations}\label{appa}

We summarize the equation of the background for the
Lagrangian~(\ref{action}) following the general equations
presented in Ref.~\cite{ru02}. The metric takes the form
\begin{equation}\label{a1}
 \dd s^2 = a^2(\eta)\left[-\dd\eta^2 + \gamma_{ij}\dd x^i\dd x^j\right]
\end{equation}
and we denote the derivation with respect to the conformal time
$\eta$ by a dot and we define the comoving Hubble parameter by
\begin{equation}\label{a2}
 {\cal H} \equiv \dot a/a.
\end{equation}

Let us start by the conservation equations. The Klein-Gordon
equation for the scalar field takes the form
\begin{eqnarray}\label{a3}
 \ddot\varphi + 2\HH\dot\varphi = -a^2 U_{,\varphi} +3 F_{,\varphi}
\left( \HH^2 +\dot\HH + K\right)
\end{eqnarray}
while the conservation equation of the matter fields is given by
\begin{equation}\label{a4}
 \dot\rho + 3\HH(\rho+P)=0.
\end{equation}
We have used the notation
\begin{equation}\label{a5}
 F_{,\varphi} = \frac{\dd F}{\dd\varphi}.
\end{equation}
Note that the matter energy density scales as $\rho \propto
a^{-3}$. This is one of the reasons for which it easier to solve
the system in Jordan frame since in Einstein frame $\rho_* =
A^4\rho$ will behave as $A[\varphi_(a_*)]a_*^{-3}$ which entangles
the evolution of the density with the one of the background.

The Einstein equations give the two Friedmann equations
\begin{eqnarray}
 3F\left(\HH^2 + K\right) &=& 8\pi G_* a^2 \rho
    +\frac{1}{2}\dot\varphi^2  \nonumber\\
   && + a^2 U - 3\HH F_{,\varphi}\dot\varphi \label{a6}\\
 2F\left(\HH^2 -\dot\HH+K\right) &=& 8\pi G_* a^2(\rho + P)
   +\dot\varphi^2 \nonumber\\
   &+& F_{,\varphi\varphi}\dot\varphi^2 +
        F_{,\varphi}\left(\ddot\varphi - 2\HH\dot\varphi\right).
\end{eqnarray}

For completeness, let us give the analog equations in Einstein
frame where the metric takes the form
\begin{eqnarray}
 \dd s_*^2 &=& a_*^2(\eta)\left[-\dd\eta + \gamma_{ij}\dd x^i\dd
 x^j\right].
\end{eqnarray}
The Klein-Gordon equation takes the form
\begin{eqnarray}\label{KG_EF}
 \ddot\varphi_* + 2\HH_*\dot\varphi_* = -a^2_* V_{,\varphi}
 -4\pi G_*\alpha(\varphi_*)(\rho_* - 3P_*)
\end{eqnarray}
and the Friedmann equations become
\begin{eqnarray}
 3\left(\HH^2_* + K\right) &=& 8\pi G_* a_*^2 \rho_*
    + \dot\varphi^2_*  + 2Va_*^2  \label{a6bis}\\
 3\dot\HH_* &=& -4\pi G_* a^2(\rho_* + 3P_*) \nonumber\\
 &&\qquad - 2 \dot\varphi^2_* +2Va_*^2.
\end{eqnarray}
The scale factors of the Jordan and Einstein metrics are related
by
\begin{equation}
 a = A(\varphi_*) a_*
\end{equation}
so that the two cosmic times are related by
\begin{equation}
 \dd t = A(\varphi_*) \dd t_*,\qquad
 \dd\eta = \dd\eta_*.
\end{equation}
This implies that redshifts in both frames are related by
\begin{equation}
 1 + z = \frac{A(\varphi_*)}{A(\varphi_*^0)}(1+z_*)
\end{equation}
and that the physical length associated with to
a comoving length $\ell_c$ are connected by
\begin{equation}
 \ell^{\rm phys}_* = \frac{A(\varphi_*^0)}{A(\varphi_*)}\ell^{\rm phys}.
\end{equation}

It will be convenient to decompose the energy-density of the
scalar field as
\begin{equation}
 \rho_\varphi = \rho_{\rm MC} + \rho_{\rm F}
\end{equation}
with
\begin{eqnarray}
 \rho_{\rm MC}a^2 &=& \frac{1}{2}\dot\varphi^2  + a^2U,\\
 \rho_{\rm F}a^2  &=& -3\HH F_\varphi\dot\varphi.
\end{eqnarray}
Then, we can define the density parameters as
\begin{equation}\label{a_defOm}
 \Omega_f = \frac{8\pi G_\eff\rho_fa^2}{3\HH^2},\quad
 \Omega_\varphi = \frac{\rho_\varphi a^2}{3F\HH^2},\quad
 \Omega_K = -\frac{K}{\HH^2}
\end{equation}
so that the the Friedman equations takes the form
\begin{equation}
 \sum_f\Omega_f + \Omega_\varphi + \Omega_K = 1.
\end{equation}
Note that we could have defined
\begin{equation}
 \bar\Omega_X = \frac{8\pi G_{\eff,0}\rho_Xa^2}{3\HH^2},\quad
 \bar\Omega_\varphi = \frac{\rho_\varphi a^2}{3F_0\HH^2},\quad
 \bar\Omega_K = -\frac{KF}{\HH^2F_0}.
\end{equation}
The two sets of density parameters agrees today and are related by
\begin{equation}
 \bar\Omega = \Omega F/F_0.
\end{equation}
We deduce that in a matter universe the Friedmann equation  can be
rewritten as
\begin{equation}
\frac{{\cal H}^2}{{\cal H}^2_0} = \Omega_{m,0}\frac{F_0}{F}(1+z)
 + \Omega_{\varphi,0}
 \frac{F_0}{F}\frac{\rho_\varphi}{\rho_{\varphi,0}}(1+z)^{-2}
 -\Omega_{K,0}
\end{equation}
To finish, let us define the function
\begin{equation}
 E(z) = (1+z){\cal H}(z)/{\cal H}_0
\end{equation}
that is explicitly  given by
\begin{equation}
 E^2 = \Omega_{m,0}\frac{F_0}{F}(1+z)^3
 + \Omega_{\varphi,0}\frac{F_0}{F}\frac{\rho_\varphi}{\rho_{\varphi,0}}
 -\Omega_K^0(1+z)^{2}.
\end{equation}

\section{Perturbation equation}\label{appb}

The general gauge invariant perturbation equations in
scalar-tensor that are being integrated in the CMB code have been
presented in Ref.~\cite{ru02}. We just summarize the scalar mode
equations in the case of a single fluid in a universe with
Euclidean spatial sections.

In Newtonian gauge, the metric takes the form
\begin{equation}\label{B1}
 \dd s^2 = a^2(\eta)\left[-(1+2\phi)\dd \eta^2 + (1-2\psi)\dd\bm{x}^2
 \right].
\end{equation}
The fluid velocity perturbation is decomposed as
\begin{equation}\label{B2}
 \delta u_\mu = a(-\phi,\partial_k V).
\end{equation}
$\deltaN$ is the density perturbation in Newtonian gauge and we
introduce the density perturbation in comoving gauge as
\begin{equation}\label{B4}
 \delta = \deltaN - 3{\cal H}(1+w) V
\end{equation}
where $w=P/\rho$.

The fluid conservation equation is given by
\begin{equation}\label{B5}
 \left(\frac{\deltaN}{1+w}\right)^. = -\Delta V + 3\dot\psi -3{\cal H}\frac{w}{1+w}\Gamma
\end{equation}
while the Euler equation takes the form
\begin{equation}\label{B6}
 \dot V + \HH V = -\phi -\frac{c_s^2}{1+w}\delta + \frac{w}{1+w}\left[\Gamma + \frac{2}{3}\Delta\bar\pi\right]
\end{equation}
where $c_s^2 = \dd P/\dd\rho$. $\Gamma$ is the entropy
perturbation and $\bar\pi$ the anisotropic stress.

Among the four independent Einstein equations, we can retain
\begin{equation}\label{B7}
 \psi - \phi = 8\pi G_*P\bar\pi +
 \frac{F_{\varphi}}{F}\delta\varphi,
\end{equation}
\begin{eqnarray}\label{B8}
 2F(\dot\psi+ \HH\phi)+\dot F\phi &=&
   -8\pi G_*\rho(1+w)a^2V \nonumber\\
   && +\dot\varphi\delta\varphi
   + \delta\dot F -\HH\delta F,
\end{eqnarray}
\begin{eqnarray}\label{B9}
 &&2F{\Delta}\phi + (\dot\varphi^2 - 3\HH\dot F)\phi -3\dot
 F\dot\phi \nonumber \\
 &&\qquad= 8\pi G_*\rho a^2\delta -8\pi G_*P{\Delta}\bar\pi  \nonumber\\
 &&\qquad - F_{\varphi}\left[{\Delta}
          +3\left(\HH^2 + \frac{\dot F^2}{F^2}\right)\right]\delta\varphi\nonumber\\
 &&\qquad +\left(U_{\varphi}a^2+3\HH\dot\varphi\right)\delta\varphi
          +\dot\varphi\delta\dot\varphi + 3\frac{\dot
          F}{F}\delta\dot F.
\end{eqnarray}
The Klein-Gordon equation for the evolution of the the scalar
field is then given by
\begin{eqnarray}\label{B10}
 && \delta\ddot\varphi + 2\HH\delta\dot\varphi
 -\left[{\Delta} + 3\dot\HH F_{\varphi\varphi}
 -U_{\varphi\varphi}a^2\right]\delta\varphi\qquad \nonumber\\
 &&\qquad = (\dot\phi+3\dot\psi)\delta\dot\varphi
 -2a^2\phi U_\varphi\nonumber\\
 &&\qquad -\left[{\Delta}(\phi-2\psi) + 3(\ddot\psi+3\HH\dot\psi+\HH\dot\phi)\right]F_\varphi.
\end{eqnarray}
In all these equations, we have
$$
 \dot F = F_\varphi\delta\dot\varphi,\qquad \ddot F =
 F_{\varphi\varphi}\dot\varphi^2 + F_\varphi\delta\ddot\varphi, \qquad
 \delta F = F_\varphi\delta\varphi.
$$

%----------------------------------------------------------------------------------------

\end{document}